\documentclass[a4paper,12pt,english]{article}

\usepackage[dvipsnames]{xcolor}
\usepackage{amsfonts,bm,amssymb,euscript,array,babel,cite}
\usepackage{mathtools}
\usepackage{tikz-cd}
\usepackage{amsmath,pifont}
\usepackage{tensor,accents}
\usepackage{hhline}
\usepackage[amsmath]{empheq}
\usepackage{hyperref}
\usepackage{cancel}
\usepackage{mathrsfs}
\usepackage{pdfsync}
\usepackage{multirow}
\usepackage[most]{tcolorbox}
\usepackage{tikz}

\usepackage{framed}

\newcommand{\rd}{\mathrm{d}}

\newsavebox{\mysaveboxM}
\newsavebox{\mysaveboxT}

\makeatletter

\newcommand{\dd}{\mathrm{d}}
\newcommand{\DD}{\mathrm{D}}

\newcommand{\w}{\wedge}

\newcommand{\be}{\begin{equation}}
\newcommand{\ee}{\end{equation}}
\newcommand{\sfrac}[2]{{\textstyle\frac{#1}{#2}}}
\def\nn{\nonumber}
\def \bea{\begin{eqnarray}} 
\def\eea{\end{eqnarray}}
\newcommand{\mf}{\mathfrak}
\def\mc{\mathcal}

\def\bi{\begin{itemize}} 
\def\ei{\end{itemize}}

\def\E{\textit{\tiny{E}}}

\newcommand{\sbullet}{%
  \hbox{\fontfamily{lmr}\fontsize{.8\dimexpr(\f@size pt)}{0}\selectfont\textbullet}}
\DeclareRobustCommand{\mathbullet}{\accentset{\sbullet}}

\makeatother

\newtheorem{theorem}[equation]{Theorem}

\newtheorem{prop}[equation]{Proposition}
\newtheorem{defn}[equation]{Definition}

\def\a{\alpha} \def\b{\beta} \def\g{\gamma} \def\G{\Gamma} \def\d{\delta} \def\D{\Delta}
\def\e{\epsilon} 
   \def\k{\kappa}
\def\l{\lambda}  \def\m{\mu}
\def\n{\nu} \def\o{\omega}   \def\r{\rho}
\def\s{\sigma} \def\S{\Sigma}

\def\one{\mbox{1 \kern-.59em {\rm l}}}

\numberwithin{equation}{section}

\begin{document}

\makeatother
\parindent=0cm
\renewcommand{\title}[1]{\vspace{10mm}\noindent{\Large{\bf #1}}\vspace{8mm}} \newcommand{\authors}[1]{\noindent{\large #1}\vspace{5mm}} \newcommand{\address}[1]{{\itshape #1\vspace{2mm}}}

\begin{titlepage}

\begin{flushright} 
RBI-ThPhys-2024-01
\end{flushright} 

\begin{center}

\title{ {\Large {Geometric BV for 
 twisted Courant sigma models \\[5pt]  and the BRST power finesse}}}

 \vskip 3mm

 \authors{\large Athanasios {Chatzistavrakidis,$^{\color{Green}{\clubsuit}}$ Noriaki Ikeda,$^{\infty}$ Larisa Jonke$\,^{{\color{Green}{\clubsuit}},\mho}$}
  }

 \vskip 3mm
 
  \address{ $^{\color{Green}{\clubsuit}}$
  	 Division of Theoretical Physics, Rudjer Bo\v skovi\'c Institute \\ Bijeni\v cka 54, 10000 Zagreb, Croatia 

    \ 

    $^{\infty}$Department of Mathematical Sciences, Ritsumeikan University \\ Kusatsu, Shiga 525-8577, Japan

    \ 

    $^{\mho}$School of Theoretical Physics, Dublin Institute for Advanced Studies \\
10 Burlington Road, Dublin 4, Ireland
  }

\vskip 2cm

\begin{abstract}
    We study twisted Courant sigma models, a class of topological field theories arising from the coupling of 3D 0-/2-form BF theory and Chern-Simons theory and containing a 4-form Wess-Zumino term. They are examples of theories featuring a nonlinearly open gauge algebra, where products of field equations appear in the commutator of gauge transformations, and they are reducible gauge systems.
    We determine the solution to the master equation using a technique, the BRST power finesse, that combines aspects of the AKSZ construction (which applies to the untwisted model) and the general BV-BRST formalism. This allows for a geometric interpretation of the BV coefficients in the interaction terms of the master action in terms of an induced generalised connection on a 4-form twisted (pre-)Courant algebroid, its Gualtieri torsion and the basic curvature tensor. It also produces a frame independent formulation of the model. We show, moreover, that the gauge fixed action is the sum of the classical one and a BRST commutator, as expected from a Schwarz type topological field theory.
\end{abstract}

\end{center}

\vskip 2cm

\end{titlepage}

\setcounter{footnote}{0}
\tableofcontents

\newpage 

\section{Introduction} 
\label{sec1}

The Courant sigma model is a three-dimensional topological field theory that was first constructed as a nontrivial coupling between 3D BF theory of $0$-form and $2$-form and Chern-Simons theory \cite{Ikeda:2002wh}. Aspects of its quantization as a model of topological open membranes were discussed in Refs. \cite{Hofman:2002jz,Hofman:2002rv} and its construction as an AKSZ theory appears in Ref. \cite{Roytenberg:2006qz}. The local gauge structure of the model is governed by the local form of the axioms of a Courant algebroid or equivalently by the homological vector field on a differential graded (dg) symplectic supermanifold of degree 2 \cite{dimaphd,Roytenberg:2002nu}. The model has appeared in several instances in physics, especially in the context of nongeometric string backgrounds and T-duality \cite{Halmagyi:2008dr,Mylonas:2012pg,Chatzistavrakidis:2015vka,Bessho:2015tkk,Arvanitakis:2023dud}.   

A direct generalization of the Courant sigma model appeared in the context of first class constrained Hamiltonian systems  together with a Wess-Zumino term supported on a 4D world volume \cite{Hansen:2009zd}. This is associated with membranes in nontrivial 4-form flux and it is the direct analogon of the transition from Poisson to Wess-Zumino-Witten or 3-form twisted Poisson sigma models in 2D \cite{Klimcik:2001vg}. The gauge structure of this model is governed by a particular type of \emph{pre-Courant} algebroids, the latter being relaxed structures originating from Courant algebroids by dropping the Jacobi identity of the Dorfman bracket \cite{Vaisman:2004msa}. Hence we refer to the corresponding 3D topological field theory as a (4-form) twisted Courant sigma model. 

Both Courant and twisted Courant sigma models feature all three characteristics of a gauge system that requires treatment within the BV-BRST formalism \cite{Batalin:1981jr}. Their gauge algebra is closed only on the stationary surface (on-shell), the coefficients of the gauge algebra are field-dependent and not constant and moreover there are reducible gauge symmetries simply because there are 2-forms in the theory. One can refer to, for instance, the textbook \cite{HTbook} for the general theory of gauge systems and the review \cite{Barnich:2000zw} for more details on local BRST cohomology. Therefore these constitute a class of examples of the simplest theories that feature all these three properties and can thus serve as prototypes to study several relations between field theory and generalised geometric or higher structures. 

On the other hand, there is an important difference between the twisted and untwisted cases. The target space of the untwisted case is a QP manifold and its master action within the BRST formalism need not be found by the usual complicated procedure; instead it acquires a geometric form within the AKSZ construction \cite{Alexandrov:1995kv}, see also the earlier fundamental papers \cite{Witten:1990wb} and \cite{Schwarz:1992nx}. This is no longer true for the twisted model. The compatibility of the Q and P structures is obstructed by the 4-form and the AKSZ construction does not apply per se. One could think that it would be enough to perform the AKSZ construction in one dimension higher and recover the twisted Courant sigma model as a boundary theory. However, as the simpler and lower-dimensional analogon of the twisted Poisson sigma model shows, this is not as straightforward as one might expect. Indeed the master action of the twisted Poisson sigma model is significantly more complicated than what one would naively get as a boundary of the Courant sigma model, notably containing quadratic terms in the components of the 3-form flux \cite{Ikeda:2019czt}. In fact there exists a much larger class of 2D topological field theories based on Dirac structures that share this feature \cite{Chatzistavrakidis:2022wdd}. 

The above discussion presents a dichotomy between the general BRST formalism and the AKSZ construction. The former is completely general and universally applicable, yet technically complicated and lacking a geometric intuition, whereas the latter is remarkably simple and with a geometric interpretation by construction, but it is not general. One would then like to develop methods that keep the positive properties and dispense with the negative ones, or at least come close to that. Twisted topological sigma models in various dimensions is one possible arena to develop and test such methods. As already mentioned, the 3D twisted Courant sigma model is perhaps the simplest class that is general enough to allow us to draw such lessons. It is also welcome that a subclass of this, called twisted R-Poisson sigma models in 3D, was completely worked out with traditional methods, albeit in a very complicated and uninspiring way \cite{ChSI}. It will be used here as a benchmark example to test the more systematic method we will propose.

According to the above, the first question we would like to answer is 
\begin{itemize}
    \item What is the master action of a 4-form twisted Courant sigma model?
\end{itemize}
Clearly it is not only the end result that we are interested in, but also the method to derive it. In this respect, a crucial role is played by particular geometrical quantities that we will encounter and use on the way. To motivate this further, we note that although for any AKSZ sigma model the full action is of course covariant, the coefficients of the various interaction terms in the master action do not exhibit manifest target space covariance term by term. Although there is nothing wrong with this, it is in fact advantageous to determine the geometric meaning of these coefficients. One reason for this is that it provides a clear path toward working out the twisted case.{\footnote{Another reason is that this approach  can be related to the recently developed theory of representations up to homotopy for Lie $n$-algebroids \cite{Jotz}, whose relation to the BRST/AKSZ formalism and twisted sigma models will be studied elsewhere. We will not discuss this further in the present paper.}} It also provides a globalization of the theory and its master action, with manifest target space covariance and without need to consider a local coordinate patch. Thus the second main question that we are interested in is: 
\begin{itemize}
    \item What is the generalised geometric meaning of the coefficients in the interaction terms of the master action for the (twisted or not) Courant sigma model?
\end{itemize}
To answer this question, we need to consider generalised connections on (twisted) Courant algebroids. We will then see that the main player that gives the sought after geometric interpretation is an induced $E$-connection on $E$, where $E$ is the vector bundle of the (twisted) Courant algebroid, together with two associated tensors. One is its torsion and the other is the so-called basic curvature. The former was introduced by Gualtieri in \cite{Gualtieri:2007bq}, after noticing that the ordinary definition of the torsion tensor is not linear in all its arguments for generalised connections. The basic curvature for Courant algebroids was introduced in   \cite{Chatzistavrakidis:2023otk} following a similar spirit, namely generalizing the notion of basic curvature for Lie algebroids \cite{Blaom,Crainic}\footnote{The concept is directly related to the notion of basic connections, see Refs. \cite{CF,GSM}.} to make it tensorial in the realm of generalised connections on Courant algebroids. It was also found earlier in \cite{Jotz} from a graded geometric viewpoint. 

Understanding the geometric meaning of these coefficients will prove important in ``finessing'' the master action without solving the master equation order by order in the number of antifields, yet in a slightly different way than the ASKZ construction such that the twisted case is also included. This approach is admittedly more complicated than AKSZ, yet simpler than the traditional one and it reveals some aspects of the formalism that remain hidden in AKSZ. Since the method we will propose involves computing powers---in the 3D case, the square and the cube---of the only on-shell closed BRST transformation (the longitudinal differential in \cite{HTbook}) in a specific way to be explained in the main text, we call this \emph{the BRST power finesse}. 

The rest of the paper is organised in the following way. In Section \ref{sec2} we recall some basic facts about Lie and pre-Courant algebroids and their description as dg manifolds. We pay special attention to the transformation of their structural data under change of coordinates and introduce $E$-connections on Lie and pre-Courant algebroids, together with their corresponding torsion and basic curvature tensors. This provides a covariant perspective that sets the stage for manifest target space covariance in the field theory realization of these structures. We also discuss twists and their particular role in each case, with emphasis in drawing a parallel between twisted Poisson and twisted Courant structures and also discussing their conceptual difference. 

In Section \ref{sec3} we turn our attention to the BRST formalism for twisted Courant sigma models. First we discuss in detail the classical action of the theory and the structure of its gauge algebra, showing also how it fits within a broader graded geometric perspective to higher gauge theories. Trivial gauge transformations are handled with care and the openness of the gauge algebra is demonstrated. Notably we reveal and elaborate on the fact that this is a prototype theory that exhibits the phenomenon of nonlinear openness, its gauge algebra containing terms quadratic in field equations. In Section \ref{sec32} we present the on-shell closed BRST transformations from the general point of view of dg manifolds. The main result is Proposition \ref{prop s0} which gives the on-shell closed BRST transformation of all fields, ghosts and ghosts for ghosts collectively in terms of the components of the homological vector field and its derivatives. Section \ref{sec33} contains a complete analysis of the tensorial version of the previously found transformations. We discuss in details the transformation of fields under changes of frame and present the fully tensorial BRST transformation in Proposition \ref{prop s0 tensorial}, which yields a direct relation between the BRST transformation and the Gualtieri torsion and Courant algebroid basic curvature tensors. This gives an answer to the second main question we posed above.

Section \ref{sec4} contains the main results of the paper and the answer to the first question we posed. After determining the square of the on-shell closed BRST transformation explicitly and introducing antifields, we explain a certain structure of higher powers of the BRST transformation that can be used to express the master action. In the present case the highest power is 3. We formulate Proposition \ref{prop AKSZ} which describes a rewritting of the AKSZ-BV master action for the untwisted Courant sigma model in terms of the elementary operations we introduced. This is completely the same as the expanded form of the AKSZ action in terms of number of antifields appearing in \cite{Roytenberg:2006qz}, but it can be used to determine the correct master action for the twisted model as well and moreover bring it in a manifestly covariant form term by term. The first statement is captured in Theorem \ref{Thm} which gives a compact form of the minimal solution of the classical master equation for the 4-form twisted Courant sigma model. The expanded (and complicated) form then follows simply and it is presented in Appendix \ref{appd} both in non manifestly and manifestly covariant forms. At the end of Section \ref{sec4} we also comment on the quantum master equation, solved by the same master action described earlier.

To complete the picture we present the form of the gauge fixed master action in Section \ref{sec5}. Notably, we show that the result can be brought in the form of the classical action plus a BRST commmutator thus proving that the model is a topological field theory of Schwarz type, as expected. Section \ref{sec6} contains a discussion on our results with an outlook to future work. There are four Appendices that contain supplementary material. Appendix \ref{appa} details aspects of the gauge invariance of the classical action of the twisted Courant sigma model and it is used to examplify some statements in Sections \ref{sec2} and \ref{sec3}.  Appendix \ref{appb} contains the local form of the on-shell closed BRST transformations and of their squares, which appear with less details and in more abstract form in Sections \ref{sec3} and \ref{sec4}. Appendix \ref{appc} is an annex to the BRST power finesse that explains several details of more technical nature. Finally, Appendix \ref{appd} contains the expanded form of the complete minimal solution of the classical master equation, also in a fully coordinate free form.

\section{Dg manifolds, twists and target space covariance} 
\label{sec2}

In field theory we often encounter target spaces which are differential non-negatively graded (dg$_+$) manifolds,---a.k.a. NQ manifolds.  These are graded (super)manifolds ${\cal M}$ endowed with a homological vector field $Q\in \mf X(\cal M)$ that satisfies 
\be \label{HVF}
Q^2:=\frac 12 \{Q,Q\}=0\,.
\ee 
We refer to \cite{Cattaneo:2010re,Qiu:2011qr,Ikeda:2012pv} for more precise definitions.
Homological vector fields offer an elegant and intuitive way to encode algebraic and geometric structures and their twists. For example, a Lie algebra $\mf{g}$, which is a vector space equipped with a skew-symmetric bilinear operation satisfying the Jacobi identity, can be viewed as a dg manifold $\mf{g}[1]$ whose homological vector field is 
\be  
Q_{\text{CE}}=-\frac 12 C^{a}_{bc}a^{b}a^{c}\frac{\partial}{\partial a^{a}}\,,
\ee 
where $a^{a}$ are odd (degree 1) coordinates on $\mf{g}[1]$. Eq. \eqref{HVF} implies that $C^{a}_{bc}$ are the structure constants of the Lie algebra. $Q_{\text{CE}}$ is the Chevalley-Eilenberg differential of ordinary Lie algebra cohomology. In the rest of this section we will take a new look to this approach for Lie, Courant and pre-Courant algebroids, emphasizing covariance and connections and elucidating the role of twists in case twisted geometric structures are encountered. 

In the following we always work with non-negative grading and therefore we shall simply refer to dg manifolds or Q manifolds without further reference to non-negative grading and moreover we shall denote as Q$n$-manifolds those with highest degree $n$.

\subsection{Revisiting Lie algebroids and Q1 manifolds} \label{sec21}

The above correspondence between a Lie algebra and a dg manifold is also established for Lie algebroids \cite{Vaintrob}. A Lie algebroid is a triple $(E,[\cdot,\cdot]_{\E},\rho)$ of a vector bundle $E$ over a smooth manifold $M$, of a Lie bracket on its sections $e\in \G(E)$ and of a smooth bundle map (the anchor) $\rho: E\to TM$ such that
\be 
[e,fe']_{\E}=f[e,e']_{\E}+\rho(e)(f)\, e'\,, \quad f\in C^{\infty}(M)\,. 
\ee 
It follows that $\rho$ is a homomorphism of bundles, 
\be 
\rho([e,e']_\E)=[\rho(e),\rho(e')]\,.
\ee 
In a local basis of $E$, say $\mf e_a$, and coordinates $x^{\m}$ on the base $M$, the bracket is given in terms of some structure functions $C_{ab}^c(x)$ and the map $\rho$ has components $\rho_{a}{}^{\m}(x)$ such that  
\be 
[\mf e_a, \mf e_b]_{\E}=C_{ab}^{c}(x)\mf e_c\,,
\quad 
\rho(\mf e_a)=\rho_{a}{}^{\m}(x)\frac{\partial}{\partial x^{\m}}\,.
\ee 
The statement that $E$ is a Lie algebroid constrains the functions $\rho_a{}^{\m}$ and $C_{ab}^c$ to satisfy certain conditions. The point now is that the very same conditions are obtained on the other side where we have the graded Q1 manifold ${\cal M}=E[1]$ by shifting the fiber degree of the vector bundle $E$ by 1 and also the homological vector field $Q\in \mf{X}(E[1])$, 
\be \label{QLAnoncov}
Q=\rho_a{}^{\m}(x) a^{a}\frac{\partial}{\partial x^{\m}} -\frac 12 C_{bc}^a(x) a^ba^{c}\frac{\partial}{\partial a^{a}}\,,
\ee 
where $x^{\m}$ and $a^{a}$ are degree 0 and 1 coordinates of the graded manifold respectively. The space of vector fields on ${\cal M}$ is spanned by the derivations $\partial/\partial x^{\m}$ and $\partial/\partial a^{a}$ of degrees $0$ and $-1$ respectively. That the vector field $Q$ is homological, equivalently that $(E[1],Q)$ is a dg manifold, is then an if and only if statement with $E$ being a Lie algebroid. 

From a geometric point of view, there is a nuisance in the above considerations. This is that the structure functions $C_{ab}^{c}(x)$ do not transform homogeneously under a change of basis of the vector bundle $E$, and therefore they are not tensorial. 
In particular, assuming a change of basis on the vector bundle 
\be 
\mf e_a=\Lambda_{a}^b(\widetilde x)\,\widetilde{\mf e}_b
\ee 
the structure functions transform according to the rule 
\be 
C_{ab}^c=(\Lambda^{-1})^c_{c'}\Lambda_a^{a'}\Lambda_b^{b'}\widetilde{C}_{a'b'}^{c'}+(\Lambda^{-1})^c_{c'}\widetilde\rho_{a'}{}^{\m}(\Lambda_a^{a'}\widetilde{\partial}_{\m}\Lambda_b^{c'}-\Lambda_{b}^{a'}\widetilde{\partial}_\m\Lambda_a^{c'})\,,
\ee 
where due to the coordinate transformation $x^{\m}=x^{\m}(\widetilde{x})$, the partial derivatives and the coefficients of the anchor map transform as 
\bea 
\frac{\partial}{\partial x^\m}&=&\frac{\partial\widetilde{x}^\n}{\partial x^\m}\frac{\partial}{\partial\widetilde{x}^\n}:=\Lambda_\m^\n \frac{\partial}{\partial\widetilde{x}^\n} \,, \\[4pt]
\rho_a{}^{\m}&=&\Lambda_a^{a'}(\Lambda^{-1})_{\m'}^\m\widetilde{\rho}_{a'}{}^{\m'}\,.
\eea 
Although there is nothing wrong with this per se, both for aesthetic reasons but also to account for the invariant geometric form of the BV/BRST action for field theories with underlying Lie algebroid structure, it is desirable to rewrite the vector field $Q$ in a different form. With some prior knowledge, we would like to express the vector field in terms of the quantity 
\be 
T_{bc}^{a}=2\omega_{[bc]}^{a}-C_{bc}^{a}\,,
\ee 
where in the non-graded differential geometric picture $\omega^{a}_{bc}$ are coefficients of an $E$-connection acting on the Lie algebroid $E$ itself. An $E$-connection on $E$ is a map 
\bea 
\nabla^{\E}: \G(E)\times \G(E)&\to& \G(E) \nn \\[4pt] (e,e')&\mapsto& \nabla^{\E}_{e}e'\,, 
\eea 
such that the following homogeneity and linearity conditions hold: 
\bea  
\nabla^\E_{fe+e'}\, e''&=&f\, \nabla^{\E}_{e}e''+\nabla^\E_{e'}e''\,, \\[4pt] 
\nabla^\E_e(fe'+e'')&=&f\nabla^\E_ee'+\nabla^\E_ee'' +\rho(e)f e'\,.
\eea 
In terms of the local basis, the connection coefficients are obtained as 
\be
\label{omegagen} 
\nabla^{\E}_{{\mf e}_b}{\mf e}_{c}= \omega_{bc}^{a}{\mf e}_a\,,
\ee 
and they transform in the usual inhomogeneous way:
\be 
\omega^{a}_{bc}=(\Lambda^{-1})^{a}_{a'}\Lambda_b^{b'}\Lambda_c^{c'}\widetilde\omega^{a'}_{b'c'}+(\Lambda^{-1})^{a}_{a'}\Lambda_b^{b'}\widetilde{\rho}_{b'}{}^{\m}\widetilde{\partial}_{\m}\Lambda_c^{a'}\,.
\ee
The transformation is such that it counterbalances the one of the structure functions, allowing for a homogeneous transformation for $T^{a}_{bc}$.
One may then immediately recognize that the quantities $T_{bc}^{a}$ are the coefficients of the Lie algebroid torsion ($E$-torsion) of this $E$-connection on $E$, which is defined as 
\be 
T^{\E}(e,e')=\nabla^{\E}_{e}e'-\nabla^{\E}_{e'}e-[e,e']_{\E}\,,
\ee 
in other words 
$T^{a}_{bc}=\langle T^{\E}(\mf e_b,\mf e_c), \mf e^a\rangle\,,$ 
where $\mf e^{a}$ is the dual basis of the vector bundle $E^{\ast}$. Clearly $T^{\E}\in \G(\w^2E^{\ast}\otimes E)$ is a tensor. 
Then we can rewrite the vector field $Q$ in a fully covariant form as 
\be \label{QLAcovgen}
Q=a^{a}\, \DD^{(0)}_a+\frac 12 \, T^{a}_{bc}\, a^ba^c\,  \DD^{(-1)}_a\,,
\ee 
where we have defined the natural degree $0$ and degree $-1$ derivations 
\bea  
\DD^{(0)}_{a}=\rho_{a}{}^{\m}\frac{\partial}{\partial x^{\m}}-\omega^{c}_{ab}a^b\frac{\partial}{\partial a^c}\,, \qquad  
\DD^{(-1)}_{a}=\frac{\partial}{\partial a^{a}}\,.
\eea  
These are natural in the sense that they satisfy 
\bea  
&& \DD^{(0)}_a x^{\m}=\rho_{a}{}^{\m}\,, \quad \DD^{(0)}_a a^{b}=-\,\o^b_{ac}a^{c}\,, \\[4pt] && \DD^{(-1)}_a x^{\m}=0\,, \quad \,\,\DD^{(-1)}_{a}a^b=\d_a^b\,.
\eea
In particular, keeping in mind that the degree $1$ coordinate $a^{b}$ on the graded side corresponds to the local basis of the dual bundle $E^{\ast}$ on the non-graded side, it should transform as a section of $E^{\ast}$, which is indeed the case since we find $-\omega^b_{ac}$ on the right-hand side.
It should be emphasized that this is a rewriting of the previous expression in covariant terms, no additional terms have been added for covariantization; in other words, the vector field of \eqref{QLAnoncov} is already covariant albeit not in a manifest way.{\footnote{When an $E$-connection on some vector bundle $V$ is considered, the graded geometric formulation goes through the dg manifold $E[1]\oplus V$ equipped with a suitable homological vector field. This is not what we do here and in the rest of this section.}} 

There exists, moreover, a special case where the above considerations simplify further. This corresponds to the choice of $E$-connection being  induced by an ordinary vector bundle connection $\nabla$ on $E$, possibly with torsion, through the anchor. Denoting this induced connection with a solid dot over it, it is 
\be \label{bullet}
\mathbullet{\nabla}^{\E}_ee'=\nabla_{\rho(e)}e'\,.
\ee 
Although special, this is the connection of interest in topological sigma models as will become transparent in the ensuing. 
It is convenient to define the difference of an arbitrary $E$-connection on $E$ from this induced one,
\be 
\phi(e,e'):=(\nabla^{\E}_e-  \mathbullet\nabla^\E_e)e'\,.
\ee 
This $\phi$ is evidently an endomorphism and its components are 
$ 
\phi^{c}_{ab}=\o^{c}_{ab}-\rho_{a}{}^{\m}\,\o^{c}_{\m b}\,.
$ 
Equivalently, we consider general $E$-connections of the form 
\be \nabla^{\E}=\mathbullet{\nabla}^{\E}+\phi\,.
\ee
The torsion tensor for the induced $E$-connection becomes 
\be 
\mathbullet{T}^{a}_{bc}=2 \rho_{[b}{}^{\m}\omega_{\m c]}^a-C_{bc}^{a}\,,
\ee  
and the homological vector field is written as 
\be 
\label{QLAcovtriv} Q=a^{a} \rho_{a}{}^{\m}\, \DD^{(0)}_{\m}+\frac 12 \, \mathbullet{T}^{a}_{bc}\, a^ba^c\,  \DD^{(-1)}_a\,,
\ee 
with 
$ 
\DD^{(0)}_{\m} =\frac{\partial}{\partial x^{\m}} -\omega_{\m b}^{c}a^{b}\frac{\partial}{\partial a^c}\,.
$ More details on this are found in \cite{Chatzistavrakidis:2023otk}.

\subsection{Poisson vs. twisted Poisson structure} \label{PvstP}

It is useful to recall a nontrivial example that can moreover be twisted. Consider the Lie algebroid on the cotangent bundle $E=T^{\ast}M$ of a Poisson manifold $(M,\Pi)$ with Poisson bivector $\Pi\in \G(\w^2TM)$. The Lie bracket is the Koszul-Schouten bracket of 1-forms, given as  
\be 
[\eta,\eta']_{\text{KS}}={\cal L}_{\Pi^{\sharp}(\eta)}\eta'-{\cal L}_{\Pi^{\sharp}(\eta')}\eta-\dd\left(\Pi(\eta,\eta')\right)\,,\quad \eta,\eta'\in \Omega^1(M)\,,
\ee 
and the anchor map   $\Pi^{\sharp}:T^{\ast}M\to TM$ is induced by the Poisson structure $\Pi$. In addition, choose a torsion-free ordinary connection $\mathring{\nabla}$ on $M$ and the induced $E$-connection on $E$ given in \eqref{bullet}. Then the $E$-torsion is given as 
\be \label{Etor}
\mathbullet{T}^{\m\n}_{\r}=-\mathring{\nabla}_{\rho}\Pi^{\m\n}\,.
\ee 
The homological vector field in covariant form turns out to be 
\be 
Q=a_{\m}\Pi^{\m\n} \left(\frac{\partial}{\partial x^{\n}}+\mathring\G^{\s}_{\r\n}a_{\s}\frac{\partial}{\partial a_{\r}}\right)-\frac 12 \mathring{\nabla}_{\rho}\Pi^{\m\n}a_{\m}a_{\n}\frac{\partial}{\partial a_{\r}}\,,
\ee 
where $a_{\m}$ is the degree $1$ coordinate and $\partial/\partial a_{\m}$ is the degree $-1$ derivation. This rewriting appeared without much explanation in Ref. \cite{Chatzistavrakidis:2021nom}. 

In presence of a closed 3-form background $H\in \Omega_{\text{cl}}^3(M)$, there exists a geometric structure that departs from vanilla Poisson to what is called twisted Poisson structure \cite{Severa:2001qm}. This is once more given by a bivector $\Pi$ but this time it satisfies the defining property 
\be \label{twisted Poisson}
\frac 12 [\Pi,\Pi]_{\text{SN}}=\langle \Pi\otimes\Pi\otimes\Pi,H\rangle\,,
\ee 
in terms of the Schouten-Nijenhuis bracket, the extension of the Lie bracket of vector fields to multivector fields. In other words, the Jacobi identity for the would-be Poisson bracket of functions is violated and the violation is precisely controlled by the 3-form $H$ three times contracted with the bivector $\Pi$ in its first slot. Nevertheless, the cotangent bundle $T^{\ast}M$ continues to bear a Lie algebroid structure, this time with the Koszul-Schouten Lie bracket being suitably twisted to  
\be 
[\eta,\eta']_{\text{HKS}}=[\eta,\eta']_{\text{KS}}+H(\Pi(\eta),\Pi(\eta'))\,,
\ee 
which satisfies the Jacobi identity by virtue of the twisted Poisson condition \eqref{twisted Poisson}. 

The fact that a Lie algebroid underlies both Poisson and twisted Poisson manifolds also means that there exists a Q1 manifold description for both. Indeed, the homological vector field that gives rise to the twisted Poisson case is  of the same form as the untwisted case, 
\be 
Q^{H}=a_{\m}\Pi^{\m\n} \left(\frac{\partial}{\partial x^{\n}}+\G^{\s}_{\r\n}a_{\s}\frac{\partial}{\partial a_{\r}}\right)-\frac 12 \mathring{\nabla}_{\rho}\Pi^{\m\n}a_{\m}a_{\n}\frac{\partial}{\partial a_{\r}}\,,
\ee 
with the important difference that the connection on $M$ that induces the connection on the Lie algebroid now has torsion given by the closed 3-form $H$ once contracted with the bivector $\Pi$  \cite{Ikeda:2019czt}. In particular, the connection coefficients are now given as 
\be 
\G_{\m\n}^{\rho}=\mathring\G_{\m\n}^{\rho}+\frac 12 \Pi^{\rho\s}H_{\m\n\s}\,.
\ee 
Note that regardless of whether the ordinary connection has torsion or not, the $E$-connection always has $E$-torsion which is independent of the 3-form \eqref{Etor}. In practice, the torsion of $\nabla$ counterbalances the twist of the Lie algebroid bracket in the definition of the $E$-torsion.

The above example teaches us two lessons. First, when the Poisson structure is twisted by a 3-form the covariant expression for the homological vector field that describes it retains its form at the expense of introducing a connection with torsion on the manifold. Second, the 3-form twists various algebraic and geometric operations in different way. Note specifically that it can be contracted once, twice or three times with the bivector $\Pi$ and each of these combinations plays a role in modifying some operation or quantity. Let us use the notation 
\be \label{geomP}
H^{\m}_{\n\rho}=\Pi^{\m\k}H_{\k\n\rho}\,,\quad H^{\m\n}_{\rho}=\Pi^{\m\k}\Pi^{\n\l}H_{\k\l\rho}\,,\quad H^{\m\n\rho}=\Pi^{\m\k}\Pi^{\n\l}\Pi^{\rho\s}H_{\k\l\s}\,.
\ee 
Then we observe that from the classical differential geometric point of view the first modifies the symmetric connection $\nabla$ on $M$ to one with torsion, the second modifies the binary bracket on $E=T^{\ast}M$ and the third modifies the Schouten-Nijenhuis bracket of the bivector with itself, i.e. the Jacobi identity for the Poisson bracket. This is of course a special case in the category of Lie algebroids, since there exists a natural symplectic form and in absence of $H$ the Q1 manifold is in fact a QP1 manifold \cite{Roytenberg:2002nu}.

\subsection{Pre-Courant algebroids and Q2 manifolds}
\label{sec23}

The picture of Lie algebroids as dg manifolds extends to Courant algebroids too, and even further to pre-Courant algebroids \cite{BG}. In the latter case it may happen that a closed 4-form is responsible for the violation of the Jacobi identity, a structure called (4-form) twisted Courant algebroid in Ref. \cite{Hansen:2009zd}. Let us give a  definition. First we recall that a Courant vector bundle is a pseudo-Euclidean anchored vector bundle $(E,\langle\cdot,\cdot\rangle,\rho:E\to TM)$  such that the transpose map $\rho^{\ast}:T^{\ast}M\to E^{\ast}\simeq E$ to the anchor satisfies $\rho\circ\rho^{\ast}=0$ \cite{Vaisman:2004msa}. 
\begin{defn} \cite{Vaisman:2004msa}
A pre-Courant algebroid is a Courant vector bundle together with a binary operation $\circ$ on the space of sections $\G(E)$ that satisfies the following three axioms:
\begin{itemize}
    \item $\rho(e\circ e')=[\rho(e),\rho(e')]\,,$
    \item $\langle e\circ e,e'\rangle=\frac 12 \rho(e')\langle e,e\rangle\,,$
    \item $\rho(e)\langle e', e'' \rangle= \langle e\circ e', e'' \rangle +\langle e', e\circ e''\rangle\,,$
\end{itemize}
for all $e, e',e'' \in \G(E)$. A Courant algebroid is a pre-Courant algebroid such that $(\G(E),\circ)$ is a Leibniz algebra, which means that the Jacobiator satisfies
\be 
\text{Jac}(e,e',e''):=e\circ(e'\circ e'')-(e\circ e')\circ e'' -e'\circ(e \circ e'')=0\,.
\ee 
\end{defn} 
For a 4-form twisted Courant algebroid the Jacobiator is instead given as 
\be \label{Jac}
\text{Jac}(e,e',e'')=\rho^{\ast}H(\rho(e),\rho(e'),\rho(e''))\,.
\ee 
The (small) difference of pre-Courant and twisted Courant algebroids was quantified in \cite{preCA}. When $\rho(E)=TM$ (transitive case) the two structures are the same. In the following sections we will be concerned with twisted Courant algebroids. We note, moreover, that one may write the definition of a pre-Courant algebroid in terms of the skew-symmetric operation 
\be 
[e,e']_{\E}:=\frac 12 (e\circ e'-e'\circ e)\,,
\ee 
see \cite{Vaisman:2004msa}. For (twisted) Courant algebroids we hence refer to the binary operations $\circ$ and $[\cdot,\cdot]_{\E}$ as the Dorfman and the Courant bracket respectively.

Below we mostly work with the alternative definition in terms of dg manifolds.
We begin with Courant algebroids. The properties of the Courant algebroid data, namely the Courant bracket $[\cdot,\cdot]_{\E}$, anchor $\rho$ and fiber metric $\langle\cdot,\cdot\rangle$ can be neatly encoded in a homological vector field $Q\in \mathfrak{X}{(\cal M)}$, this time with the dg manifold $\cal M$ being a Q2 manifold (in fact QP2, namely equipped with a degree 2 symplectic form $\omega$, which is compatible with $Q$ in the sense that $L_{Q}\,\omega=0$) \cite{dimaphd}.{\footnote{A more recent approach to shifted symplectic structures is \cite{shifted1}, where a version of the AKSZ construction in derived algebraic geometry is proposed. The relation of this approach to theories such as twisted Courant sigma models was studied in \cite{shifted2}.}} In general, $\cal M$ is obtained as the symplectic submanifold of $T^{\ast}[2]E[1]$ that corresponds to the isometric embedding of $E$ in the direct sum $E\oplus E^{\ast}$ 
with respect to the canonical pairing of $E$ and $E^{\ast}$. 
Local coordinates on this graded manifold are $(x^{\m},a^{a}, b_{\m})$ of respective degrees $(0,1,2)$.   The space of vector fields on ${\cal M}$ is spanned by the corresponding derivations of opposite degrees $(0,-1,-2)$.  
Then the most general homological vector field compatible with the  graded symplectic form on $\mc M$ has the form{\footnote{If one works with general Q2 manifolds without necessarily having a symplectic structure, then the most general case proliferates to semi-strict Lie 2-algebroids. This was discussed in detail in Ref. \cite{Grutzmann:2014hkn}.}} \cite{Ikeda:2012pv}
\bea \label{QCAnoncov}
Q=\rho_{a}{}^{\m}a^{a}\frac{\partial}{\partial x^{\m}} - \big(\eta^{ab}\rho_b{}^{\m}b_{\m}+\frac 12 C^{a}_{bc}a^{b}a^{c}\big)\frac{\partial}{\partial a^{a}}-\big(\partial_{\m}\rho_{a}{}^{\n}b_{\n}a^{a}+\frac 1{3!}\partial_{\m}C_{abc} a^{a}a^{b}a^{c}\big)\frac{\partial}{\partial b_{\m}}\,.
\eea 
Thinking in terms of ordinary differential geometric terms, we have introduced again a basis $\mf e_a$ for the vector bundle $E$ of a Courant algebroid. In this basis, $C^{a}_{bc}(x)$ are the structure functions of the Courant bracket and $\eta_{ab}$ are the components of the fiber metric, in our conventions 
\be \label{fibermetric}
\eta_{ab}=2\langle\mf e_a,\mf e_b\rangle_{\E}\,.
\ee 
The $\eta^{ab}$ are the inverse metric components and Latin indices are raised and lowered with this metric, for instance $C_{abc}=\eta_{cd}C^{d}_{ab}$. Finally, $\rho_{a}{}^{\m}(x)$ are once more the components of the anchor in this basis. 
The statement now is that the condition $Q^2=0$ is equivalent to the local coordinate form of the axioms of a Courant algebroid laid down in \cite{Liu:1995lsa}, as given in \cite{Ikeda:2002wh} in suitable conventions. The homological vector field given in \eqref{QCAnoncov} requires covariantization in the same spirit as the Lie algebroid homological vector field of the previous section. Indeed, there are two obvious sources of noncovariance, the structure functions $C_{bc}^{a}(x)$ and the partial derivative on the components of the anchor. This evidently calls for the introduction of a suitable connection. Moreover, this is once more a case of non-manifest covariance. Our goal is to make covariance manifest and show the relation to appropriate Courant algebroid tensors. 

An important difference to the case of Lie algebroids regards the transformation of the structure functions under a change of basis on the bundle. In the present case this transformation reads 
\bea 
&&C_{ab}^c=(\Lambda^{-1})^c_{c'}\Lambda_a^{a'}\Lambda_b^{b'}\widetilde{C}_{a'b'}^{c'}+(\Lambda^{-1})^c_{c'}\widetilde\rho_{a'}{}^{\m}(\Lambda_a^{a'}\widetilde{\partial}_{\m}\Lambda_b^{c'}-\Lambda_{b}^{a'}\widetilde{\partial}_\m\Lambda_a^{c'})-\nn\\[4pt] 
&&\qquad \qquad-\, \sfrac 12 \widetilde\eta_{a'b'}\widetilde\eta^{c'd'}\widetilde{\rho}_{d'}{}^\mu (\Lambda^{-1})^c_{c'}(\Lambda_a^{a'}\widetilde{\partial}_{\m}\Lambda_b^{b'}-\Lambda_{b}^{a'}\widetilde{\partial}_\m\Lambda_a^{b'}) \,,
\eea 
with the new term appearing due to the unconventional, anomalous Leibniz rule that the Courant bracket satisfies:
\be 
[e,fe' ]_{\E}=f[e,e']_{\E}+\rho(e)f \, e'-\dd_{\E}f\langle e,e'\rangle\,,
\ee 
where $\dd_{\E}: C^{\infty}(M)\to \G(E)$ is the Courant algebroid differential defined via $\langle\dd_{\E}f,e\rangle=\frac 12 \rho(e)f$. We remark that  $\eta_{ab}$  transforms homogeneously, namely 
\bea 
\eta_{ab}=\Lambda_a^{a'}\Lambda_b^{b'}\widetilde{\eta}_{a'b'}\,. 
\eea  
Note that the unconventional transformation of the structure functions does not improve for the Dorfman bracket, since it regards both entries in the bracket and even though the Leibniz rule is standard in the second entry it is not so in the first one. 

Then the first thing to note is that we need an $E$-connection on $E$, where $E$ is the Courant algebroid at hand. We use the same notation as for Lie algebroids and in the local basis the connection coefficients are as in Eq. \eqref{omegagen}. Famously, for Courant algebroid connections the naive torsion tensor is not suitable since it fails to exhibit linearity in all its arguments. A suitable Courant algebroid $E$-torsion tensor $\mc T^{\E}\in \G(\w^3E^{\ast})$ was defined in \cite{Gualtieri:2007bq} and it reads as 
\bea 
\mc T^{\E}(e_1,e_2,e_3)=\langle \nabla^{\E}_{e_1}e_2- \nabla^{\E}_{e_2}e_1 - [e_1,e_2]_{\E}, e_3\rangle_{\E}+\frac 12 \left(\langle  \nabla^{\E}_{e_3}e_1,e_2\rangle_{\E} -\langle \nabla^{\E}_{e_3}e_2,e_1\rangle_{\E}\right)
\eea 
in terms of the skew-symmetric Courant bracket and analogously for the Dorfman bracket. Other possible options for a good definition of a torsion tensor for Courant algebroid connections were suggested in \cite{Boffo:2019zus}. In a chosen local basis, the components of the Gualtieri $E$-torsion are 
\be \label{scte}
\mc T_{abc}= \omega^d_{[ab]}\eta_{cd}-\frac 12 C_{ab}^{d}\eta_{dc}+\frac 12 \omega^d_{c[a}\eta_{b]d}\,.
\ee 
We could read this formula in an upside down way by solving in terms of the structure functions of the bracket, 
\be \label{CintermsofT}
C^d_{ab}=\big(-2 \mc T_{abc}+2\omega^{e}_{[ab]}\eta_{ec}+\omega^e_{c[a}\eta_{b]e}\big)\eta^{cd}\,.
\ee 
This can be used to substitute the structure functions with the $E$-torsion in the vector field $Q$, plus additional terms that must and will be accounted for. Secondly, since there is a differentiation on functions on $M$ in the components of $Q$, we need a vector bundle connection $\nabla$ on $TM \oplus E$ as well. This connection can be extended to act on differential forms in the usual way and therefore we have an analog of the tetrad connection for Vielbeins, now on the components of the anchor.{\footnote{To avoid misunderstanding of this analogy, we mention that no ``tetrad postulate'' is imposed here.}} Specifically, 
\be \label{nablarho}
\nabla_{\m}\rho_{a}{}^{\n}=\partial_{\m}\rho_a{}^{\n}+\Gamma^{\n}_{\m\rho}\rho_{a}{}^{\rho}-\omega_{\m a}^{b}\rho_{b}{}^{\n}\,.
\ee 
We shall denote the torsionless part of this connection by $\mathring{\nabla}$ as before. 

We are now ready to express $Q$ in a different form using \eqref{CintermsofT} and \eqref{nablarho}. To reach a  manifestly covariant result, it is necessary to perform a coordinate transformation (corresponding to a nonlinear field redefinition in the gauge theory to be discussed below) as follows 
\be \label{bredef}
b_{\m}^{\scriptscriptstyle\nabla} := b_{\m} +\frac 12\o_{\m ab}a^{a}a^b\,,
\ee
mixing the
degree $2$ and degree $1$ coordinates. Let us pause to further justify this transformation. First note that the coordinate and bundle transformations of the Courant algebroid in the classical differential geometric side correspond to graded coordinate transformations in the Q2 manifold side. Specifically, these coordinate transformations are 
\bea 
x^{\m}&=&x^\m(\widetilde{x})\,\\[4pt]
a^a&=&(\Lambda^{-1})_{a'}^a \widetilde a^{a'}\,\\[4pt]
b_\m&=&\Lambda_\m^{\m'}\,\widetilde b_{\m'}+\frac 12 \partial_\m(\Lambda^{-1})_a^{a'}\Lambda^c_{a'}\widetilde{\eta}_{cb}\widetilde a^{a}\widetilde a^{b} \label{b trafo}
\eea 
They are obtained by integrating the infinitesimal counterparts which are generated by a quadratic Hamiltonian and the canonical Poisson bracket on $\mc M$ which exists because it is a cotangent bundle \cite{Roytenberg:2002nu}. Observing the complicated transformation of the degree 2 coordinate we can ask whether it can be undone by a suitable redefinition. Indeed this is the redefinition of \eqref{bredef}. The fact that 
\be \label{b nabla trafo}
b_\m^{\scriptscriptstyle\nabla}=\Lambda_\m^{\m'}\,\widetilde b_{\m'}^{\scriptscriptstyle\nabla}
\ee 
justifies it. In the following we will work with this ``covariant'' degree 2 coordinate and the corresponding degree $-2$ derivation $\partial/\partial b^{\scriptscriptstyle\nabla}_{\m}$.

Taking all these into account, a straightforward calculation leads to the following alternative form of the homological vector field:
\bea
&&Q= a^{a} \rho_{a}{}^{\m} \DD^{(0)}_{\m} 
-\bigg(\eta^{ac}\rho_{a}{}^{\m}b_\m^{\scriptscriptstyle\nabla}+\big(-\mc T_{ab}{}^c+\phi_{ab}{}^c+\frac 12\phi^c{}_{ab}\big)a^aa^b\bigg)\DD^{(-1)}_{c}+ \nn\\[4pt] && \qquad +\,\bigg( -\nabla_{\n}\rho_{a}{}^{\m}b^{\scriptscriptstyle\nabla}_{\m}a^{a}+\big(\frac 13 \nabla_{\n}\mc T_{abc}-\frac 12\nabla_\n\phi_{abc}+\frac 12 \rho_{a}{}^{\m}\eta_{cd}R^d{}_{b\n\m}\big)a^aa^ba^c\bigg)\DD^{\n}_{(-2)}\,, \qquad \label{QCAcovgen}
\eea
where we have defined the natural degree $0$, degree $-1$ and degree $-2$ derivations in the present case, which have the form 
\bea  
\label{D0} \DD^{(0)}_{\m}&=& \frac{\partial}{\partial x^{\m}}-\frac 12\frac{\partial \omega_{\n bc}}{\partial x^{\m}} a^ba^c\frac{\partial}{\partial b_{\n}}+\Gamma_{\n\m}^\s b_\s^{\scriptscriptstyle\nabla} \frac{\partial}{\partial b_{\n}}-\o_{\m b}{}^c a^b\DD_c^{(-1)} \,, \\[4pt] 
\label{D-1} \DD_a^{(-1)}&=&\frac{\partial}{\partial a^{a}}+\o_{\m ba}a^b\frac{\partial}{\partial b_{\m}}\,, \\[4pt] 
\label{D-2} \DD^{\m}_{(-2)}&=&\frac{\partial}{\partial b_{\m}^{\scriptscriptstyle\nabla}}\,.
\eea 
Once more these are natural in the sense of the following relations when they act on the coordinates,
\bea 
\DD_\m^{(0)}x^{\n}&=&\d_{\m}{}^{\n}\,, \qquad\,\,\quad\qquad\,\, \DD^{(-1)}_{a}x^{\m}=0\,,\quad\,\quad \,\,\,\, \DD^{\n}_{(-2)}x^{\m}=0\,,\\[4pt] 
\DD_\m^{(0)}a^c&=& -\,\o_{\m b}{}^ca^b\,,\quad \quad\quad\,\, \DD_a^{(-1)}a^c=\d_a^c\,, \quad\quad\,\,\,\, \DD^{\n}_{(-2)}a^c=0\,, \\[4pt] 
\DD_{\n}^{(0)}b_{\m}^{\scriptscriptstyle\nabla} &=&\G^{\s}_{\m\n}b_{\s}^{\scriptscriptstyle\nabla} \,,\,\quad\qquad\, \quad \DD_a^{(-1)}b^{\scriptscriptstyle\nabla}_{\m}=0\,,\,\,\,\, \quad\,\quad \DD^{\m}_{(-2)}b^{\scriptscriptstyle\nabla}_{\n}=\d^{\m}_{\n}\,.\label{DDsCA3}
\eea 
In case the induced connection $\mathbullet\nabla^{\E}$ is chosen, the endomorphism $\phi$ vanishes and the vector field takes the simpler form
\bea
&&Q= a^{a} \rho_{a}{}^{\m} \DD^{(0)}_{\m} 
-\bigg(\eta^{ac}\rho_{a}{}^{\m}b_\m^{\scriptscriptstyle\nabla}-\mathbullet{\mc T}_{ab}{}^ca^aa^b\bigg)\DD^{(-1)}_{c}+ \nn\\[4pt] && \qquad +\,\bigg( -\nabla_{\n}\rho_{a}{}^{\m}b^{\scriptscriptstyle\nabla}_{\m}a^{a}+\big(\frac 13 \nabla_{\n}\mathbullet{\mc T}_{abc}+\frac 12 \rho_{a}{}^{\m}\eta_{cd}R^d{}_{b\n\m}\big)a^aa^ba^c\bigg)\DD^{\n}_{(-2)}\,. \qquad 
\eea

There is a further geometrical association to make. 
Recall that from the non-manifestly covariant expression \eqref{QCAnoncov} we observe that a derivative on the structure functions appears. This turned to a derivative on the components of the $E$-torsion $\mc T^{\E}$ through \eqref{CintermsofT} and in turn this derivative got covariantized in \eqref{QCAcovgen}. However, the covariant derivative on the Courant algebroid torsion has an independent geometrical essence. As proven in \cite{Chatzistavrakidis:2023otk}, it is related to the tensor that measures the compatibility between an (ordinary) connection on $E$ and the bracket on the Courant algebroid---see \cite{Blaom} for introducing this notion for Lie algebroids and \cite{Crainic,Kotov:2016lpx} for more detailed explanations. This tensor is called the basic curvature tensor ${\mc S}^{\E}\in\Omega^1(\w^2E^{\ast}\otimes E^{\ast})$ and for Courant algebroids it is defined as \cite{Jotz,Chatzistavrakidis:2023otk} 
\bea 
\mc S^{\E}(e_1,e_2,e_3)X&=&\langle \nabla_{X}[e_1,e_2]_{\E}-[\nabla_Xe_1,e_2]_{\E}-[e_1,\nabla_Xe_2]_{\E}-\nabla_{\overline{\nabla}^{\E}_{e_2}X}e_1+\nabla_{\overline{\nabla}^{\E}_{e_1}X}e_2 , e_3\rangle_{\E} \,\,\nn\\[4pt] 
&+&\frac 12 \left(\langle\nabla_{\overline{\nabla}^{\E}_{e_3}X}e_1,e_2\rangle_{\E}-\langle\nabla_{\overline{\nabla}^{\E}_{e_3}X}e_2,e_1\rangle_{\E}\right)\,,
\label{basiccurvaturecourant}
\eea 
in terms of an $E$-connection on $TM$ (the opposite to an ordinary vector bundle connection on $E$) given as 
\be 
\overline{\nabla}^{\E}_{e}X=\rho(\nabla_{X}e)+[\rho(e),X]\,.
\ee 
It is also proven in \cite{Chatzistavrakidis:2023otk} that it may be expressed as 
\bea  
\mc S^{\E}(e_1,e_2,e_3)X&=&-\nabla_{X}\mc T^{\E}(e_1,e_2,e_3)+\langle \mc C(e_2,X)e_1-\mc C(e_1,X)e_2,e_3\rangle_{\E} \nn\\[4pt]
&& -\, \frac 12 \bigg(\langle\mc C(e_3,X)e_1,e_2\rangle_{\E}-\langle \mc C(e_3,X)e_2,e_1\rangle_{\E}\bigg)\,,
\eea  
 in terms of the Gualtieri torsion and a tensor $\mc C$  defined through
\be \label{ctensor}
{\cal C}(e_1,X)e_2:=[\nabla^\E_{e_1},\nabla_X]e_2-\nabla_{[\rho(e_1),X]}e_2+\phi(\nabla_Xe_1,e_2)\,,
\ee 
where we recall that $\phi$ is the difference of the $E$-connection we work with from the induced $E$-connection. Note that $\phi$ obviously vanishes when the induced connection is chosen. However, in the general case the last term in the definition of $\mc C$ is absolutely necessary to guarantee that it transforms tensorially in all arguments.
 
 In a local basis the components of the Courant algebroid basic curvature tensor read 
\bea 
\mc S_{\m abc}= -\nabla_{\m}\mc T_{abc} -\mc C_{\m[ab]c} -\frac 12 \mc C_{\m c[ab]}\,,
\eea 
with 
\bea 
\mc C_{\m ab}{}^c= \rho_a{}^{\n}\partial_{\n}\o^c_{\m b}-\partial_{\m}\o^c_{ab}+2\omega^d_{\m b}\o^c_{(ad)}- \o^d_{ab}\o^{c}_{\m d}+ \partial_{\m}\rho_a{}^{\n}\o^c_{\n b}- \rho_{d}{}^{\n}\o^d_{\m b}\o^{c}_{\n a}\,.
\eea 
Caution is drawn to the fact that the coefficients of two different connections appear in this formula, both denoted as $\o$ and distinguished by their indices. One is the ordinary vector bundle connection and one is the $E$-connection, which can be completely general. In the case of induced $E$-connection, the tensor simplifies greatly to 
\be 
\mathbullet{\mc C}_{\m ab}{}^{c}= \rho_{a}{}^{\n}R^{c}{}_{b\n\m }\,,
\ee 
thus being the anchored curvature tensor. Note that the usual expression for the curvature for Courant algebroid connections is a tensor if and only if the induced connection is chosen, which explains this relation.

We are now ready to express $Q$ in an alternative manifestly covariant form using the definitions related to the Courant algebroid basic curvature tensor $\mc S$. For the induced $E$-connection the result is remarkably simple,
\begin{tcolorbox}[ams nodisplayskip, ams align]\label{QCAcovinduced}
&&Q= a^{a} \rho_{a}{}^{\m} \DD^{(0)}_{\m} 
-\big(\eta^{ac}\rho_{a}{}^{\m}b_\m^{\scriptscriptstyle\nabla}-\mathbullet{\mc T}_{ab}{}^ca^aa^b\big)\DD^{(-1)}_{c}- \big( \nabla_{\n}\rho_{a}{}^{\m}b^{\scriptscriptstyle\nabla}_{\m}+\frac 13 \mathbullet{\mc S}_{\n abc}a^ba^c\big)a^{a}\DD^{\n}_{(-2)}\,. \,\,\,\,\qquad 
\end{tcolorbox}
We have thus expressed the homological vector field corresponding to a Courant algebroid in terms of the Gualtieri torsion and the basic curvature of the induced Courant algebroid connection on itself. In the following we will always refer to the connection $\mathbullet\nabla$ and hence we denote $\mathbullet{\mc T}$ and $\mathbullet{\mc S}$ simply as $\mc T$ and $\mc S$.

Let us now turn to twisted Courant algebroids. As discussed earlier, the main difference to Courant algebroids is that the Jacobi identity for the Dorfman bracket \eqref{Jac} is modified.
The description of this structure in terms of a Q2 manifold is as follows. We consider the same graded manifold $\mc M$ and the homological vector field 
\bea \label{QHCA}
&&Q^H=Q-\frac 1{3!} \rho_a{}^\nu\rho_b{}^\s\rho_c{}^\l H_{\m\n\s\l}a^a a^ba^c\frac{\partial}{\partial b_{\m}}\,,
\eea 
where $Q$ is the one of the Courant algebroid. In other words, the vector field \eqref{QHCA} is homological if and only if the following conditions hold:
\begin{align}
& \eta^{ab} \rho_a{}^\m \rho_b{}^\n =0\,,
\label{CAidentity21}
\\[4pt]
& \rho_b{}^\m \partial_\m \rho_a{}^\n - \rho_a{}^\m \partial_\m \rho_b{}^\n
+ \eta^{ef} \rho_e{}^\n C_{fab} =0\,,
\label{CAidentity22}
\\[4pt]
& \rho_d{}^\m \partial_\m C_{abc} - \rho_a{}^\m \partial_\m C_{bcd}
+ \rho_b{}^\m \partial_\m C_{cda} - \rho_c{}^\m \partial_\m C_{dab}\,+
\nonumber \\[4pt]  & \qquad 
+ \eta^{ef} C_{eab} C_{cdf} + \eta^{ef} C_{eac} C_{dbf} + \eta^{ef} C_{ead} C_{bcf} 
= -\, \rho_d{}^\m \rho_a{}^\m \rho_b{}^\k \rho_c{}^\l H_{\m\n\k\l}\,.
\label{CAidentity23}
\end{align}
These are nothing else but the local coordinate form of the axioms of a twisted Courant algebroid. The third, $H$-dependent equation is the local coordinate form of the Jacobiator \eqref{Jac}. Note that compatibility of $Q^{H}$ with the symplectic form is now obstructed by the 4-form, since $L_{Q^{H}}\omega\ne 0$.

It is useful to write down the covariant form of these conditions in terms of the ordinary connection $\nabla$ on $E$ and of the basic curvature and Gualtieri torsion tensors. For \eqref{CAidentity22}, the covariant expression in terms of the components of the Gualtieri torsion is 
\be\label{id2 cov}
\rho_a{}^\m\nabla_\m \rho_b{}^\n - \rho_b{}^\m\nabla_\m \rho_a{}^\n
+2\eta^{ef} \rho_e{}^\n {\mc T}_{abf} =0\,. 
\ee
Similarly, the covariant expression for \eqref{CAidentity23} can be written in terms of the Gualtieri torsion and the basic curvature tensor, yielding the following elegant formula: 
\begin{tcolorbox}[ams nodisplayskip, ams align]
\rho_{[d}{}^\m {\mc S}_{\m abc]} 
+\sfrac 32  \eta^{ef} {\mc T}_{ea[b} {\mc T}_{cd]f} 
= -\sfrac 18\r_d{}^\m\r_a{}^\n\r_b{}^\r\r_c{}^\s H_{\m\n\r\s}\,,
\label{CAidentity23cs}
\end{tcolorbox} 
which is essentially an algebraic Bianchi identity for the Courant algebroid basic curvature.
Furthermore, prompted by Eq. \eqref{QHCA} we define
\be \label{ternary}
\widetilde{\mc S}_{\m abc}:=\mc S_{\m abc}+\frac 12 H_{\m abc}\,,
\ee 
having introduced the notation 
\bea  
&& H_{\m\n\rho a}=\rho_a{}^{\k}H_{\m\n\rho\k}\,,
\nn \\[4pt] 
&& H_{\m\n ab}=\rho_a{}^{\k}\rho_b{}^{\l}H_{\m\n\k\l}\,,\nn\\[4pt]  && H_{\m abc}=\rho_{a}{}^{\k}\rho_b{}^{\l}\rho_c{}^{\n}H_{\m\k\l\n}\,, \nn\\[4pt]  
&& H_{abcd}=\rho_a{}^{\m}\rho_b{}^{\n}\rho_c{}^{\k}\rho_d{}^{\l}H_{\m\n\k\l}\,.
\eea
As usual, the bundle indices in these quantities are raised with the (inverse) metric $\eta^{ab}$, and $H$ remains totally antisymmetric in all indices, e.g $H_{\m abc}=-H_{a\m bc}=H_{ ab\m c}=-H_{abc\m}$.
Similar to the discussion of the twisted Poisson case following Eq. \eqref{geomP}, these quantities have a definite geometric interpretation. The last quantity, $H_{abcd}$, modifies the Jacobi identity  for the Dorfman bracket \eqref{Jac}.  The next to last, $H_{\m bcd}$, modifies the basic curvature \eqref{ternary}. 
 In different terms, akin to the $L_{\infty}$ description of a Courant sigma model \cite{GJ}, this term twists the ternary bracket.{\footnote{We recall that the $L_{\infty}$ description of a Courant algebroid was given in Ref. \cite{RW}. More generally, the relation of $L_{\infty}$ to perturbative (gauge) field theories was explored and reviewed in \cite{Hohm:2017pnh,Grigoriev:2023lcc}.}} We shall see later that  $H_{\m \n cd}$ modifies the ordinary curvature of the connection $\nabla$ on $E$, while the first term $H_{\m\n\r a}$ controls the non-exact part of $H$, as discussed in App. \ref{appa}.
Then we observe that the homological vector field for a twisted Courant algebroid may be written as 
\bea  
Q^{H}= a^{a} \rho_{a}{}^{\m} \DD^{(0)}_{\m} 
-\big(\eta^{ac}\rho_{a}{}^{\m}b_\m^{\scriptscriptstyle\nabla}-{\mc T}_{ab}{}^ca^aa^b\big)\DD^{(-1)}_{c}- \big( \nabla_{\n}\rho_{a}{}^{\m}b^{\scriptscriptstyle\nabla}_{\m}+\frac 13 \widetilde{\mc S}_{\n abc}a^ba^c\big)a^{a}\DD^{\n}_{(-2)}\,.  \label{QH final}
\eea 
Notice that the form is the same as for the usual Courant algebroid with the difference that $\mc S$ is replaced by the $H$-dependent $\widetilde{\mc S}$. 

\section{BRST structure of twisted Courant sigma models} 
\label{sec3} 

\subsection{Action, gauge symmetry \& nonlinearly open gauge algebra}
\label{sec31}

The Courant sigma model can be described as a membrane sigma model with target space $\mc M$, a symplectic dg manifold of degree 2 \cite{Ikeda:2002wh,Hofman:2002jz,Roytenberg:2006qz}. As such, its gauge structure is encoded in the axioms of a Courant algebroid and vice versa; or, in other words, the solution to the classical master equation in the BV/BRST formulation of the theory is encoded in the graded symplectic structure and the homological vector field on $\mc M$, which is the original spirit of the AKSZ construction \cite{Alexandrov:1995kv}. 
Here we take the above logic one step further and slightly depart from the usual AKSZ construction by considering topological membrane sigma models with a Wess-Zumino term. This was originally suggested in \cite{Hansen:2009zd} in the context of Hamiltonian systems with first class constraints and also implicitly in \cite{Chatzistavrakidis:2018ztm} in studies related to fluxes in double field theory. We refer to them as twisted Courant sigma models here because it turns out that their gauge structure is governed by a twisted Courant algebroid.

In this section we revisit the classical action of the model using dg manifold language and filling in some gaps regarding the structure of its gauge algebra. The source space (world volume) of the model is the dg manifold $T[1]\S$, where $\S$ is a three-dimensional manifold without boundary (the closed membrane). For convenience, we introduce coordinates on $T[1]\S$ denoted as $\s^{m}$ of degree 0 and $\theta^m$ of degree 1 for the fibre coordinates. In accord with the notation introduced in the previous section, the classical fields are real functions $X^{\m}(\s)$, $X^{\ast}E$-valued 1-forms $A=A^{a}_{m}(\s)\theta^m e_a$, where $e_a$ is a local basis of the pullback bundle $X^{\ast}E$ and $X^{\ast}T^{\ast}M$-valued 2-forms $B=\frac 12 B_{\m mn}(\s)\theta^m\theta^n\dd X^{\m}$. Note that the 2-form is not global, see the way that the degree 2 coordinate transforms in \eqref{b trafo}. As discussed in Section \ref{sec23}, a connection is needed to account for this. We address this issue in Section \ref{sec33}. The latter two fields can also be viewed as pullback coordinates for a suitable \emph{degree-preserving} map $\phi: T[1]\S\to \mc M$, at least locally. As suggested in \cite{Grutzmann:2014hkn}, eventually it is necessary to extend this to a map 
\be 
\phi_{\times}:T[1]\S\to T[1]\S\times  \mc M\,,
\ee 
which is the identity on the first factor, in order to be able to treat $\Sigma$-dependent gauge parameters. Then locally $A^{a}=\phi^{\ast}_{\times}(a^{a})$ and $B_{\m}=\phi^{\ast}_{\times}(b_{\m})$.  
The classical action functional of maps in our conventions reads
\be\label{Scl}
S_{\text{cl}} = \int_{T[1]\Sigma} 
\left(-B_\m  \rd X^\m + \frac{1}{2} \eta_{ab} A^a  \rd A^b
+\rho^\m_a(X) B_\m  A^a 
+ \frac{1}{3!} C_{abc}(X) A^a  A^b  A^c
\right)+\int_{\widehat{\S}}X^{\ast}H\,
\ee
where $H\in\Omega_{\text{cl}}^{4}(M)$ and we make the usual assumptions the guarantee the existence of the extension of the world volume to a 4D one $\widehat{\S}$ with $\partial\widehat{\S}=\S$ and the independence of the action functional of this extension, see e.g. \cite{Figueroa-OFarrill:2005vws} for a clear exposition in any dimension.

The topological membrane action \eqref{Scl} is invariant under a prescribed set of infinitesimal gauge symmetries provided that $\rho_a^\m, \eta_{ab}, C_{abc}$ and $H_{\m\n\rho\s}$ are the components of the anchor map, fibre metric, binary bracket and 4-form of a twisted Courant algebroid $E$. This means that they satisfy the set of algebraic and differential equations \eqref{CAidentity21}, \eqref{CAidentity22} and \eqref{CAidentity23}. 
These gauge symmetries may be neatly encoded in compact form using the homological vector field $Q^{H}$, Eq.\eqref{QH final}. Denote the fields of the model collectively as $\phi^{\a}=(X^{\m}, A^{a}, B_{\m})$, understood as pull-backs via the map $\phi_{\times}$ defined above of the coordinates $x^{\a}=(x^{\m},a^{a},b_{\m})$, namely
\be 
\phi^{\a}=\phi^{\ast}_{\times}(x^{\a})\,.
\ee 
Note that $\phi^{\a}=\phi^{\a}(\sigma,\theta)$ are dependent on the coordinates of the source Q manifold $T[1]\S$. The homological vector field on $T[1]\S$ is simply the de Rham differential $\dd=\theta^m\partial/\partial \s^{m}$. Define a collective gauge parameter as  $\e^{\a}=(0,\e^{a},-\psi_{\m})$ with components of degrees $0$ and $1$ respectively---we denoted explicitly the obvious absence of gauge parameter for the lowest field which is a scalar. Then we can rewrite 
\be 
\dd \e^{\a}=[\dd,\e]^\a=(\mc L_{\dd}\e)^{\a}\,,
\ee 
in terms of the (graded) Lie bracket of vector fields, provided we introduce the vector field $\e=\e^{\a}\partial_{\a}$ on $T[1]\S\times\mc M$. Then it becomes advantageous to introduce the following distinguished vector field on the product manifold $T[1]\S\times \mc M$, 
\be 
\widehat{Q}=\dd +Q^{H}\,.
\ee 
The gauge transformations of the fields are then given in terms of this vector field as 
\be \label{delta phi noncov}
\d \phi^{\a}=(\phi_{\times}^{\ast}\circ \mc L_{\widehat Q}\e)(x^{\a})+\d_{\text{triv}}\phi^{\a}\,,
\ee 
in the spirit of \cite{Grutzmann:2014hkn} as revisited in \cite{Chatzistavrakidis:2023otk}.
Here $\mc L_{\widehat Q}$ is the Lie derivative along the homological vector field $\widehat{Q}$. The second term denotes the contributions that are proportional to the field equations $F^{\m}$ and $G^{a}$ obtained by varying with respect to the 2-form and the 1-form respectively---the remaining field equation has degree 3, one higher than the top form field in the model and does not contribute. It is useful to mention that the field strengths of the various fields take the form 
\be \label{field strengths Q}
{\mc F}^{\a}=\dd\phi^{\a}-\phi^{\ast}(Q^{\a})\,,
\ee 
in terms of the components of the homological vector field $Q^{H}$ \cite{Grutzmann:2014hkn}. 

Note that as long as $\e$ is expressed in the canonical basis $(\partial_{\a})$ these gauge transformations are not in manifestly covariant form. We are not going to present the expanded form of this set of gauge transformations at this stage. For Courant algebroids it is found in various works in the literature, for example \cite{Ikeda:2002wh}. For twisted Courant algebroids it is instructive to isolate the $H$-dependence in this gauge transformations in order to understand its geometric significance. Note that only the top form field acquires an $H$-dependence in its gauge transformation, see App. \ref{appa}, therefore we only write this one: 
\be 
\d B_{\m}=\d B_{\m}|_{H=0}\, -\frac{1}{2} H_{\m abc}\,\e^aA^bA^c- \frac 12 H_{\m\n ab}\,\e^aA^bF^\n-\frac{1}{3!} H_{\m \n\r a}\,\e^aF^\n F^\r
\ee 
where $\d B_{\m}|_{H=0}$ is the transformation of the 2-form in the Courant sigma model and $F$ is given as
\be 
F^{\m}=\dd X^{\m}-\rho_a{}^{\m}A^a\,.
\ee 
Then $F^{\m}=0$ is the field equation of the 2-form field. Thus we observe that in presence of the 4-form $H$, there are field-equation dependent (trivial) gauge transformations in $\d B_{\m}$. What is more, these include a somewhat unorthodox nonlinear term in the field equations. 

Two important remarks are in order. First, the algebra of gauge transformations is both open and soft, in other words it only closes on the stationary surface and its structure ``constants'' are field dependent. Moreover, since the theory features a 2-form field, it is a first stage reducible Hamiltonian system. Let us take a closer look to the gauge algebra. 
We focus on the most complicated commutator of two gauge transformations on the highest degree field $B_{\m}$. This is 
\bea \label{gauge algebra}
[\d_1,\d_2]B_{\m}=\d_{12}B_{\m}+U_{\m\n}\w F^{\n}+V_{\m\n\r}F^\n\w F^{\rho}+W_{\m a}G^a\,,
\eea 
where $G^a$ is the field equation for the field $A^{a}$ given in non-covariant form as 
\be 
G^{a}:=\dd A^a +\eta^{ab}\r_b{}^\m B_\m+\frac 12{C}^{a}_{bc}A^b\w A^c=0\,,
\ee 
$\d_{12}$ is the gauge transformation with gauge parameters $\e^{a}_{12}$ and $\psi_{\m 12}$, which are found to be 
\bea 
\e_{12}^{a}&=& C_{bc}^a \e_1^b\e_2^c\,, \\[4pt] 
\psi_{\m 12}&=&2\partial_\m\r_a{}^\n\e_{[1}^a\psi_{\n 2]}-(\partial_\m C_{abc}+H_{\m abc})\e_1^a\e_2^b A^c-\sfrac 12  H_{\m\n ab}\e_1^a\e_2^b F^\n\,,
\eea 
and the three coefficients $U, V$ and $W$ are given by{\footnote{We note in advance that tensors such as the basic curvature appear in the covariant version of these coefficients. That the basic curvature appears as a coefficient in the gauge algebra of the simpler case of Lie algebroid gauge theories was first noticed in \cite{Mayer:2009wf}, even though the geometric meaning of the tensor was not yet identified there.}} 
\bea 
U_{\m\n}&=& 2\partial_\m\partial_\n\r_a{}^\s\psi_{\s[1}\e^a_{2]}+\e_1^a\e_2^b A^c (\partial_\m\partial_\n C_{abc}+\partial_{(\m}H_{\n)abc})\,,\\[4pt] 
\label{Vmnr}
V_{\m\n\rho}&=& -\sfrac 13 \e^a_1\e^b_2\, \partial_{(\m} H_{\n)\r a b}\,,\label{Vsym}\\[4pt] 
W_{\m a}&=&- \e_1^b\e_2^c (\partial_\m C_{abc}+\sfrac 12 H_{\m abc})\,.
\eea 
We observe that a term quadratic in the field equations is generated in the gauge algebra \eqref{gauge algebra}. This term is absent for vanishing 4-form. Thus we see that twisted Courant algebroids feature a new property compared with standard gauge theories, that of nonlinear openness of their gauge algebra. This was also identified before in the context of twisted R-Poisson models \cite{Chatzistavrakidis:2021nom}, further studied in \cite{Ikeda:2021rir,ChSI}. Related to this is the fact that the 1-form gauge parameter $\psi_{\m 12}$ receives a field equation contribution, one more unconventional feature of the model. In principle one should further calculate the Jacobi identity for the gauge algebra and higher identities thereof and identify all the structural quantities that appear on the way. We refrain from embarking in this complicated task for the moment, since an equivalent way of addressing this goes through the BRST transformation and its various powers, as we discuss below. Another feature of the gauge algebra which is worth highlighting is that the coefficient of the term which is nonlinear in the field equations is explicitly symmetric in its first two indices, see Eq. \eqref{Vmnr}. This will also play a crucial role in understanding the systematics of the BRST formalism for the model.

The second remark is that the gauge transformation $\d\phi^{\a}$ is not tensorial. To express it in a basis-independent form one must take into account changes of frame in the target space. This can be parametrized by a connection on the twisted Courant algebroid, which is one of the reasons we introduced such objects in the previous section. We will also account for this directly within the BRST formalism.   

\subsection{On-shell closed BRST transformations}\label{sec32}

Following the standard steps of the BRST procedure, we now promote the gauge transformations discussed previously to BRST transformations which will be nilpotent on-shell, namely on the constrained surface. The two gauge parameters $\e$ and $\psi$ are assigned a ghost degree 1 (we do not introduce new notation for them) and moreover due to the first stage reducibility of the system we must introduce an additional ghost of degree 2 which we denote as $\widetilde{\psi}$. This is the ghost for the ghost $\psi$. We summarize the fields and ghosts, which are the same as for the usual Courant sigma model, in Table \ref{table1}. 

		\begin{table}
	\begin{center}	\begin{tabular}{| c | c | c | c | c | c | c |}
			\hline 
			\multirow{3}{5.2em}{Field/Ghost} &&&&&& \\ & $X^{\m}$ & $A^a$  & $B_{\m}$ & $\epsilon^a$ & $\psi_{\m}$ & $\widetilde\psi_\m$ \\ &&&&&& \\ \hhline{|=|=|=|=|=|=|=|}
			\multirow{3}{6.0em}{Ghost degree} &&&&&& \\ & $0$ & $0$  & $0$ & $1$ & $1$ & $2$ \\ &&&&&& \\\hline 
			\multirow{3}{5.5em}{Form degree} &&&&&& \\  & $0$ & $1$  & $2$ & $0$ & $1$ & $0$ \\ &&&&&&
			\\\hline  \end{tabular}\end{center}\caption{The eight fields and ghosts of the twisted Courant sigma model (due to the index $a$ the fields $A$ and $\epsilon$ are doubled). The ghosts $\e^a, \psi_\m$ correspond to the scalar and 1-form gauge parameters and the degree 2 ghost $\widetilde{\psi}_{\m}$ is the ghost for the ghost $\psi_{\m}$. }\label{table1}\end{table}

 We denote the on-shell closed BRST transformations  as $s_0$ to avoid confusion with previous and later notation. We have not introduced antifields yet but we can already mention that $s_0$ is the ``longitudinal differential'' in the sense of \cite{HTbook} and therefore it annihilates antifields. According to our earlier discussion, the BRST transformation of the fields in non manifestly covariant form is
\be 
s_0\phi^{\a}=\dd \e^{\a}+\e^{\b}\partial_{\b}Q^{\a}+s_{\text{triv}}\phi^{\a}\,,
\ee 
where $Q^{\a}$ are the components of $Q^{H}$. The coefficients are controlled by the first derivatives of the homological vector field, as expected by the general form that contains the Lie derivative along it. Anticipating a generalization of the present situation to any higher gauge theory in the sense of Ref. \cite{Grutzmann:2014hkn}, we move on to determine the analogous general collective expressions for the ghosts and ghost for ghost, a task that was not performed in \cite{Grutzmann:2014hkn}.  
 To determine the general form of the BRST transformation on the ghosts we make the following degree 2 Ansatz, neglecting the pull-back map for the time being:
\be \label{s epsilon ansatz}
s_0\e^{\a}=\dd \widetilde\e^{\,\a}+\widetilde{\e}^{\,\b}\k^{\a}_{\b}(\phi)+\frac 12\e^\b\e^\g \l^{\a}_{\b\g}(\phi)+s_{\text{triv}}\e^{\a}\,,
\ee
where $\widetilde{\e}^{\,\a}=(0,0,-\widetilde{\psi}_{\m})$ is an alternative notation for the ghost for ghost that aligns with the previously introduced notation. 
Consistency of the BRST formalism imposes constraints on the undetermined coefficients $\k$ and $\l$. 
To determine $\k$ and $\l$ in Eq. \eqref{s epsilon ansatz} it is enough to require that the square of the BRST transformation on the fields vanishes weakly,  
\be 
s_0^{2}\phi^{\a}\approx 0\,, 
\ee 
with $\approx$ denoting weak equality. Note that $s_{\text{triv}}\phi^{\a}\approx 0$ anyway and it does not influence the rest of the calculation. The result of imposing this condition is that the coefficients $\k$ and $\l$ are given as 
\bea  
\k^{\a}_{\b}&=&-\partial_{\b}Q^{\a}\,, \\[4pt] 
\l^{\a}_{\b\g}&=& 
-\partial_\b\partial_\g Q^{\a}\,.
\eea   
We observe that they are controlled by the first and second derivatives of the homological vector field $Q^{H}$. We note in passing that second derivatives of the homological vector field are related to the Atiyah cocycle on the dg manifold $\mc M$, here for the trivial connection. A relation of the Atiyah cocycle to gauge theory was suggested in Ref. \cite{Chatzistavrakidis:2023otk}. 

To complete this local analysis, we need to determine the BRST transformation on the ghosts for ghosts $\widetilde\e^{\,\a}$. Since in the present case the single component of this quantity is a scalar field, we do not need to include any exterior derivatives. We make the degree 3 Ansatz
\be 
s_0\widetilde\e^{\,\a}=\widetilde\e^{\,\b}\e^\g\widetilde \k^\a_{\b\g}(\phi)+\sfrac{1}{3!}\e^\b\e^\g\e^\d\widetilde{\l}^{\a}_{\b\g\d}(\phi)\,.
\ee 
Consistency requires that 
\be 
s_0^2 \e^{\a}\approx 0\,,
\ee 
which fixes the new undetermined coefficients to be 
\bea 
\widetilde{\k}^{\a}_{\b\g}&=&\partial_{\b}\partial_{\g}Q^{\a}\,,\\[4pt] 
\widetilde{\l}^{\a}_{\b\g\d}&=&\partial_\b\partial_\g\partial_\d Q^{\a}\,.
\eea
It is thus found that also third derivatives of the homological vector field appear in the BRST transformation of the ghost for ghost. This is also expected in view of the relation of these transformations to the Kapranov $L_{\infty}[1]$ algebra that governs the BRST structure of any higher gauge theory as proposed in Ref. \cite{Chatzistavrakidis:2023otk} to which we refer for more details. 
Summarizing the above discussion and ignoring pullbacks for the time being, we have proven that consistency of the BRST formalism leads to the following:
\begin{prop}\label{prop s0}
    The BRST transformations of the fields $(\phi^{\a})=(X^{\m},A^{a},B_{\m})$, ghosts $(\e^{\a})=(0,\e^a,-\psi_{\m})$ and ghost for ghost $(\widetilde{\e}^{\,\a})=(0,0,-\widetilde\psi_{\m})$ for a 4-form twisted Courant sigma model take the following form in terms of the components $(Q^{\a})$ of the homological vector field $Q^{H}$  of a twisted Courant algebroid (Eq. \eqref{QH final}) and their derivatives:
    \begin{tcolorbox}[ams nodisplayskip, ams align]
    s_0\phi^{\a} &= \dd \e^{\a}+\e^{\b}\partial_{\b}Q^{\a}+s_{\text{triv}}\phi^{\a}\,,\\[4pt] 
    s_0\e^{\a}&=\dd \widetilde\e^{\,\a}-\widetilde{\e}^{\,\b}\partial_{\b}Q^{\a}-\frac 12\e^\b\e^\g \partial_{\b}\partial_{\g}Q^{\a}+s_{\text{triv}}\e^{\a}\,, \\[4pt] 
    s_0\widetilde{\e}^{\,\a}&=\widetilde\e^{\,\b}\e^\g \partial_{\b}\partial_{\g}Q^{\a}+\frac{1}{3!}\e^\b\e^\g\e^\d\partial_{\b}\partial_{\g}\partial_{\d}Q^{\a}\,,
   \end{tcolorbox}
    with $s_{\text{triv}}$ vanishing on the stationary surface.
\end{prop}

\subsection{Manifestly covariant BRST transformations}\label{sec33}

 We would now like to understand the global meaning of the various coefficients in Proposition \ref{prop s0} and reach a manifestly target space covariant formulation. 
 Let us first explain what it would mean to implement manifest target space covariance and obtain inherently tensorial expressions for the gauge symmetries. One may think of this as the following three-step algorithm: (i) Consider the field $\phi^{\a}$ and its non-tensorial gauge transformation $\d \phi^{\a}$ and rewrite it in terms of covariant quantities using the connection $\nabla$; (ii) consider the index-free field $\phi=\phi^{\a}\otimes {e}_{\a}$ and isolate the transformation of the pull-back basis ${e}_{\a}$ in $\d\phi^{\a}$; (iii) define a \emph{different}, tensorial transformation $\d^{\scriptscriptstyle\nabla}\phi$ by absorbing suitable field equation dependent terms such that the final expression is manifestly covariant and tensorial. 
 
 To illustrate this procedure let us consider an example. In the twisted Courant sigma model take the gauge transformation of the 1-form as given from Eq. \eqref{delta phi noncov}\footnote{A gauge transformation always has ambiguity of trivial gauge transformations, $s_{\text{triv}}\phi^{\a}$. We fix this ambiguity of the gauge transformation of $A^{a}$ to the standard form. Such ambiguities of gauge transformations in the BV and BFV formalisms have been discussed for the twisted PSM in \cite{Ikeda:2020eft}.}:  
 \be 
\d A^{a}=\dd \e^{a}+C_{bc}^{a}A^b\e^c +\eta^{ab}\rho_b{}^{\m}\psi_{\m}\,.
 \ee 
 Step (i) of the above algorithm instructs us to rewrite this transformation in the alternative form  
 \be \label{delta Aa covariantized}
\d A^{a}=\DD\e^{a}-2\eta^{ad}\mc T_{bcd}A^{b}\e^{c}+\eta^{ab}\rho_b{}^{\m}\psi^{\scriptscriptstyle\nabla}_{\m}-\omega^{a}_{\m b}F^{\m}\e^{b}-\rho_{b}{}^{\m}\omega_{\m c}^{a}A^{c}\e^{b}\,,
 \ee 
 where $\DD$ is the exterior covariant derivative induced by $\nabla$, $\mc T$ is the Gualtieri torsion tensor and $\psi^{\scriptscriptstyle\nabla}$ is a redefined 1-form gauge parameter. We take this opportunity to introduce the following redefined field and parameters in accord with the analysis of Section \ref{sec2}:{\footnote{Of course it is no accident that these three quantities appear together in the same superfield of degree 2 in the AKSZ construction of Courant sigma models.}}
\bea \label{nb}
B_\m^{\scriptscriptstyle\nabla}&=&B_\m+\sfrac 12 \o_{\m ab}A^aA^b\,, \\[4pt] 
{\psi}_{\m}^{\scriptscriptstyle\nabla}&=&{\psi}_{\m}+\o_{\m ab}A^a\e^b\,,\label{np}\\[4pt] 
\widetilde{\psi}_{\m}^{\scriptscriptstyle\nabla}&=&\widetilde{\psi}_{\m}+\sfrac 12\o_{\m ab}\e^a\e^b\,\label{ntp}.
\eea 
Note now that this completely equivalent form of the gauge transformation for the 1-form $A^{a}$ contains two additional terms generated during the covariantization process. If we now consider the pull-back basis $e_a$ of the vector bundle $X^{\ast}E$, this transforms under a change of frame according to 
\be 
\d  e_a=\omega_{\m a}^{b}\d X^{\m}  e_b\,.
\ee 
We now observe that for the field $A=A^{a}\otimes e_a$ we can apply the Leibniz rule to obtain 
\be 
\d A=\d A^{a}\otimes e_a+A^{a}\otimes \d e_{a} : =\bar{\d} A^{a}\otimes e_a\,,
\ee
where we defined the tensorial transformation (see e.g. \cite{Bojowald:2004wu})
\be \label{delta bar}
\bar{\delta}A^a=\d A^{a}+\omega_{\m b}^{a}A^b\d X^{\m}\,.
\ee 
It is a matter of mere inspection to see that the second term in this Leibniz rule corresponds precisely to the last term in the transformation $\d A^{a}$ in the form of Eq. \eqref{delta Aa covariantized}. This isolates the transformation of the pull-back basis as instructed in step (ii) of the algorithm. Finally we define the new transformation 
\be \label{forApp}
\d^{\scriptscriptstyle\nabla}A^{a}:=\d A^{a}+\omega^{a}_{\m b}F^{\m}\e^{b}\,,
\ee 
and the associated tensorial one 
\be \bar\d^{\scriptscriptstyle\nabla}A^{a}:=\d^{\scriptscriptstyle\nabla} A^{a}+\omega_{\m b}^{a}A^b\d X^{\m}\,.
\ee 
This is step (iii) of the algorithm and together with the previous steps leads to the final tensorial and manifestly covariant result 
\be 
\d^{\scriptscriptstyle\nabla} A=(\d^{\scriptscriptstyle\nabla}A)^{a}\otimes e_a=\bar{\d}^{\scriptscriptstyle\nabla}A^a\otimes e_a=\DD \e-2 \mc {T}(A,\epsilon)+\rho^{\ast}(\psi^{\scriptscriptstyle\nabla})\,,
\ee 
where $\rho^{\ast}$ is the transpose map to the anchor $\rho$. In the BRST setting we should replace $\d$ and $\d^{\scriptscriptstyle\nabla}$ with  operators $s_0$ and $s_0^{\scriptscriptstyle\nabla}$ and the gauge parameters with the corresponding ghosts. 

Even though the analysis above illustrates the general procedure, it hides the advantages of the covariantized homological vector field for twisted Courant algebroids introduced in Section \ref{sec2}. Indeed this rewriting obviates the need to apply the above algorithm. Instead one obtains \emph{automatically} the correct tensorial transformations on the fields of the model 
as follows. We start by revisiting the example of the field $A^{a}$ that was detailed above. 
Note first that there are two expansions of the homological vector field $Q^{H}$, one in the basis $(\partial_{\a})$ and one in the basis $(\DD_{\a})$
 introduced by the differential operators $\DD^{(0)}$, $\DD^{(-1)}$ and $\DD^{(-2)}$ in Eqs. \eqref{D0}, \eqref{D-1} and \eqref{D-2}. The coefficients in each expansions are of course different:
 \be 
Q^{H}=Q^{\a}\partial_{\a}=Q^{\a}_{\scriptscriptstyle\nabla}\DD_{\a}\,.
 \ee 
 According to Eq. \eqref{QH final} the ``covariant'' components are given as 
 \bea 
Q_{\scriptscriptstyle\nabla}^{\m}&=&a^{a}\rho_a{}^{\m}\,, \\[4pt] 
Q_{\scriptscriptstyle\nabla}^{a}&=&-\eta^{ab}\rho_b{}^{\m}b^{\scriptscriptstyle\nabla}_{\m}+{\mc T}_{bc}{}^{a}a^ba^c\,, \\[4pt] 
Q^{\scriptscriptstyle\nabla}_{\m}&=&-\nabla_{\m}\rho_a{}^{\n}b^{\scriptscriptstyle\nabla}_{\n}a^a-\frac 13 \widetilde{\mc S}_{\m abc}a^a a^b a^c\,.
 \eea 
 Notice now that 
 \bea 
\DD^{\m}_{(-2)}Q_{\scriptscriptstyle\nabla}^{a}&=&-\eta^{ab}\rho_{b}{}^{\m}\,, \\[4pt]
\DD_b^{(-1)}Q_{\scriptscriptstyle\nabla}^{a}&=&2{\mc T}_{bc}{}^{a}a^c\,.
 \eea 
 Putting these together we directly observe that 
 \be 
\bar\d^{\scriptscriptstyle\nabla}A^{a}=\DD\e^{a}+\e_{\scriptscriptstyle\nabla}^{\a}\DD_{\a}Q_{\scriptscriptstyle\nabla}^{a}\,,
 \ee 
 where we used the coefficients in the alternative expansion of the collective gauge parameter:
 \be 
\e=\e^{\a}\partial_{\a}=\e^{\a}_{\scriptscriptstyle\nabla}\DD_{\a}\,,
 \ee 
 which are nothing but $(\e^{\a}_{\scriptscriptstyle\nabla})=(0,\e^a,-\psi^{\scriptscriptstyle\nabla}_{\m})$. 
 Doing the same analysis for all fields and ghosts in the classical basis of the model, it  is straightforward to show that the tensorial version of Proposition \ref{prop s0} (with implicit pull-backs) is:
 \begin{prop}\label{prop s0 tensorial}
      The BRST transformations on the fields $\phi=(\phi_{\scriptscriptstyle\nabla}^{\a})=(X^{\m},A^{a},B^{\scriptscriptstyle\nabla}_{\m})$, the ghosts $\e=(\e_{\scriptscriptstyle\nabla}^{\a})=(0,\e^{a},-\psi^{\scriptscriptstyle\nabla}_{\m})$ and the ghost for ghost $\widetilde{\e}=(\widetilde{\e}_{\scriptscriptstyle\nabla}^{\,\a})=(0,0,-\widetilde\psi^{\scriptscriptstyle\nabla}_{\m})$ of a 4-form twisted Courant sigma model with source $T[1]\S$ and target $\mc M$ with coordinates $(x^{\a})=(x^{\m},a^{a},b^{\scriptscriptstyle\nabla}_{\m})$ of degrees $(0,1,2)$ are given in terms of the homological vector field $Q^{H}$ of a 4-form twisted Courant algebroid given in \eqref{QH final} and in tensorial form as $s_0^{\scriptscriptstyle\nabla}\phi=s_0^{\scriptscriptstyle\nabla}\phi^{\a}_{\scriptscriptstyle\nabla}\otimes e_{\a}$ (and similarly for $\epsilon$ and $\widetilde{\epsilon}$) with
   \begin{tcolorbox}[ams nodisplayskip, ams align]
   s_0^{\scriptscriptstyle\nabla} \phi_{\scriptscriptstyle\nabla}^{\a}&=\DD\e^{\a}_{\scriptscriptstyle\nabla}+\e^{\b}_{\scriptscriptstyle\nabla}\DD_{\b}Q^{\a}_{\scriptscriptstyle\nabla}+s^{\scriptscriptstyle\nabla}_{\text{triv}}\phi^{\a}_{\scriptscriptstyle\nabla}\,, \\[4pt] s_0^{\scriptscriptstyle\nabla}\e_{\scriptscriptstyle\nabla}^{\a}&=\DD\widetilde{\e}^{\,\a}_{\scriptscriptstyle\nabla}-\widetilde{\e}^{\,\b}_{\scriptscriptstyle\nabla}\DD_{\b}Q^{\a}_{\scriptscriptstyle\nabla}-\frac 12 \e_{\scriptscriptstyle\nabla}^{\b}\e_{\scriptscriptstyle\nabla}^{\g}\DD_{\b}\DD_{\g}Q^{\a}_{\scriptscriptstyle\nabla}+s^{\scriptscriptstyle\nabla}_{\text{triv}}\e^{\a}_{\scriptscriptstyle\nabla}\,, \\[4pt] 
s_0\widetilde{\e}_{\scriptscriptstyle\nabla}^{\,\a}&=\widetilde{\e}^{\,\b}_{\scriptscriptstyle\nabla}\e^{\g}_{\scriptscriptstyle\nabla}\DD_{\b}\DD_{\g}Q_{\scriptscriptstyle\nabla}^{\a}+\frac 1{3!}\e^{\b}_{\scriptscriptstyle\nabla}\e^{\g}_{\scriptscriptstyle\nabla}\e^{\d}_{\scriptscriptstyle\nabla}\DD_{\b}\DD_{\g}\DD_{\d}Q^{\a}_{\scriptscriptstyle\nabla}\,,
   \end{tcolorbox}
   where $\DD$ is the exterior covariant derivative on $\S$, the differential operators $\DD_{\a}$ are as in Eqs. \eqref{D0}, \eqref{D-1} and \eqref{D-2} and $s_{\text{triv}}^{\scriptscriptstyle\nabla}$ vanish on the stationary surface. 
 \end{prop}
The proof of the proposition is completed using the following set of equations for first derivatives
\begin{align*}
&\DD^\n Q_{\scriptscriptstyle{\nabla}}^\m=0 ~,   \qquad\qquad \qquad \DD_aQ_{\scriptscriptstyle{\nabla}}^\m=\r_a{}^\m~,\\[4pt]
&\DD^\n Q_{\scriptscriptstyle{\nabla}}^a=-\eta^{ab}\r_b{}^\n~, \quad \qquad \,  \DD_b Q_{\scriptscriptstyle{\nabla}}^a=2{\cal T}^a_{bc}a^c\,, \\[4pt]
&\DD^\n Q^{\scriptscriptstyle{\nabla}}_{\m}=-\nabla_\m\r_a{}^\n  a^a ~,  \qquad \DD_aQ^{\scriptscriptstyle{\nabla}}_{\m}=-\nabla_\m\r_a{}^\n b^{\scriptscriptstyle{\nabla}}_\n-\widetilde{\cal S}_{\m abc}a^ba^c~,
\end{align*}
and the following set for second and third derivatives
\begin{align*}
&\DD_a \DD^\n Q^{\scriptscriptstyle{\nabla}}_{\m}=-\nabla_\m\r_a{}^\n   ~,   \qquad \DD_c\DD_b Q_{\scriptscriptstyle{\nabla}}^a=2{\cal T}^a_{bc}~,\\[4pt]
&\DD_b\DD_aQ^{\scriptscriptstyle{\nabla}}_{\m}=-2\widetilde{\cal S}_{\m abc} a^c~, \quad \, \DD^\n\DD_aQ^{\scriptscriptstyle{\nabla}}_{\m}=-\nabla_\m\r_a{}^\n\,,\\[4pt]
& \DD_c\DD_b\DD_aQ_{\scriptscriptstyle{\nabla}\m}=-2\widetilde{\cal S}_{\m abc} ~.
\end{align*}
These allow to express the tensorial form of the BRST transformations on the fields and ghosts in terms of the classical differential geometric data on a twisted Courant algebroid, namely the anchor map $\rho$ and its transpose $\rho^{\ast}$, the Gualtieri torsion $\mc T$ and the 4-form twisted basic curvature $\widetilde{\mc S}$. We present some additional  details and expressions in a coordinate basis as supplementary material in Appendix \ref{appb}; here we report the final manifestly tensorial and basis-independent result---with suitable grouping of terms: 
\bea \label{snablaX}
s_0^{\scriptscriptstyle\nabla} X&=& \rho(\e)\,, \\[4pt]
\nn \\[4pt]
s_0^{\scriptscriptstyle\nabla} \e &=& \,\, \quad -\,\rho^{\ast}(\widetilde{\psi}^{\scriptscriptstyle\nabla})+\mc T(\e,\e)\,, \label{snablae}\\[4pt]
 s_0^{\scriptscriptstyle\nabla}A&=& \DD\e+\rho^{\ast}(\psi^{\scriptscriptstyle\nabla})-2{\mc T}(A,\e)\,, 
\label{snablaA}\\[4pt] 
\nn \\[4pt]
 s_0^{\scriptscriptstyle\nabla}\widetilde{\psi}^{\,\scriptscriptstyle\nabla}&=& \hspace{106pt} -\,\nabla\rho\,(\widetilde{\psi}^{\,\scriptscriptstyle\nabla},\e) -\sfrac 13 \widetilde{\mc S}(\e,\e,\e)\,, \label{snablawp}\\[4pt] 
 s_0^{\scriptscriptstyle\nabla}\psi^{\scriptscriptstyle\nabla} &=& \hspace{10pt} {\rm D}\widetilde{\psi} ^{\,\scriptscriptstyle\nabla}+\nabla\r(\widetilde{\psi}^{\,\scriptscriptstyle\nabla}, A)+\nabla\r({\psi}^{\scriptscriptstyle\nabla},\e)+\widetilde{\mc S}(A,\e,\e)+s_{\text{triv}}^{\scriptscriptstyle\nabla}\psi^{\scriptscriptstyle\nabla}\,,\label{snablap}
\\[4pt] 
s_0^{\scriptscriptstyle\nabla} B^{\scriptscriptstyle\nabla} &=& -\DD\psi^{\scriptscriptstyle\nabla}+\nabla\rho(\psi^{\scriptscriptstyle\nabla},A)-\nabla\rho(B^{\scriptscriptstyle\nabla},\e)-\widetilde{\mc S}(A,A,\e)+s_{\text{triv}}^{\scriptscriptstyle\nabla}B^{\scriptscriptstyle\nabla}\,,\label{snablaB}
\eea 
where the final terms in the BRST transformation of the 1-form ghost and the 2-form field are given by 
\bea\label{strivPsi}
s_{\text{triv}}^{\scriptscriptstyle\nabla}\psi^{\scriptscriptstyle\nabla}&=&\frac 14( H(\r(\e),\r(\e))-R(\e,\e))(-,F) \,,\\[4pt]
s_{\text{triv}}^{\scriptscriptstyle\nabla}B^{\scriptscriptstyle\nabla}&=&-  \frac{1}{2}( H(\r(\e),\r(A))-R(\e,A))(-,F)-\frac{1}{3!} H(-,\r(\e),F,F)\label{strivB}
\eea
To completely clarify and explain our notation, first note that all geometric objects mentioned below are understood in composition with the map $X$ and their components in a coordinate basis are therefore functions of the pull-back coordinates $X^{\m}$. One can understand this better by noting that for example the scalar parameter $\e\in \G(X^{\ast}E[1])$ (which should not be confused with the collective gauge parameter introduced earlier) is also a section of the pull-back bundle, and thus $\rho$ cannot just act on it since its domain is the bundle $E$; a composition with $X$ is due. The same remark holds in all other cases.  

The covariant BRST transformation $s_0^{\scriptscriptstyle\nabla}$ (obtained directly from Proposition \ref{prop s0 tensorial} and corresponding to $\bar\d^{\scriptscriptstyle\nabla}$ on the classical fields) is \emph{different} off-shell than the BRST transformation $s_0$ (corresponding to $\bar\d$ on fields) for all non-scalar fields, but remains unchanged for the scalar fields $X, \e$ and $\widetilde{\psi}^{\,\scriptscriptstyle\nabla}$. To explain this, first note that the field equations of the fields $B_{\m}$ and $A^{a}$ written in covariant form are 
\bea \label{HPSMeom1}
F^{\m}&:=&\dd X^{\m}-\r_a{}^\m A^a=0\,,
\\[4pt] \label{HPSMeom2}
G_{{\scriptscriptstyle\nabla}}^{a}&:=&G^{a}+\o_{\m b}^aF^\m A^b={\rm D} A^a +\eta^{ab}\r_b{}^\m B_\m^{\scriptscriptstyle\nabla}-\eta^{ab}{\mc T}_{cdb}A^cA^d=0\,.
\eea  
Our definition of covariant BRST transformation is then given for the three non-scalar fields as: 
\bea \label{snA}
s_0^{\scriptscriptstyle\nabla}A^a&=&s_0A^a+\omega^{a}_{\m b}F^{\m}\e^b\,,\\[4pt] 
s_0^{\scriptscriptstyle\nabla}\psi^{\scriptscriptstyle\nabla}_{\m}&=& s_0{\psi}_{\m}^{\scriptscriptstyle\nabla} -\mathring{\G}_{\m\s}^\n{\widetilde\psi}_{\n}^{\,\scriptscriptstyle\nabla}F^\s+\sfrac 12 (\partial_{(\m}\o_{\n) ab}+\o_{[\m a}^d\o_{\n] bd})F^\n\e^a \e^b\,, \label{snp}\\[4pt] 
s_0^{\scriptscriptstyle\nabla}{B}_{\m}^{\scriptscriptstyle\nabla} &=& s_0{B}_{\m}^{\scriptscriptstyle\nabla} -\mathring{\G}_{\m\s}^\n{\psi}_{\n}^{\scriptscriptstyle\nabla}F^\s- (\partial_{(\m}\o_{\n) ab}+\o_{[\m a}^d\o_{\n] bd})F^\n A^a \e^b- \o_{\m a b}G^a\e^b\,,\label{snB}
\eea 
where $s_0\phi^{\a}$ is the tensorial transformation corresponding to the gauge transformation $\bar\d\phi^{\a}$.
Obviously on scalar fields there is no difference since there are no scalar  equations of motion and also the field equation of $X^{\m}$ plays no role in this since it is of too high degree (a 3-form). 

\section{The BRST power finesse and the master action}
\label{sec4}

The BRST transformation is nilpotent on-shell, a fact that does not depend on whether a manifest target space covariant formulation is employed or not---we shall mostly be working on a local patch in this section but the globalisation can be performed in a fairly straightforward way as in previous sections. Off-shell its square contains contributions proportional to the field equations of the model, a fact we already used in order to fix it. It is instructive and in fact very useful to determine these expressions, see Appendix \ref{appb} for more technical details. In terms of the components $Q^{\a}$ of the homological vector field $Q^{H}$, using definition \eqref{field strengths Q} for the various field strengths that correspond to the field equations of the model, first we find that\footnote{The consistent choice of grading is dictated by the form degree of the field in question: $|\phi|=0,\,|\e|=-1,\,|\widetilde{\e}|=-2,\,|\mc F|=1$. Note also that we use Deligne's sign convention for fields with multiple grading, e.g. for differential form degree $f$ and ghost degree $g$: $\alpha_1\alpha_2=(-1)^{f_1f_2+g_1g_2}\a_2\a_1$.} 
\be \label{sF}
s_0{\cal F}^{\a}=(-1)^{|\e^\b|}\e^\b{\mc F}^\g\partial_\g\partial_\b Q^\a+\dd(s_{\rm triv} \phi^\a)-s_{\rm triv} \phi^\b\partial_\b Q^\a
~,
\ee 
where ${\cal F}^{\a}=\{F^\m,G^a\}$ are the field equations defined in \eqref{HPSMeom1} and \eqref{HPSMeom2}.
 This is simply the statement that the field strengths transform covariantly. Furthermore, the Bianchi identities have the simple form
 \be\label{BI}
 \dd{\mc F}^\a+{\mc F}^\b\partial_\b Q^\a=0~.
 \ee
 We need these transformations in order to proceed with determining the explicit form of the square of the BRST transformation on the classical basis. Indeed it is straightforward to show that 
 \bea \label{s2phi}
s_0^2\phi^{\a}&=& -\widetilde{\e}^\b {\mc F}^\g\partial_\g\partial_\b Q^\a-\sfrac 12 (-1)^{|\e^\b|+|\e^\g|}\e^\b \e^\g{\mc F}^\d\partial_\d\partial_\b\partial_\g Q^\a \nn\\[4pt]
&& +\,\dd(s_{\rm triv} \e^\a)+s_0(s_{\rm triv} \phi^\a)+ s_{\rm triv} \e^\b\partial_\b Q^\a-\e^\b s_{\rm triv} \phi^\g\partial_\g\partial_\b Q^\a \,,\\[4pt] 
s_0^2\epsilon^{\a}&=&(-1)^{|\e^\g|}\widetilde{\e}^\b \e^\g{\mc F}^\d\partial_\d\partial_\b\partial_\g Q^\a+\sfrac {1}{3!} (-1)^{|\e^\b|+|\e^\g|+|\e^\d|}\e^\b \e^\g\e^\d{\mc F}^\k\partial_\k\partial_\b\partial_\g\partial_\d Q^\a \nn\\[4pt]
&&+\,s_0(s_{\rm triv} \e^\a)-s_{\rm triv} \e^\b \e^\g\partial_\b\partial_\g Q^\a-\sfrac 12\e^\b \e^\g s_{\rm triv} \phi^\d\partial_\d\partial_\b\partial_\g Q^\a \label{s2e}\,. 
 \eea 
The square of the BRST transformation on all scalar fields vanishes identically, which also explains the absence of $s_0^2\widetilde{\e}^{\,\a}$. 
 Using \eqref{sF} and \eqref{BI} we can easily  see that both $s_0^2\phi^{\a}$ and $s_0^2\e^{\a}$ vanish weakly. Additionally,   expression \eqref{s2phi} is used to find the trivial transformation of the ghosts from the trivial transformations of the physical fields by demanding the mutual cancellation of the $\dd\e^\a$-dependent terms arising from the first two terms in the second line of \eqref{s2phi}.
This results in the general expressions
\bea
s_{\text{triv}}\phi^{\a}&=&-(-1)^{(|Q^\b|+1){\mc F}^\a}\frac{1}{3!}H^\a{}_{\b\g\d}\e^\varepsilon\partial_\varepsilon Q^\b\left(3{\mc F}^\g Q^\d +{\mc F}^\g {\mc F}^\d\right)~,\\[4pt]
s_{\text{triv}}\e^{\a}&=&(-1)^{(|Q^\b|+1){\mc F}^\a}\frac{1}{4}H^\a{}_{\b\g\d}\e^\varepsilon\partial_\varepsilon Q^\b\e^{\varepsilon'}\partial_{\varepsilon'} Q^\g{\mc F}^\d ~,
\eea
with only nonvanishing component of $H^{\a}{}_{\b\g\d}$ being the 4-form $H_{\m\n\k\l}$. Although this is a somewhat redundant presentation of the already found ones in \eqref{strivPsi} and \eqref{strivB}, it shows the appearance of the derivatives of the homological vector field in the trivial transformations.  
 Furthermore, the explicit form of the non-vanishing squares on the non-scalar fields is:
\bea\label{ncovBRST2}
s_0^2 A^a &=& -\eta^{ab} F^\m\bigg(\partial_\m\r_b{}^\n\widetilde{\psi}_\n +\sfrac 12(\partial_\m C_{bcd}+\sfrac 12 H_{\m bcd})\e^c\e^d\bigg)~,\nn\\[4pt]
s_0^2\psi_\m &=&-F^\l\bigg(\partial_\m\partial_\l \r_a{}^\n\widetilde{\psi}_\n\e^a+\sfrac 16(\partial_\m\partial_\l  C_{abc }+\partial_{(\m}H_{\l) abc})\e^a\e^b\e^c\bigg)
~,\nn\\[4pt]
s_0^2 B_\m &=&-F^\l\bigg(\partial_\m\partial_\l \r_a{}^\n(\psi_\n \e^a+\widetilde{\psi}_\n A^a)+\sfrac 12(\partial_\m\partial_\l  C_{abc }+\partial_{(\m}H_{\l) abc}) A^a\e^b\e^c\bigg)\nn\\[4pt]
&& -\, G^a\bigg(\partial_\m\r_a{}^\n\widetilde{\psi}_\n +\sfrac 12(\partial_\m C_{abc}+\sfrac 12 H_{\m abc})\e^b\e^c\bigg)- \sfrac 16  \partial_{(\m}H_{\l) \s ab}\e^a\e^b F^\l F^\s\,.
\eea
These reflect the openness of the gauge algebra of the model, even more so the nonlinear openness as obvious from the last  term  in Eq. \eqref{ncovBRST2}, where we note again the symmetrization in the first two indices in the derivative of the 4-form components.

At the end of the day we should have a BV-BRST transformation which is nilpotent off-shell. The usual procedure invokes antifields such that they transform to equations of motion. In that way the contributions of the square $s_0^2$ will appear in the solution of the classical master equation together with the prescribed combination of antifields.  
We shall return to this in more detail below. 
What is important to mention here already is that besides the square of the BRST transformation, higher powers of it which are also weakly vanishing can (and in the present case will) in principle play the same role. 
In other words, one should also compute all necessary $s_0^n$ for $n>2$, in the precise way we will explain below. 
In practice it is not necessary (or possible) to compute all these powers. However, there is a maximum power $n$ relevant in each case. To determine this power we need to introduce the antifields, even though we will continue working with the longitudinal differential (which annihilates them) and not with the Koszul-Tate one. There is one antifield for each field, ghost and ghost for ghost of the model. In the present case of twisted Courant sigma models the set of antifields together with their ghost number and form degree appear in Table \ref{table2}.

 \begin{table}
	\begin{center}	\begin{tabular}{| c | c | c | c | c | c | c |}
			\hline 
			\multirow{3}{4em}{Antifield} &&&&&& \\ & $X^{\dagger}_{\m}$ & $A^{\dagger}_a$  & $B^{\dagger\,\m}$ & $\epsilon^{\dagger}_a$ & $\psi^{\dagger \,\mu}$ & $\widetilde{\psi}^{\dagger\,\m}$ \\ &&&&&& \\ \hhline{|=|=|=|=|=|=|=|}
			\multirow{3}{6.0em}{Ghost degree} &&&&&& \\ & $-1$ & $-1$ & $-1$ & $-2$ & $-2$ & $-3$ \\ &&&&&& \\\hline 
			\multirow{3}{5.5em}{Form degree} &&&&&& \\  & $3$ & $2$ & $1$ & $3$ & $2$ & $3$ \\ &&&&&& 
			\\\hline  
	\end{tabular}\end{center}\caption{The set of antifields of the twisted Courant sigma model.}\label{table2}\end{table} 

 We recall the main properties of these antifields. For a given field or ghost $\varphi$ the corresponding antifield $\varphi^{\dagger}$ has ghost number $\text{gh}(\cdot)$ and form degree $\text{deg}(\cdot)$  such that 
 \bea 
\label{ghost sum} &&\text{gh}(\varphi)+\text{gh}(\varphi^{\dagger})=-1\,, \\[4pt] 
&&\text{deg}(\varphi)+\text{deg}(\varphi^{\dagger})=3\,.
 \eea 
 In words, the sum of the form degrees of any field and its antifield must be equal to the spacetime dimension where the model is defined, here 3, because they appear as a pair in the sector of the master action with one antifield, whereas the ghost degrees sum to $-1$ since the BRST formalism introduces the extended cotangent bundle $T^{\ast}[-1]\mc M$ with negative degree shift that accomodates the fields and antifields together. The BV-BRST transformation{\footnote{Our terminology is such that the off-shell nilpotent transformation is called BV-BRST and the only on-shell nilpotent one is called just BRST.}} of each antifield is an equation of motion for its associated field, in our conventions
 \be\label{s antifield}
s \varphi^{\dagger}= (-1)^{\rm deg(\varphi^\dagger)}{\cal F}_{\varphi}+\dots\,,
 \ee 
 where the ellipses denote further terms that are antifield dependent and guarantee that the BV-BRST transformation is nilpotent. Recall that the $s_0$ (or $s_0^{\scriptscriptstyle\nabla}$) part of $s$ annihilates the antifields, 
 \be 
 s_0\varphi^{\dagger}=0\,,
 \ee 
 and the part of it that is responsible for Eq. \eqref{s antifield} is the Koszul-Tate differential \cite{HTbook}. 
 
 In the present work we will not have to deal with the full BV-BRST transformation. The above properties of the antifields already dictate which powers of the BRST transformation before the introduction of antifields are sufficient to determine the master action. To find this all we need to know is that the master action must have vanishing ghost degree. Obviously the classical action $S_{\text{cl}}$ already satisfies this requirement. Furthermore the generic term $\varphi^{\dagger} s_0\varphi$ also satisfies it for any field and ghost due to \eqref{ghost sum} and the fact that the BRST transformation has ghost degree $1$. This is the well known term in the master action with one antifield. Taking this logic one step further, let us consider the square of the BRST transformation on some field or ghost $\varphi$. This is either zero or it contains terms proportional to one or more equations of motion, including products thereof in the case of nonlinear openness. To include these terms in the action we need to replace  equations of motion by corresponding antifields that transform to them under the Koszul-Tate differential. Then one must consider higher powers of the BRST transformation and do the same as long as the result is of vanishing ghost degree. 
 To remain general, assume that we include in the action a term of the type 
 \be \label{phidagger O phi}
\varphi^{\dagger} \mc O_{(n)} \varphi\,, 
 \ee 
of ghost degree zero, and with $\mc O_{(1)}\varphi=s_0\varphi$. 
 Assuming absence of scalar antifields and of terms in the action that would contain a Hodge star, the lowest form degree of an antifield is 1 and we conclude that there is a bound on the integer $n$, equivalently on the power $n-1$:
 \be 
n-1\le \text{deg}(\varphi)\,.
 \ee 
 Thus the total power of the BRST transformation we must consider is field dependent and the absolute maximum power in a model is equal to the highest form degree. For instance in the case of the twisted Courant sigma model this power amounts to $n=3$, whereas for the twisted Poisson sigma model it is not necessary to go beyond $n=2$. 
 
Based on the above discussion and disregarding scalar fields for which this is obviously irrelevant, we see that for the 1-forms $A^{a}$ and $\psi_{\m}$ we only need to know the square of their BRST transformation, which we have already found, see \eqref{ncovBRST2}. These will contribute to the action as
\bea
\varphi^{\dagger}{\mc O}_{(2)}\varphi=\varphi^{\dagger} \mf{f}\left(s_0(\mc O_{(1)}\varphi)\right)~,
\eea
where we defined the operation of taking any expression $\varphi$ (here of the form $(s_0)^2\varphi$) that contains an equation of motion and replacing the equation of motion with the corresponding antifield: 
\be 
\mf{f}(\varphi)=\varphi|_{{\cal F}\to (-1)^{{\rm deg} {\cal F}+1}\varphi^\dagger_{\cal F}}\,.
\ee 
Specifically, 
\bea\label{o2}
A^{\dagger}_{a}{\mc O}_{(2)} A^a &=& \eta^{ab} A^{\dagger}_{a}B^{\dagger\m}(-\partial_\m\r_b{}^\n\widetilde{\psi}_\n -\sfrac 12(\partial_\m C_{bcd}+\sfrac 12 H_{\m bcd})\e^c\e^d)~,\nn\\[4pt]
\psi^{\dagger\,\m}{\mc O}_{(2)}\psi_\m &=&\psi^{\dagger\,\m}B^{\dagger\,\l}(-\partial_\m\partial_\l \r_b{}^\n\widetilde{\psi}_\n\e^a-\sfrac 16(\partial_\m\partial_\l  C_{abc }+\partial_{(\m}H_{\l) abc})\e^a\e^b\e^c)
~.
\eea
Moving on, for the 2-form $B_{\m}$ there is a subtlety due to the nonlinear openness of the gauge algebra. Recall that in the commutator of gauge transformations on the 2-form, Eq. \eqref{gauge algebra}, we encountered a nonlinear term with symmetrization in the first two indices of its coefficient. The same structure appears in the square of the BRST transformation, the last term in Eq. \eqref{ncovBRST2}.  Since this will have to appear together with a product of two commuting antifields, we are prompted to replace only the equation of motion whose index is within the symmetrization. In other words the operator $\mf{f}$ acts only on one of the two field equations in the product as specified. This gives the term
\bea
B^{\dagger\,\m}{\mc O}_{(2)} B_\m &=&B^{\dagger\,\m}B^{\dagger\l}(-\partial_\m\partial_\l \r_a{}^\n(\psi_\n \e^a+\widetilde{\psi}_\n A^a)-\sfrac 12(\partial_\m\partial_\l  C_{abc }+\partial_{\m}H_{\l abc}) A^a\e^b\e^c)\nn\\[4pt]
&&-\,\eta^{ad}B^{\dagger\,\m}A^\dagger_d(-\partial_\m\r_a{}^\n\widetilde{\psi}_\n -\sfrac 12(\partial_\m C_{abc}+\sfrac 12 H_{\m abc})\e^b\e^c)\nn\\[4pt]
&&-\, \sfrac 16  B^{\dagger\,\m}B^{\dagger\l}\partial_{\m}H_{\l \s ab}\e^a\e^b  F^\s~.
\eea
Furthermore, for the 2-form field $B_\m$ we expect a cubic antifield contribution of ghost degree zero\footnote{The cubic contributions of higher ghost degree contribute to potential anomalies controlled by Bianchi identities \cite{Barnich:2000me}. We shall not analyse these issues here.}.
The cubic contribution is obtained by replacing the remaining equation of motion in $O_{(2)}\varphi$ with the corresponding antifield, performing a (third) BRST transformation and again replacing the newly generated field equation with the corresponding antifield. The result contains terms of various ghost degrees and to isolate the correct term we should project to degree 0. What we explained in words takes the form: 
\be
\varphi^{\dagger}{\mc O}_{(3)}\varphi=\varphi^{\dagger}\mf{f}\left(s_0\left(\mf{f}\left({\mc O}_{(2)}\varphi\right)\right)\right)|_{_{\text{gh}=0}}\,.
\ee
This is the right way in which the cube of the BRST transformation contributes. 
Specializing to the only field for which this general procedure is relevant in the present case, the contribution to the master action is
\bea\label{O3B}
B^{\dagger \m}{\mc O}_{(3)} B_\m &=&B^{\dagger \m}B^{\dagger \l}B^{\dagger \s}\bigg(\partial_\m\partial_\l\partial_\s \r_a{}^\n\widetilde{\psi}_\n\e^a +\sfrac 16 \partial_\m\partial_\l\partial_\s C_{abc} \e^a\e^b\e^c\nn\\[4pt]
 &&\qquad\qquad\quad +\, \sfrac 16\big(\partial_\s\partial_{\m}H_{\l abc}-\sfrac 32 \partial_\m\partial_\l\r_a{}^\n H_{\n\s bc}\big)\e^a\e^b\e^c\bigg)\,.
\eea
The procedure is described in more technical detail in App.\ref{appc}.
Then as a preliminary, calibrating result which is simple to prove we state the following for the case of vanishing 4-form: 
\begin{prop} \label{prop AKSZ}
The AKSZ-BV master action for the (untwisted) Courant sigma model can be expressed as 
\bea 
S_{\text{\tiny{\rm BV,0}}}=S_{\rm cl}|_{_{H=0}}+\sum_{n=1}^{3}\frac{(-1)^{gh \varphi+n}}{\beta_{\dagger}!\beta_n}\,\varphi^{\dagger}\cdot \mc O_{(n)}\varphi|_{_{H=0}}\,, \label{S BV noH}
\eea 
where $S_{\rm cl}$ is the classical action, $\varphi=(\phi,\e,\widetilde\e)$ is the collection of fields, ghosts and ghost for ghost, and the  $\mc O_{(n)}\varphi$ is defined as above with $H=0$. The factor $\beta_{\dagger}$ counts the number of antifields of the same kind in any of the terms in the sum and the symmetry factor $\beta_n$ counts the terms that appear more than once for a fixed $n$.
\end{prop}
The proof is a direct comparison{\footnote{Up to irrelevant differences in conventions and notation.}} of each term in the sum of Eq. \eqref{S BV noH} with the corresponding terms with 1, 2 and 3 antifields in Ref. \cite{Roytenberg:2006qz}. This is then nothing but an alternative rewriting of the AKSZ theory in 3D, in which the expansion in the antifields is obtained at face value. This slightly alternative perspective is useful for two reasons. First, it can be upgraded relatively easily to a manifestly target space covariant form where the various coefficients in the master action acquire a geometric interpretation in terms of the Courant algebroid structures and tensors, namely the Gualtieri torsion and the basic curvature. The second advantage, in which we focus here, is that the above form of the master action can be used to account for the \emph{4-form twisted} Courant sigma model for which there is no three-dimensional AKSZ construction.{\footnote{Even though one can consider that it can be obtained as a boundary theory of the AKSZ construction in four dimensions, see e.g. the approach developed in Ref. \cite{Ikeda:2013wh}, it would be difficult to find the right boundary conditions for this to happen, especially to do this in a systematic way.}} This is accounted for by the following:
\begin{theorem}\label{Thm}
 The minimal solution of the classical master equation for the 4-form twisted Courant sigma model is
\begin{tcolorbox}[ams nodisplayskip, ams align]
S_{\text{\tiny{\rm BV}}}=S_{\rm cl}+\sum_{n=1}^{3}\frac{(-1)^{gh \varphi+n}}{\beta_{\dagger}!\beta_n}\,\varphi^{\dagger}\cdot \mc O_{(n)}\varphi\,, \label{S BVH}
\end{tcolorbox} 
where $S_{\rm cl}$ is the classical action, $\varphi=(\phi,\e,\widetilde\e)$ is the collection of fields, ghosts and ghost for ghost, and the  $\mc O_{(n)}\varphi$ is defined as above including the 4-form contributions. The factor $\beta_{\dagger}$ counts the number of antifields of the same kind in any of the terms in the sum and the symmetry factor $\beta_n$ counts the terms that appear more than once for a fixed $n$.
\end{theorem}
We call this procedure for obtaining the master action the ``BRST power finesse'' because instead of the standard approach of homological perturbation theory where the master equation is solved order by order by adding terms to the master action with increasing number of antifields, here all essential information is obtained directly from the on-shell closed BRST transformation. This is of course completely equivalent, it just avoids some technically challenging steps at the price of computing the higher powers of the BRST transformation as explained above. In cases where the AKSZ construction is not directly applicable or when a global geometric interpretation of the coefficients in the master action is desired, this procedure is technically advantageous.

The detailed expression for the BV master action is given in App.\ref{appd}. Obviously it reduces to the AKSZ-BV action when the 4-form vanishes. Moreover, it is directly checked that for the special case of $M$ being a twisted R-Poisson manifold of order 3 in the sense of Ref. \cite{Chatzistavrakidis:2021nom} it reduces to the master action of the three-dimensional twisted R-Poisson sigma model, which was tediously determined in Ref. \cite{ChSI}, see in particular Section 4 in that paper. The agreement of the presently more general case to the special case there is established with the following identifications, 
\be 
\rho=\Pi^{\sharp}\,, \quad C_{\r}^{\m\n}=\partial_{\r}\Pi^{\m\n}\,,\quad C^{\m\n\r}=R^{\m\n\r}\,,
\ee 
where $\Pi^{\sharp}:T^{\ast}M\to TM$ is the map induced by a Poisson structure $\Pi$ and $R^{\m\n\r}$ are the components of a 3-vector that satisfies 
\be 
[\Pi,R]_{\text{SN}}+\langle\Pi^{\otimes 4},H\rangle\,,
\ee 
which is the defining condition of a 4-form twisted R-Poisson structure of order 3, giving rise to a special case of a 4-form twisted Courant algebroid. 

The general result may also be written as 
\be
S_{\text{\tiny{\rm BV}}}=S_{\text{\tiny{\rm BV,0}}}+ S_{\text{\tiny{\rm BV,H}}}~,
\ee
where the AKSZ action is what we denoted above as $S_{\text{\tiny{BV,0}}}$ and $ S_{\text{\tiny{\rm BV,H}}}$ contains only $H$-dependent terms. This is useful in isolating the new contributions and eventually proving that 
\bea  
&&\left(S_0,S_3\right)_{\text{\tiny{\rm BV}}}+\left(S_1,S_2\right)_{\text{\tiny{\rm BV}}}=0\,, \\[4pt] 
&&2(S_1,S_3)_{\text{\tiny{\rm BV}}}+(S_2,S_2)_{\text{\tiny{\rm BV}}}=0\,,
\eea  
where $S_i$ is the sector of $S_{\text{\tiny{BV}}}$ with $i$ antifields and $(\cdot,\cdot)_{\text{\tiny{\rm BV}}}$ is the odd BV bracket. Using the axioms of the pre-Courant algebroid in the covariant form \eqref{id2 cov} and \eqref{CAidentity23cs}, their derivatives and the following identity,
\begin{tcolorbox}[ams nodisplayskip, ams align]
\r_{[a}{}^\m\nabla_\m \widetilde{\mc S}_{\n bcd]}=-\nabla_\n\r_{[a}{}^\m\widetilde{\mc S}_{\m bcd]}-3\,\widetilde{\mc S}_{\n e[ab}{\mc T}^e_{cd]}~,
\end{tcolorbox}
we have validated by direct computations
that these nontrivial conditions hold. Together with the remaining, simpler identities, this establishes that this is a solution to the classical master equation and that the above theorem is valid. 

\paragraph{Comment on the quantum master equation.} Before closing this section, let us turn to the question of the solution to the quantum master equation. Recall that the BV Laplacian is defined as 
\be 
\D=\sum_{\a}(-1)^{\text{gh}(\varphi^{\a})}\frac {\d^{2}}{\d\varphi^{\a}\varphi^{\dagger}_{\a}}\,,
\ee 
where the sum is taken over all fields and ghosts and the functional derivative acts from left to right.
Then the quantum master equation is 
\be 
i\hbar \D W=\frac 12 (W,W)_{\text{BV}}\,,
\ee 
where $W$ is the quantum action whose loop expansion starts with the classical master action $S_{\text{\tiny{\rm{BV}}}}$. 
It was shown in Ref. \cite{Cattaneo:1999fm} that the minimal solution to the classical master equation of the Poisson sigma model satisfies the quantum master equation too. Note, moreover, that this remains true for the twisted Poisson sigma model, as can be directly seen via its master action in Eq. (4.36) of Ref. \cite{Ikeda:2019czt}.

Let us check what happens in the Courant or twisted Courant sigma model. 
 There are six terms in the minimal solution to the classical master action on which the BV Laplacian does not vanish, namely those with simultaneous appearance of a field and its antifield: 
 \be
 S_{h}=\int \big(-X^{\dagger}_{\m}\rho_{a}{}^{\m}(X)\e^{a}-(A^{\dagger}_{a}A^{b}+\frac 12 \e^{\dagger}_{a}\e^{b})C^{a}_{bc}\e^{c}  
 +  (B^{\dagger \m}B_{\n}+\psi^{\dagger\m}\psi_{\n}+\widetilde{\psi}^{\dagger\m}\widetilde{\psi}_{\n})\partial_{\m}\rho_{a}{}^{\n}\e^{a}\big)\,.\label{Sh}
 \ee
 This is easier to see in the expanded form of Appendix \ref{appb}. The rest of the terms are in the kernel of the Laplacian. All terms in \eqref{Sh} are independent of the 4-form and therefore they are the same for both the twisted and the untwisted models. A simple inspection shows that the $X^{\dagger}X$ and $\widetilde\psi^{\dagger}\widetilde\psi$ terms cancel out after the action of the BV Laplacian. Noting moreover that a 2-form and an 1-form have the same number of components in three dimensions, it turns out that the $B^{\dagger}B$ term cancels out with the $\psi^{\dagger}\psi$ term. What remains are the $A^{\dagger}A$ and $\e^{\dagger}\e$ terms, which vanish separately under the action of $\D$ because the result is proportional to $C^{a}_{ab}$.  Thus the classical master action we determined for the (twisted) Courant sigma model is also a solution of the quantum master equation
 \footnote{Note that the calculation of the $\D S$ includes divergences since the BV Laplacian includes the trace of the functional derivative. In fact, $\D S_{h} = C \cdot 0$ with an infinite constant $C$ similar to the case of the Poisson sigma model \cite{Cattaneo:1999fm}. We can prove that the proper gauge invariant regularization exists and the (twisted or not) Courant sigma model  satisfies the quantum master equation under this regularization.}. We clarify that by this we mean the naive quantum master equation in the sense of Ref. \cite{Bonechi:2010tbl}, since in the infinite dimensional setting of field theory the odd Laplacian is not well defined.

\section{The gauge fixed master action} \label{sec5}

To complete the picture painted in the previous sections we embark in gauge fixing the large gauge symmetry of the twisted Courant sigma model. This is a necessary task if one wants to compute correlation functions, even though we shall not do this in the present paper---see \cite{Hofman:2002jz} for computations of bulk/boundary correlators in the open membrane case. It also serves as an illustrating example for the general theory of gauge fixing within the BRST formalism \cite{HTbook,Gomis:1994he} and the AKSZ construction in particular \cite{Ikeda:2001fq}. 

The first step toward gauge fixing is to further enlarge the minimal set of fields and ghosts to a non-minimal set. A non-minimal set contains more fields than it is sufficient to solve the classical master equation. However the classical master equation does not have a unique solution and indeed the larger set still generates a solution. These additional fields are usually called trivial pairs and they contain antighosts{\footnote{We remind here that antighosts are not the same fields as the antifields of the ghosts.}} and Lagrange multipliers. We will use the general notation $\xi$ for antighosts and $\zeta$ for Lagrange multipliers. They also come together with their corresponding antifields so that $\xi^{\dagger}$ will be antifields for the antighosts and $\zeta^{\dagger}$ antifields for the Lagrange multipliers. 

Let us briefly recall the general properties of these additional fields. For an ordinary 1-form gauge field such as $A^{a}$, we have a scalar gauge parameter such as $\e^{a}$ which we replaced by a ghost of ghost number 1 with the same name in the BRST treatment. To each such ghost we assign a trivial pair $(\xi_{a},\zeta_{a})$. The ghost number of the antighost $\xi_{a}$ is the opposite of one for its ghost, in the present case $\text{gh}(\xi_{a})=-1$. We also require that the ghost number of the Lagrange multiplier is in general
\be \label{ghost number zeta}
\text{gh}(\zeta)=\text{gh}(\xi)+1\,.
\ee 
The reason for this is that the Lagrange multiplier will eventually pair with the antifield of the antighost in the non-minimal solution of the classical master equation and therefore the two must have opposite ghost number. Since $\text{gh}(\xi^{\dagger})=-1-\text{gh}(\xi)$, this means that we require 
$\text{gh}(\zeta)=-\text{gh}(\xi^{\dagger})$ which is nothing but Eq. \eqref{ghost number zeta}. This means in particular that for a usual gauge field the ghost number of the Lagrange multipler vanishes. Counting degrees of freedom, an 1-form in $d$ dimensions has $d$ components and we reduce $1$ for the ghost and $1$ for the antighost to reach the correct number of $d-2$ polarizations for a massless gauge field. 

For 2-form gauge fields the situation is a little more involved due to the reducibility of the gauge theory. An example is the 2-form{\footnote{In this section we consider the original fields before the redefinition.}} $B_{\m}$ in the twisted Courant sigma model but the discussion is general. In this case we have a 1-form gauge parameter $\psi_{\m}$ which we turn into a ghost of the same name and ghost number 1. To this we assign an 1-form antighost $\xi^{\m}$ of ghost number $-1$ and the Lagrange multiplier $\zeta^{\m}$ of ghost number $0$. We also have the scalar ghost for ghost $\widetilde{\psi}_{\m}$ of ghost number $2$ to which we assign a scalar antighost $\widetilde{\xi}^{\,\m}$ of ghost number $-2$ and a Lagrange multiplier $\widetilde{\zeta}^{\,\m}$ of ghost number $-1$. This is however not enough to yield the correct degree of freedom count. There is one additional field to introduce, what is usually called an extraghost that we denote as $\bar\xi_{\m}$. This is an auxiliary scalar field of vanishing ghost number together with the Lagrange multiplier $\bar\zeta_{\m}$ of ghost number $+1$. To justify this proliferation of fields, recall that a 2-form in $d$ dimensions has $d(d-1)/2$ components from which we subtract $2d$ from the two 1-form (anti)ghosts and we add $3$ at the next level from the ghost for ghost, antighost for ghost and extraghost. The result is $(d-2)(d-3)/2$ which is the correct number of polarizations for a massless 2-form.  
For the field content of the (twisted or not) Courant sigma model we summarize the trivial pair content in Table \ref{table3} and their antifields in Table \ref{table4}. Although in the case at hand the situation is not exceedingly complicated, we also present the rather simple BV triangles for clarity: 
	\be\begin{tikzcd}
		&	A^{a} \arrow[dr,blue,dash] & &&& B_{\m} \arrow[dr,blue,dash] & &
		 \\
		\xi_{a} \arrow[ur,blue,rightarrow] 
		& & \e^{a} &&\xi^{\m}\arrow[ur,blue,rightarrow]& &\psi_{\m} \arrow[dr,blue,dash]& 
  \\ 
  &&&\bar\xi_{\m}\arrow[ur,blue,leftrightarrow]&&\widetilde\xi^{\,\m}\arrow[ur,blue,rightarrow]&&\widetilde{\psi}_{\m}
	\end{tikzcd}\ee

	\begin{table}
	\begin{center}	\begin{tabular}{| c | c | c | c | c | c | c | c | c |}
			\hline 
			\multirow{3}{9.2em}{Antighost/Multiplier} &&&&&&&& \\ & $\xi_{a}$ & $\xi^{\m}$  & $\widetilde\xi^{\,\m}$ & $\bar\xi_{\m}$ & $\zeta_{a}$ & $\zeta^{\m}$ & $\widetilde{\zeta}^{\,\m}$ & $\bar\zeta_{\m}$ \\ &&&&&&&& \\ \hhline{|=|=|=|=|=|=|=|=|=|}
			\multirow{3}{6.0em}{Ghost degree} &&&&&&&& \\ & $-1$ & $-1$  & $-2$ & $0$ & $0$ & $0$ & $-1$ & $+1$ \\ &&&&&&&& \\\hline 
			\multirow{3}{5.5em}{Form degree} &&&&&&&& \\  & $0$ & $1$  & $0$ & $0$ & $0$ & $1$ & $0$ & $0$ \\ &&&&&&&&
			\\\hline  \end{tabular}\end{center}\caption{Antighosts/extraghost $\xi$ and Lagrange multipliers $\zeta$ for the twisted Courant sigma model.}\label{table3}\end{table}

\begin{table}
	\begin{center}	\begin{tabular}{| c | c | c | c | c | c | c | c | c |}
			\hline 
			\multirow{3}{3.5em}{Antifield} &&&&&&&& \\ & $\xi^{\dagger\,a}$ & $\xi^{\dagger}_{\m}$  & $\widetilde\xi^{\,\dagger}_{\m}$ & $\bar\xi_{\dagger}^{\m}$ & $\zeta^{\dagger\,a}$ & $\zeta^{\dagger}_{\m}$ & $\widetilde{\zeta}^{\,\dagger}_{\m}$ & $\bar\zeta_{\dagger}^{\m}$ \\ &&&&&&&& \\ \hhline{|=|=|=|=|=|=|=|=|=|}
			\multirow{3}{6.0em}{Ghost degree} &&&&&&&& \\ & $0$ & $0$  & $+1$ & $-1$ & $-1$ & $-1$ & $0$ & $-2$ \\ &&&&&&&& \\\hline 
			\multirow{3}{5.5em}{Form degree} &&&&&&&& \\  & $3$ & $2$  & $3$ & $3$ & $3$ & $2$ & $3$ & $3$ \\ &&&&&&&&
			\\\hline  \end{tabular}\end{center}\caption{Antifields of the antighosts/extraghost $\xi$ and of the Lagrange multipliers $\zeta$ for the twisted Courant sigma model.}\label{table4}\end{table}

Now we are ready to write down a non-minimal solution to the classical master equation which is 
\be 
S^{+}_{\text{\tiny{\rm BV}}}=S_{\text{\tiny{\rm BV}}}+\int \xi^{\dagger\,\a}\zeta_{\a}\,,
\ee 
where the collective notation $\xi^{\dagger\a}=(\xi^{\dagger\,a},\xi^{\dagger}_{\m},\widetilde\xi^{\,\dagger}_{\m},\bar\xi_{\dagger}^{\m})$ and $\zeta_{\a}=(\zeta_{a},\zeta^{\m},\widetilde{\zeta}^{\,\m},\bar\zeta_{\m})$ was introduced.
This is the action that we would like to gauge fix. To do so we must introduce a Lagrangian of degree $-1$ usually called the gauge fixing fermion and denoted by $\Psi$. This is a functional independent of all antifields and it is used to fix all antifields as 
\be 
\varphi^{\dagger}=\frac{\partial \Psi}{\partial \varphi}
\ee 
and thus eliminate them from the action as 
\be 
S_{\text{\tiny{\rm GF}}}= S^{+}_{\text{\tiny{\rm BV}}}\bigg[\varphi,\varphi^{\dagger}=\frac{\partial \Psi}{\partial \varphi}\bigg]\,.
\ee 
The gauge fixing fermion is of course not unique and its choice implements different gauges. Nevertheless there are some basic requirements that have to be satisfied, stemming from the necessity to have properly defined propagators in the theory. We refer to \cite{Batalin:1983ggl} and to the review \cite{Gomis:1994he} for more details on this general point. For a $\d$-function gauge fixing the gauge fixing fermion is independent of the Lagrange multipliers and an admissible choice for a first stage reducible system has the following form in our notation: 
\be 
\Psi_{\d}=\int\bigg(\xi_a f_1^{a}(\phi)+\xi^{\m}f_{2\,\m}(\phi)+\widetilde{\xi}^{\,\m}f_{3\,\m}^{\n}(\phi)\psi_{\n}+\xi^{\m}f_{4\,\m}^{\n}(\phi)\bar\xi_{\n}\bigg)\,,\label{GF fermion delta}
\ee 
where all four $f_i$ are functionals only of the fields and they are required to have proper rank \cite{Batalin:1983ggl} so that propagators exist. As remarked e.g. in \cite{Bonechi:2022aji} a standard gauge fixing in the AKSZ construction is to consider a Riemannian metric that defines a Hodge star operator and a gauge fixing fermion with fields in the image of the codifferential $\dd^{\dagger}=-\ast \dd \,\ast$. This is in line with Eq. \eqref{GF fermion delta} and it was already used in \cite{Hofman:2002jz} to consider the following gauge fixing fermion for the Courant sigma model
\be 
\Psi_{\d,\text{\tiny{\rm{CSM}}}}=\int \bigg(\dd\xi_a\ast A^{a}+\dd\xi^{\m}\ast B_{\m}+\dd\widetilde{\xi}^{\,\m}\ast\psi_{\m}+\dd\bar\xi_{\m}\ast\xi^{\m}\bigg)\,,
\ee 
which up to total derivatives is of the same form as \eqref{GF fermion delta} for suitable choice of $f_{i}$, that is $f_1^{a}=\ast\,\dd^{\dagger}A^{a}$, $f_{2\,\m}=\ast\,\dd^{\dagger}B_{\m}$, $f_{3\,\m}^{\n}=\ast\,\dd^{\dagger}\d_{\m}^{\n}$ and $f_{4\,\m}^{\n}=\dd^{\dagger}\ast \d_{\m}^{\n}$. We can use the same one for the twisted Courant sigma model too. This choice fixes 
the antifields as
\bea 
&& X^{\dagger\,\m}=0\,,\qquad \quad \,\,A^{\dagger}_{a}=\ast\,\dd\xi_a\,, \qquad \qquad \,\,\, \,\,B^{\dagger\,\m}=\ast\,\dd\xi^{\m}\,, \\[4pt] 
&& \e^{\dagger}_{a}=0\,,\qquad \,\,\,\quad\,\,\,\,\,\,\psi^{\dagger\,\m}=-\ast\,\dd\widetilde{\xi}^{\,\m}\,, \,\qquad\,\,\,\,\widetilde{\psi}^{\,\dagger\,\m}=0\,, \\[4pt] 
&& \xi^{\dagger\, a}=-\dd\ast A^{a}\,, \quad \xi^{\dagger}_{\m}= \dd\ast B_{\m}+\ast\,\dd\bar\xi_{\m}\,,\,\, \,\,\,\,\widetilde{\xi}^{\,\dagger}_{\m}=-\dd\ast\psi_{\m}\,, \qquad \bar\xi^{\m}_{\dagger}=-\dd\ast\xi^{\m}\,,\qquad \quad  
\eea
and all antifields of the Lagrange multipliers vanish: $\zeta^{\dagger\,\a}=0$.
Then we can write the complete form of the gauge fixed master action as 
\bea \label{gfaction}
S_{\text{\tiny{GF}}}=S_{\text{\tiny{kin}}}+S_{\text{\tiny{int,0}}}+S_{\text{\tiny{int,H}}}\,,
\eea 
where the kinetic, $H$-independent interaction and $H$-dependent interaction terms are given in non manifestly covariant form as:
\bea 
S_{\text{\tiny{kin}}}\!\!&=&\!\!\!\!\!\int \bigg(-B_\m(\rd X^\m-\ast\rd \zeta^{\m})+A^a(\sfrac 12\eta_{ab}\rd A^a+\ast\rd\zeta_a)+\psi_\m(\ast\rd\widetilde\zeta^{\,\m}-\rd\ast\rd\xi^{\m})\nn\\[4pt]
&&\qquad\,\,\,\,-\e^a\rd\ast\rd\xi_a+\widetilde{\psi}_\m\rd\ast\rd\widetilde{\xi}^{\,\m}+\zeta^{\m}\ast\rd\bar{\xi}_\m+\bar{\zeta}_\m\rd\ast\xi^{\m}\bigg)~,
\eea 
\bea 
S_{\text{\tiny{int,0}}}\!\!&=&\!\!\!\!\!\int\bigg(\r_a{}^\m B_\m A^a+\sfrac{1}{3!}C_{abc}\e^a\e^b\e^c +(\eta^{ab}\r_b{}^\m\psi_\m+C^a_{bc}A^b\e^c)\ast\rd\xi_a\nn\\[4pt]
&&\qquad\,\,\,\,+(\partial_\m\r_a{}^\n(B_\n \e^a-\psi_\n A^a)+\sfrac 12\partial_\m C_{abc}A^aA^b\e^c)\ast\rd\xi^{\m}\nn\\[4pt]
&&\qquad\,\,\,\,-(\partial_\m\r_a{}^\n(\widetilde{\psi}_\n A^a+\psi_\n\e^a)+\sfrac 12\partial_\m C_{abc}A^a\e^b\e^c)\ast\rd\widetilde{\xi}^{\,\m}\nn\\[4pt]
&&\qquad\,\,\,\,-\eta^{ab}(\partial_\m\r_b{}^\n\widetilde{\psi}_\n+\sfrac 12\partial_\m C_{bcd}\e^c\e^d)\ast\rd\xi_a\ast\rd\xi^{\m}\nn\\[4pt]
&&\qquad\,\,\,\,+(\partial_\m\partial_\l\r_a{}^\n\widetilde{\psi}_\n\e^a+\sfrac 16\partial_\m\partial_\l C_{abc}\e^a\e^b\e^c)\ast\rd\widetilde{\xi}^{\,\m}\ast\rd\xi^{\l}\nn\\[4pt]
&&\qquad\,\,\,\,-\sfrac 12(\partial_\m\partial_\l\r_a{}^\n(\psi_\n\e^a+\widetilde{\psi}_\n A^a)+\sfrac 12\partial_\m\partial_\l C_{abc}A^a\e^b)\ast\rd\xi^{\m}\ast\rd\xi^{\l}\nn\\[4pt]
&&\qquad\,\,\,\,+\sfrac 16(\partial_\m\partial_\l\partial_\s \r_a{}^\n\widetilde{\psi}_\n\e^a +\sfrac 16\partial_\m\partial_\l\partial_\s C_{abc}\e^a\e^b\e^c)\ast\rd\xi^{\m}\ast\rd\xi^{\l}\ast\rd\xi^{\s}\bigg)~,\quad
\eea 
\bea 
S_{\text{\tiny{int,H}}}\!\!&=&\!\!\!\!\!\int \bigg(-\sfrac 12( H_{\m abc}A^aA^b\e^c+ H_{\m\n a b}\e^a A^b F^\n+\sfrac 13 H_{\m\n\l a}\e^a F^\n F^\l)\ast\rd\xi^{\m}\nn\\[4pt]
&&\qquad\,\,\,\,-\sfrac 12(H_{\m abc}A^a\e^b\e^c+\sfrac 12 H_{\m\n ab}\e^a\e^b F^\n)\ast\rd\widetilde{\xi}^{\,\m}\nn\\[4pt]
&&\qquad\,\,\,\,-\sfrac 14\eta^{ab}H_{\m bcd}\e^c\e^d\ast\rd\xi_a\ast\rd\xi^{\m}+\sfrac 16\partial_{(\m}H_{\l)abc}\e^a\e^b\e^c\ast\rd\widetilde{\xi}^{\,\m}\ast\rd\xi^{\l}\nn\\[4pt]
&&\qquad\,\,\,\,-\sfrac 14(\partial_{\m}H_{\l abc}A^a\e^b+\sfrac 13\partial_{\m}H_{\l \s bc}\e^a\e^bF^\s)\ast\rd\xi^{\m}\ast\rd\xi^{\l}\nn\\[4pt]
&&\qquad\,\,\,\,+\sfrac {1}{36}(\partial_\m\partial_\l H_{\s abc}-\sfrac 32\partial_\m\partial_\l\r_{a}{}^\n H_{\n \s bc})\e^a\e^b\e^c\ast\rd\xi^{\m}\ast\rd\xi^{\l}\ast\rd\xi^{\s}\bigg)\nn\\[4pt]
&&+\int_{\widehat{\S}}X^{\ast}H~.
\eea 

The gauge fixed master action can be written as the sum of the classical action plus a BRST exact contribution, as expected for a topological field theory of Schwarz type \cite{Birmingham:1991ty}, where the classical action is completely metric independent. In order to show this, we can first write the gauge fixed BRST transformations:
\be 
s_{\text{\tiny{\rm{GF}}}}\varphi=(S_{\text{\tiny{\rm{BV}}}},\varphi)_{\text{\tiny{\rm{BV}}}}|_{\varphi^{\dagger}=\frac{\partial\Psi}{\partial\varphi}}\,.
\ee 
The precise form of these for the relevant fields is
\bea
s_{\text{\tiny{\rm GF}}}A^a &=& s_0 A^a+\eta^{ab}\big(\partial_\m\r_b{}^\n\widetilde{\psi}_\n+\sfrac 12(\partial_\m C_{bcd}+\sfrac 12 H_{\m bcd})\e^c\e^d\big)\ast\rd\xi^{\mu}~,\\[4pt]
s_{\text{\tiny{\rm GF}}}\psi_\m &=& s_0 \psi_\m+\big(\partial_\m\partial_\l\r_a{}^\n\widetilde{\psi}_\n\e^a+\sfrac 16(\partial_\m\partial_\l C_{abc}+\partial_{(\m}H_{\l) abc})\e^a\e^b\e^c\big)\ast\rd\xi^{\l}\,,\\[4pt]
s_{\text{\tiny{\rm GF}}}\xi^\a &=&\zeta^\a~,
\eea
and finally a longer one for the 2-form which is 
\bea 
s_{\text{\tiny{\rm GF}}}B_\m &=& s_0 B_\m-\eta^{ab}\big(\partial_\m\r_b{}^\n\widetilde{\psi}_\n+\sfrac 12(\partial_\m C_{bcd}+\sfrac 12 H_{\m bcd})\e^c\e^d\big)\ast\rd\xi_a\nn\\[4pt]
&+&\big(\partial_\m\partial_\l\r_a{}^\n\widetilde{\psi}_\n\e^a+\sfrac 16(\partial_\m\partial_\l C_{abc}+\partial_{(\m}H_{\l) abc})\e^a\e^b\e^c\big)\ast\rd\widetilde\xi^{\,\l}\nn\\[4pt]
&-&\big(\partial_\m\partial_\l\r_a{}^\n(\psi_\n\e^a+\widetilde{\psi}_\n A^a)+\sfrac 12(\partial_\m\partial_\l C_{abc}+\partial_{(\m}H_{\l) abc})A^a\e^b\e^c+\nn\\[4pt]
& & \qquad\qquad \qquad\qquad+\sfrac 16\partial_{(\m}H_{\l)\s ab}\e^a\e^b F^\s\big)\ast\rd\xi^{\l}\nn\\[4pt]
&+&\big( \partial_\m\partial_\l\partial_\s \r_a{}^\n\widetilde{\psi}_\n\e^a +\sfrac 16\partial_\m\partial_\l\partial_\s C_{abc}\e^a\e^b\e^c\nn\\[4pt]
& &  \quad\qquad +\,\big( \sfrac 16\partial_{(\m}\partial_{\l}H_{\s) abc}+\sfrac 14 \partial_{(\m}\partial_\l\r_a{}^\n H_{\s)\n bc}\big)\e^a\e^b\e^c\big)\ast\rd\xi^{\s}\ast\rd\xi^{\l}~.
\eea 
Using these BRST transformations, the gauge fixed action \eqref{gfaction} for the 4-form twisted Courant sigma model can be written in the simple form 
\begin{tcolorbox}[ams nodisplayskip, ams align]
S_{\text{\tiny{\rm GF}}}=S_{\rm cl}+\int s_{\text{\tiny{\rm GF}}}(A^a\ast\rd\xi_a+B_\m\ast\rd\xi^{\m}-\psi_\m\ast\rd\widetilde\xi^{\,\m}+\xi^{\m}\ast\rd\bar\xi_\m)\,.
\end{tcolorbox}
This proves that the twisted Courant sigma model is a topological field theory of Schwarz type, as expected. 

 \section{Conclusions and outlook}\label{sec6}

Motivated by the geometric approach to the BRST formalism mainly represented by the celebrated AKSZ construction, we have studied in depth the BRST structure of 4-form twisted Courant sigma models. These are topological membrane sigma models in three dimensions with a Wess-Zumino term that corresponds to a 4-form flux. Their gauge structure is controlled by pre-Courant algebroids where the violation of the Jacobi identity of their binary bracket is controlled by a closed 4-form. Our approach offers simultaneously a global perspective to ordinary Courant sigma models that constitute one of the main examples of the AKSZ construction and a method to solve the master equation for twisted models where the AKSZ construction is not directly applicable. 

Specifically, we showed that there is a manifestly frame independent formulation of the master action of the Courant sigma model with all coefficients in the interaction terms acquiring a geometric meaning. This is achieved by the introduction of an auxiliary (generalised) $E$-connection on the Courant algebroid with vector bundle $E$. Two tensors associated to this connection play a distinguished role. One is the Gualtieri torsion tensor and the other is the so-called basic curvature tensor which measures whether an ordinary connection on the Courant algebroid from which the generalised connection is induced is compatible with the Courant bracket---(in the sense of being a derivation of it).
Although the full AKSZ master action is of course covariant in any form, our approach reveals that covariantizing term by term the expanded action the interaction terms come together with these tensors and their derivatives. 

Apart from the aesthetic appeal of a geometric interpretation of the BV coefficients, an advantage of this geometric observation is that it shows a clear path in finding the master action of the twisted Courant sigma model, which is not of the AKSZ type in three dimensions, without residing on four-dimensional AKSZ and boundary conditions or to direct order by order computations, both of which can be very complicated. In applying the method we proposed, it is enough to work at the level of the on-shell closed BRST transformation and compute higher (in the present case third) powers of it  beyond the usual square, at least in a certain way that involves a specific replacement of field equations with antifields before each higher power is taken. What happens then is that we obtain a formally identical expanded expression of the master action for the untwisted (AKSZ) and twisted (non AKSZ) cases. This BRST power finesse of the master action utilizes the masterful AKSZ construction to go slightly beyond it, being admittedly more complicated that vanilla AKSZ but still simpler and more geometric than fully fledged BRST. 

Having identified all the 4-form-dependent terms in the master action, we also discussed its gauge fixing and in particular we showed that the complete quantum action can be written as the classical action plus a ``BRST commutator'' term. This confirms that the theory is topological and in particular that it takes precisely the form of a Schwarz type theory, as expected since this is the case for Chern-Simons theory which is the backbone of the twisted Courant sigma model. 

 Although the model we studied is particular, the approach we have followed is expected to have broader applicability in the realm of higher gauge theory, especially in the graded geometric approach of Ref. \cite{Grutzmann:2014hkn}. One obvious desideratum would be to extend the spirit of the AKSZ construction beyond the topological field theory territory. An example of this has been suggested in Refs. \cite{Grigoriev:2020xec,Grigoriev:2022zlq}. Another path toward more universal results would be to develop a precise correspondence between the BRST formulation of the twisted Courant sigma model and representations up to homotopy (ruths) for Lie 2-algebroids. The reason to expect  a direct correspondence is that the main example of ruth for Lie algebroids is their adjoint representation constructed in \cite{Crainic} in terms of the basic connection and the basic curvature tensor. Then one could expect that the appearance of the basic curvature tensor for Courant algebroids in the Courant sigma model and its twisted version is a signal of a ruth. Ruths beyond Lie algebroids were discussed in \cite{Jotz} and precise constructions inspired by the BRST formalism for higher dimensional Hamiltonian mechanics will be reported in future work. 

 \paragraph{Acknowledgements.} This work was supported by the Croatian Science Foundation project IP-2019-04-4168
“Symmetries for Quantum Gravity”, and JSPS Grants-in-Aid for Scientific Research Number 22K03323.
This article is based upon work from the COST Action
21109 CaLISTA, supported by COST (European Cooperation in
Science and Technology).

   \newpage 
   
\appendix 

\paragraph{Supplementary material} 

\section{Classical gauge invariance}\label{appa}

The classical action functional of the twisted Courant sigma model \eqref{Scl} is
\be
S_{\text{cl}} = \int_{T[1]\Sigma} 
\left(-B_\m  \rd X^\m + \frac{1}{2} \eta_{ab} A^a  \rd A^b
+\rho^\m_a(X) B_\m  A^a 
+ \frac{1}{3!} C_{abc}(X) A^a  A^b  A^c
\right)+\int_{\widehat{\S}}X^{\ast}H\,\nn
\ee
where $H$ is the pullback of closed 4-form on the body $M$ of the target. If $H$ is also exact, $H=\rm d G$, the resulting action functional 
can be rewritten as an untwisted Courant sigma model with redefined 2-form $\widetilde B_\m$ and redefined structure functions $\widetilde{C}_{abc}$ 
\bea
\widetilde B_\m &=& B_\m-\sfrac{1}{3!}G_{\m\n\s}F^\n F^\s+\sfrac{1}{2}G_{\m\n\s}\r_a{}^\s A^a F^\n - \sfrac{1}{2}G_{\m\n\s}\r_a{}^\n \r_b{}^\s A^aA^b~,\nn\\[4pt]
\widetilde C_{abc}&=& C_{abc}+\r_{a}{}^\m\r_b{}^\n\r_{c}^\s G_{\m\n\s}~,\nn
\eea
for $G=\sfrac 16 G_{\m\n\s}(X)\rd X^\m\rd X^\n\rd X^\s$ and $F^\m=\rd X^\m-\r_a{}^\m A^a$. Thus we see that the only relevant part of the twist originates from the non-exact part of the 4-form $H$ \cite{Hansen:2009zd}.

The gauge variation of the Wess-Zumino term drops to the boundary:
\be\label{gaugeH}
\delta \int_{\widehat{\S}} \sfrac{1}{4!} H_{\m\n\s\l}\rd X^\m\rd X^\n\rd X^\s\rd X^\l=\int_{T[1]\S} \sfrac{1}{3!} \e^a\r_a{}^\m H_{\m\n\s\l}\rd X^\n\rd X^\s\rd X^\l ~. 
\ee
Now we can use the symmetry $H\to H+\rd \Lambda$, for some 3-form $\Lambda$, in order to remove the exact part of $H$ from further discussion. Namely, 
\be\label{gfH}
\delta \int_{\widehat{\S}} X^\ast (H+\rd \Lambda)=\int_{T[1]\S} \sfrac{1}{3!} \e^a(H_{a\n\s\l}+4\r_a{}^\m\partial_{[\m}\Lambda_{\n\s\l]})\rd X^\n\rd X^\s\rd X^\l ~, \nn
\ee
so we can choose $\Lambda$ to remove the exact part of $H$, i.e., to set 3-form on the target to zero.  

One can now proceed with finding the gauge transformations that leave the classical action invariant. The only new term in comparison to the untwisted case is \eqref{gaugeH}. The simplest way to obtain the sought-after result is to substitute $\rd X^\m=F^\m+\r_a{}^\m A^a$ in all its three appearances on the right hand side of \eqref{gaugeH}. All $H$-dependent terms that contain $F^\m$ will go to the redefinition of the gauge transformation of $B_\m$, and the only one that does not contain $F^\m$ and is proportional to $H_{abcd}$ twists the Jacobi identity \eqref{CAidentity23}.  Using the classical gauge transformations for the untwisted case, which we recall here,
\be
\d X^\m=\e^a\r_a{}^\m~, \quad \d A^a=\rd \e^a+C^a_{bc}\e^b\e^c+\eta^{ab}\r_b{}^\m\psi_\m~,\nn
\ee
and Eqs. \eqref{CAidentity21} and \eqref{CAidentity22} we obtain
\bea
&& \d S_{\rm cl} = \int \left( - F^\m(\d B_\m+\rd  {\psi}_\m +\partial_\m\r_a{}^\n(B_\n\e^a-\psi_\n A^a)+\frac 12(\partial_\m C_{abc}+H_{\m abc})A^a A^b\e^c+\right.\nn\\[4pt]
 && \qquad\qquad \quad + \frac 12 H_{\m\n ab}\e^a A^b F^\n +\frac 16 H_{\m\n\l a}\e^a F^\n F^\l)+\nn\\[4pt]
 && \qquad\qquad \quad \left. +\frac 16\e^d A^a A^b A^c(\r_d{}^\m\partial_\m C_{abc}-3\r_a{}^\m\partial_\m C_{ bcd}+3C_{ebc}C^e_{ad}-H_{abcd})\right) ~. 
\eea
The vanishing of the first two lines fixes the classical gauge variation of $B_\m$, and the last line gives the Jacobi identity for a twisted Courant algebroid, \eqref{CAidentity23}.

\section{BRST transformations in local form}\label{appb}

In Proposition \ref{prop s0} we gave a set of general expressions for the BRST transformation of the fields $(\phi^{\a})=(X^{\m},A^{a},B_{\m})$:
\bea\label{gtApp}
s_0\phi^{\a} = \dd \e^{\a}+\e^{\b}\partial_{\b}Q^{\a}+s_{\text{triv}}\phi^{\a}\,,
\eea
The invariance of the classical action \eqref{Scl} fixes the transformations $s_{\text{triv}}\phi^{\a}$. Using the explicit expression for the homological vector $Q^{H}$ given in \eqref{QHCA} one can show that the transformations \eqref{gtApp} which explicitly read
\bea\label{ncBRSTf}
s_0 X^\m &=&\r_a{}^\m \e^a~,\nn\\[4pt]
s_0 A^a &=&\rd \e^a+\eta^{ab}\r_b{}^\m\psi_\m+C^a_{bc}A^b\e^c~,\nn\\[4pt]
s_0 B_\m &=&-\,\rd  {\psi}_\m -\partial_\m\r_a{}^\n(B_\n\e^a-\psi_\n A^a)-\sfrac 12(\partial_\m C_{abc}+H_{\m abc})A^a A^b\e^c+\nn\\[4pt]
&&-\,\sfrac 12 H_{\m\n ab}\e^a A^b F^\n -\sfrac 16 H_{\m\n\l a}\e^a F^\n F^\l~,
\eea
indeed leave the classical action invariant. As explained in Section \ref{sec32}, the BRST transformations of the ghosts $(\e^{\a})=(0,\e^a,-\psi_{\m})$ and ghost for ghost $(\widetilde{\e}^{\,\a})=(0,0,-\widetilde\psi_{\m})$ are determined imposing that $(s_0)^2\phi^\a$  and $(s_0)^2\e^\a$ vanish weakly. Thus the general expressions
\bea
s_0\e^{\a}&=&\dd \widetilde\e^{\,\a}-\widetilde{\e}^{\,\b}\partial_{\b}Q^{\a}-\frac 12\e^\b\e^\g \partial_{\b}\partial_{\g}Q^{\a}+s_{\text{triv}}\e^{\a}\,, \nn\\[4pt] 
s_0\widetilde{\e}^{\,\a}&=&\widetilde\e^{\,\b}\e^\g \partial_{\b}\partial_{\g}Q^{\a}+\frac{1}{3!}\e^\b\e^\g\e^\d\partial_{\b}\partial_{\g}\partial_{\d}Q^{\a}\,,
\eea
lead to
\bea\label{ncBRSTg}
s_0\e^a &=& -\,\eta^{ab}(\r_b{}^\m\widetilde{\psi}_\m+\sfrac 12 C_{bcd}\e^c\e^d)~,\nn\\[4pt]
s_0\widetilde{\psi}_\m &=& -\,\partial_\m\r_a{}^\n\widetilde{\psi}_\n\e^a-\sfrac 16(\partial_\m C_{abc}+H_{\m abc})\e^a\e^b\e^c~,\nn\\[4pt]
s_0\psi_\m &=&\rd \widetilde{\psi}_\m +\partial_\m\r_a{}^\n(\widetilde{\psi}_\n A^a+\psi_\n\e^a)+\sfrac 12(\partial_\m C_{abc}+H_{\m abc})A^a\e^b\e^c+\sfrac 14 H_{\m\n ab}\e^a e^b F^\n~.\,\,
\eea
It is straightforward to verify that \eqref{ncBRSTf} and \eqref{ncBRSTg} indeed  give \eqref{ncovBRST2} when squared.

Similarly, in Proposition \ref{prop s0 tensorial} we defined the tensorial BRST transformations of (redefined) fields and ghosts (\ref{nb}-\ref{ntp}). The explicit coordinate expressions of the covariant BRST transformations with the basis transformations defined in analogy to \eqref{forApp} are:
\begin{eqnarray}
s_0^{\scriptscriptstyle\nabla}X^{\m} &=&\rho_a{}^{\m} \e^a\,,\nn\\[4pt]
s_0^{\scriptscriptstyle\nabla}\e^a &=&-\eta^{ab}\r_b{}^\m\widetilde{\psi}_{\m}^{\scriptscriptstyle\nabla}+\eta^{ab}{\mc T}_{cdb}\e^c\e^d-\r_b{}^\m\e^b\o_{\m c}^a\e^c\,,\nn\\[4pt]
s_0^{\scriptscriptstyle\nabla}\widetilde{\psi}_{\m}^{\scriptscriptstyle\nabla} &=& -\nabla_\m\r_a{}^\n \widetilde{\psi}_{\n}^{\scriptscriptstyle\nabla}\e^a-\sfrac 13\widetilde{S}_{\m abc}\e^a\e^b\e^c+\r_a{}^\s\e^a\mathring{\G}_{\m\s}^\n\widetilde{\psi}_{\n}^{\scriptscriptstyle\nabla}\,,\nn \\[4pt]
 s_0^{\scriptscriptstyle\nabla}A^a &=& {\rm D} \e^a+\eta^{ab}\r_b{}^\m{\psi}_{\m}^{\scriptscriptstyle\nabla}-2\eta^{ab}{\mc T}_{cdb}A^c\e^d-\r_b{}^\m\e^b\o_{\m c}^aA^c,\,,\nn \\[4pt]
 s_0^{\scriptscriptstyle\nabla}{\psi}_{\m}^{\scriptscriptstyle\nabla} &=& {\rm D}\widetilde{\psi}_{\m}^{\scriptscriptstyle\nabla}+\nabla_\m\r_a{}^\n(\widetilde{\psi}_{\n}^{\scriptscriptstyle\nabla}A^a+ {\psi}_{\n}^{\scriptscriptstyle\nabla}\e^a)+\widetilde{S}_{\m abc}A^a\e^b\e^c+\r_a{}^\s\e^a\mathring{\G}_{\m\s}^\n{\psi}_{\n}^{\nabla}\nn\\[4pt] &&+\,\sfrac 14(H_{\m\n ab}-R_{ab\m\n})\e^a\e^b F^\n~,  \nn\\[4pt]
 s_0^{\scriptscriptstyle\nabla}{B}_{\m}^{\scriptscriptstyle\nabla} &=& - {\rm D}{\psi}_{\m}^{\scriptscriptstyle\nabla}+\nabla_\m\r_a{}^\n ({\psi}_{\n}^{\scriptscriptstyle\nabla}A^a- {B}_{\n}^{\scriptscriptstyle\nabla}\e^a)-\widetilde{S}_{\m abc}A^aA^b\e^c+\r_a{}^\s\e^a\mathring{\G}_{\m\s}^\n{B}_{\n}^{\nabla}\nn\\[4pt] 
&&-\,\sfrac 12 (H_{\m\n ab}-R_{ab\m\n})\e^a A^b F^\n -\sfrac 16 H_{\m\n\s a}\e^a F^\n F^\s~.
\end{eqnarray}
As discussed in Section \ref{sec3}, for $A,\psi^{\scriptscriptstyle\nabla},B^{\scriptscriptstyle\nabla}$ these transformations, defined  in Eqs. (\ref{snA}-\ref{snB}),  differ off-shell from the non-tensorial ones.
Finally, the non-vanishing squares are:
\bea
(s_0^{\scriptscriptstyle\nabla})^2A^a &=& \eta^{ab}F^\m(-\nabla_\m \r_b{}^\n\widetilde{\psi}_{\n}^{\scriptscriptstyle\nabla}-\widetilde S_{\m cdb}\e^c\e^d+\sfrac 14 (H_{\m b cd}-\r_b{}^\n R_{cd\m\n})\e^c\e^d)
~,  \nn\\[4pt]
(s_0^{\scriptscriptstyle\nabla})^2{\psi}_{\m}^{\scriptscriptstyle\nabla} &=& F^\l((\r_a{}^\s R^\n_{(\m\l)\s}-\nabla_{(\m}\nabla_{\l)} \r_a{}^\n)\widetilde{\psi}_{\n}^{\scriptscriptstyle\nabla}\e^a-\sfrac 13 \nabla_{(\m}\widetilde S_{\l) abc}\e^a\e^b\e^c)
 ~,  \nn\\[4pt]
(s_0^{\scriptscriptstyle\nabla})^2{B}_{\m}^{\scriptscriptstyle\nabla} &=& F^\l((\r_a{}^\s R^\n_{(\m\l)\s}-\nabla_{(\m}\nabla_{\l)} \r_a{}^\n)({\psi}_{\n}^{\scriptscriptstyle\nabla}\e^a+\widetilde{\psi}_{\n}^{\scriptscriptstyle\nabla}A^a)-\nabla_{(\m}\widetilde S_{\l) abc}A^a\e^b\e^c)\nn\\[4pt]
&+&G^{{\scriptscriptstyle\nabla} a}(-\nabla_\m\r_a{}^\n\widetilde{\psi}_{\n}^{\scriptscriptstyle\nabla}-\widetilde S_{\m abc}\e^b\e^c+\sfrac 14 (H_{\m ab c}-\r_a{}^\n R_{bc\m\n})\e^b\e^c))\nn\\[4pt]
&+&F^\l F^\s(\sfrac 23 R^\n_{(\m\l)\s}{\widetilde\psi}_{\n}^{\scriptscriptstyle\nabla}-\sfrac 16(\nabla_{(\m}H_{\l)\s a b}-\nabla_{(\m}R_{ab\l)\s})\e^a\e^b)
~.
\eea

\section{Technical annex to the BRST power finesse}\label{appc}

As discussed in Section \ref{sec3}, we can use just the on-shell closed BRST  transformations to obtain the minimal solution to the classical master equation. When the 4-form $H$ vanishes  we can fix ${\mc O}_{(n)}\varphi$ using only the homological vector field $Q$, as one expects from the AKSZ construction. For $n=1$, by definition we have 
\bea
{\mc O}_{(1)}\phi^\a=s_0\phi^\a=\rd \e^a+\e^\b\partial_\b Q^\a ~.\nn
\eea
As explained in the main text, ${\mc O}_{(2)}\phi^\a$ is obtained from $(s_0)^2\phi^\a$ by substituting a field equation with the corresponding antifield. Because of the grading we must carefully define how this substitution is done. We postulate that the field equation is first brought to the left-most position of each expression and then the substitution is performed. Thus we rewrite Eq. \eqref{s2phi} as   
\bea
(s_0)^2\phi^\a=-{\mc F}^\d\left(\widetilde{\e}^\b \partial_\d\partial_\b Q^\a+\frac 12 (-1)^{s}\e^\b \e^\g\partial_\d\partial_\b\partial_\g Q^\a\right) ~,\nn
\eea
where $s=(|\e^\b|+|\e^\g|)(|{\mc F}^\d|+1)$ are signs obtained from commuting graded variables.
After substituting ${\cal F}\to (-1)^{\rm{deg} {\cal F}+1}\phi^\dagger_{\cal F}$
we obtain
\bea
{\mc O}_{(2)}\phi^\a=(-1)^{|{\cal F}^\d|}\phi^{\dagger \d}_{\cal F}\left(\widetilde{\e}^\b \partial_\d\partial_\b Q^\a+\frac 12 (-1)^{s}\e^\b \e^\g\partial_\d\partial_\b\partial_\g Q^\a\right) ~.
\eea
Similarly, for the ghosts we have
\bea
{\mc O}_{(1)}\e^{\a}&=&\dd \widetilde\e^{\,\a}-\widetilde{\e}^{\,\b}\partial_{\b}Q^{\a}-\frac 12\e^\b\e^\g \partial_{\b}\partial_{\g}Q^{\a}\,, \nn\\[4pt]
{\mc O}_{(2)}\e^{\a}&=&\phi^{\dagger \d}_{\cal F}\left((-1)^{s_1}\widetilde{\e}^\b \e^\g \partial_\d\partial_\b\partial_\g Q^\a+\frac {1}{3!} (-1)^{s_2}\e^\b \e^\g\e^\varepsilon\partial_\d\partial_\b\partial_\g\partial_\varepsilon Q^\a\right)~,\nn
\eea
where the signs are  $s_1=(|\e^\g|+1)|({\mc F}^\d|+1)$ and $s_2=(|\e^\b|+|\e^\g|+|\e^\varepsilon|+1)(|{\mc F}^\d|+1)$.

For the Courant sigma model we also get a cubic contribution from the highest-degree field $B_\m$. Thus we need to calculate $s_0({\mc O}_{(2)}\phi^\a)$ and collect contributions proportional to the field equations. The only relevant contributions come from $\rd \widetilde{\e}$ and $\rd \e$ terms which are in turn obtained from $s_0\e^a$ and $s_0\phi^\a$:
\bea
s_0({\mc O}_{(2)}\phi^\a) &\supset & (-1)^{|{\cal F}^\d|+1}\phi^{\dagger \d}_{\cal F}\bigg(\widetilde{\e}^\b \rd\e^\g\partial_\g \partial_\d\partial_\b Q^\a+ (-1)^{s}\rd \widetilde\e^\b \e^\g\partial_\d\partial_\b\partial_\g Q^\a\nn\\[4pt]
&& +\,  \frac 12 (-1)^{s'}\e^\b \e^\g\rd \e^\l\partial_\l\partial_\d\partial_\b\partial_\g Q^\a\bigg) ~;\nn
\eea
we used $\supset$ to denote that the terms on the right hand side are included to the quantity on the left hand side together with additional terms that we do not focus on presently, and $s'=(|\e^\b|+|\e^\g|)|{\mc F}^\d|$.
We find that the  first two terms on the right-hand side collect into a term proportional to $\rd (\widetilde{\e}^\b\e^\g)$, while the last one is rewritten as  a term proportional to $\rd(\e^\b\e^\g\e^\l)$.
Thus we have
\bea
s_0(O_{(2)}\phi^\a) &\supset & -\rd\left(\phi^{\dagger \d}_{\cal F}\big(\widetilde{\e}^\b \e^\g\partial_\g \partial_\d\partial_\b Q^\a+ \frac 16 (-1)^{s'}\e^\b \e^\g \e^\l\partial_\l\partial_\d\partial_\b\partial_\g Q^\a\big)\right) \nn\\
&+& \rd \phi^{\dagger \d}_{\cal F}\bigg(\widetilde{\e}^\b \e^\g\partial_\g \partial_\d\partial_\b Q^\a+ \frac 16 (-1)^{s'}\e^\b \e^\g \e^\l\partial_\l\partial_\d\partial_\b\partial_\g Q^\a\bigg)\nn\\
&+& (-1)^{|{\cal F}^\d|} \phi^{\dagger \d}_{\cal F}\bigg(\widetilde{\e}^\b \e^\g\rd \phi^\k\partial_k\partial_\g \partial_\d\partial_\b Q^\a+ \frac 16 (-1)^{s'}\e^\b \e^\g \e^\l\rd \phi^\k\partial_k\partial_\l\partial_\d\partial_\b\partial_\g Q^\a\bigg)\nn~. 
\eea
The first two lines do not contribute to the ghost number  zero action. We focus on the last line and  use $\rd \phi^\a={\mc F}^\a +\phi^\ast (Q^\a)$  keeping only the field equation part to get
\bea
s_0(O_{(2)}\phi^\a) &\supset & (-1)^{|{\cal F}^\d|} \phi^{\dagger \d}_{\cal F}\left(\widetilde{\e}^\b \e^\g{\cal F}^\k\partial_\k\partial_\g \partial_\d\partial_\b Q^\a+ \frac 16 (-1)^{s'}\e^\b \e^\g \e^\l{\cal F}^\k\partial_\k\partial_\l\partial_\d\partial_\b\partial_\g Q^\a\right)\nn~. 
\eea
Now we commute the field equation to the left-most position and replace it with the corresponding antifield
\bea
O_{(3)}\phi^\a &=& (-1)^{s_1} \phi^{\dagger \k}_{\cal F}\phi^{\dagger \d}_{\cal F}\bigg(\widetilde{\e}^\b \e^\g\partial_\k\partial_\g \partial_\d\partial_\b Q^\a+ \frac 16 (-1)^{s_2}\e^\b \e^\g \e^\l\partial_\k\partial_\l\partial_\d\partial_\b\partial_\g Q^\a\bigg)\nn~,
\eea
where the signs are $s_1=1+|{\cal F}^\d|+|{\cal F}^\k|(|\e^\g|+|{\cal F}^\d|+1)$ and $s_2=|{\cal F}^\d|(|\e^\b|+|\e^\g|)+|{\cal F}^\k|(|\e^\b|+|\e^\l|)$. Evaluating the relevant derivatives of the homological vector field $Q$ we obtain
\bea
B^{\dagger\,\m}O_{(3)}B_\m=B^{\dagger\,\m}B^{\dagger\,\l}B^{\dagger\,\s}\left(\partial_\m\partial_\l\partial_\s\r_a{}^\n \widetilde\psi_\n\e^a+\frac 16 \partial_\m\partial_\l\partial_\s C_{abc}\e^a\e^b\e^c\right)~.\nn
\eea

\section{Complete minimal solution of the master equation}\label{appd}

Based on the results of Theorem \ref{Thm} we write down the expanded form of the minimal solution of the classical master equation for the twisted Courant sigma model, separating sectors by number of antifields. This comprises the usual sectors found e.g. in Ref. \cite{Roytenberg:2006qz}---in different notation and conventions---plus additional contributions that contain the 4-form components and it reads:
\be  \label{S BV as sum}
S_{\text{\tiny{\rm BV}}} =S_0+S_1+S_2+S_3\,,
\ee 
where $S_0=S_{\text{cl}}$ is the classical action \eqref{Scl} and the rest of the terms are given as 
\bea
S_1 &=& \int\biggl\{  -\, A^\dagger_a(\rd \e^a+\eta^{ab}\r_b{}^\m\psi_\m+C^a_{bc}A^b\e^c)-X^\dagger_\m\r_a{}^\m\e^a-\e_{a}^\dagger(\eta^{ab}\r_b{}^\m\widetilde\psi_\m+\sfrac 12 C^a_{bc}\e^b\e^c)\nn\\[4pt]
&&\qquad +\, B^{\dagger \m}\bigg(\rd{\psi}_\m+\partial_\m\r_a{}^\n(B_\n \e^a-\psi_\n A^a)+\sfrac 12(\partial_\m C_{abc}+H_{\m abc})A^aA^b\e^c +\nn\\[4pt] 
& & \qquad\qquad\qquad 
 + \sfrac 12  H_{\m\n ab}\e^a A^b F^\n+\sfrac 16 H_{\m\n\l a}\e^a F^\n F^\l\bigg)\nn\\[4pt]
&&\qquad +\,\widetilde{\psi}^{\dagger \m}\big(\partial_\m\r_a{}^\n\widetilde{\psi}_\n\e^a+\sfrac 16 (\partial_\m C_{abc}+H_{\m abc})\e^a\e^b\e^c\big)\nn\\[4pt]
&&\qquad +\, \psi^{\dagger \m}\bigg(\rd \widetilde{\psi}_\m+\partial_\m\r_a{}^\n(\widetilde{\psi}_\n A^a+\psi_\n\e^a)+ \nn\\[4pt] &&\qquad\qquad\qquad +\, \sfrac 12(\partial_\m C_{abc}+H_{\m abc})A^a\e^b\e^c+\sfrac 14  H_{\m\n ab}\e^a\e^b F^\n\bigg)\biggl\}\,,
\eea 
\bea 
S_2&=&\int \biggl\{ -\, A^\dagger_a B^{\dagger \m}\eta^{ab}\big(\partial_\m\r_b{}^\n\widetilde{\psi}_\n+\sfrac 12(\partial_\m C_{bcd}+\sfrac 12 H_{\m bcd})\e^c\e^d\big)\nn\\[4pt]
&&\qquad +\,\psi^{\dagger \m}B^{\dagger \l}\big(\partial_\m\partial_\l\r_a{}^\n\widetilde{\psi}_\n\e^a+\sfrac 16(\partial_\m\partial_\l C_{abc}+\partial_{(\m}H_{\l) abc})\e^a\e^b\e^c\big)\nn\\[4pt]
&&\qquad -\sfrac 12 B^{\dagger \m}B^{\dagger \l}\big(\partial_\m\partial_\l\r_a{}^\n(\psi_\n\e^a+\widetilde{\psi}_\n A^a)+\sfrac 12(\partial_\m\partial_\l C_{abc}+\partial_{\m}H_{\l abc})A^a\e^b\e^c\nn\\[4pt]
& & \qquad\qquad\qquad\quad  +\, \sfrac 16\partial_{\m}H_{\l\s ab}\e^a\e^b F^\s\big)\biggl\}\,, 
\eea 
\bea S_3&=&
\int -\sfrac 16 B^{\dagger \m}B^{\dagger \l}B^{\dagger \s}\bigg( \partial_\m\partial_\l\partial_\s \r_a{}^\n\widetilde{\psi}_\n\e^a +\sfrac 16\partial_\m\partial_\l\partial_\s C_{abc}\e^a\e^b\e^c\nn\\[4pt]
& & \qquad\qquad \qquad\qquad +\,\big( \sfrac 16\partial_\s\partial_{\m}H_{\l abc}-\sfrac 14 \partial_\m\partial_\l\r_a{}^\n H_{\n\s bc}\big)\e^a\e^b\e^c\bigg)
\eea 
The manifestly covariant expression where all coefficients have a clear geometric meaning is 
\be \label{S BV as sum cov}
S_{\text{\tiny{\rm BV}}}=S_0'+S_1'+S_2'+S_3'\,,
\ee 
where the primed sectors are
\bea
S'_{0}=\int
\left( -B_\m^{\scriptscriptstyle\nabla} (\dd X^\m-\r_a{}^\m A^a) +\sfrac 12 \eta_{ab}A^a{\rm D}A^b-\sfrac{1}{3}{\mc T}_{abc}A^aA^bA^c\right)+\int_{\widehat{\S}}X^{\ast}H\,,
\eea 
\bea 
S'_1&=& \int\biggl\{
 - X^{\dagger{\scriptscriptstyle\nabla}}_\m \r_a{}^\mu\e^a-A^\dagger_a({\rm D}\e^a+\eta^{ab}\r_{b}{}^\m\psi_\m^{\scriptscriptstyle\nabla}-2\eta^{ab}{\mc T}_{bcd} A^c\e^d)\nn\\[4pt]
&&\qquad
 +\,B^{\dagger\m}\big({\rm D}\psi_\m^{\scriptscriptstyle\nabla}-\nabla_\m\r_a{}^\n(\psi_\n^{\scriptscriptstyle\nabla}A^a-B_\n^{\scriptscriptstyle\nabla}\e^a)+\widetilde{\mc S}_{\m abc}A^a A^b\e^c+ \nn\\[4pt]
& & \qquad\qquad\qquad
 + \sfrac 12  (H_{\m\n ab}-R_{ab\m\n})\e^a A^b F^\n+\sfrac 16 H_{\m\n\l a}\e^a F^\n F^\l\big)\nn\\[4pt]
&&\qquad +\,
\e^{\dagger}_a\eta^{ab}(-\r_b{}^\m\widetilde\psi_\m^{\scriptscriptstyle\nabla}+{\mc T}_{cdb}\e^c\e^d)+\widetilde\psi^{\dagger\m}(\nabla_\m\r_a{}^\n\widetilde\psi_\n^{\scriptscriptstyle\nabla}\e^a+\sfrac 13 \widetilde{\mc S}_{\m abc}\e^a\e^b\e^c)\nn\\[4pt]
&&\qquad +\, 
\psi^{\dagger \m}\left({\rm D}\widetilde\psi_\m^{\scriptscriptstyle\nabla}+\nabla_\m\r_a{}^\n(\widetilde\psi_\n^{\scriptscriptstyle\nabla}A^a+\psi_\n^{\scriptscriptstyle\nabla}\e^a)+\widetilde{\mc S}_{\m abc}A^a \e^b\e^c+ \right.\nn\\[4pt] && \left.\qquad\qquad\qquad +\,\sfrac 14  (H_{\m\n ab}-R_{ab\m\n})\e^a\e^b F^\n\right)\biggl\}\,, 
\eea 
\bea S'_2&=&
\int\biggl\{A^\dagger_aB^{\dagger \m}\eta^{ab}\left(-\nabla_\m \r_b{}^\n\widetilde{\psi}_{\n}^{\scriptscriptstyle\nabla}-\widetilde{\mc S}_{\m cdb}\e^c\e^d+\sfrac 14 \r_b{}^\n(H_{\m \n cd}- R_{cd\m\n})\e^c\e^d\right)\nn\\[4pt]
&&\qquad -\,\psi^{\dagger \m}B^{\dagger \l}\left((\r_a{}^\s R^\n_{(\m\l)\s}-\nabla_{(\m}\nabla_{\l)} \r_a{}^\n)\widetilde{\psi}_{\n}^{\scriptscriptstyle\nabla}\e^a-\sfrac 13 \nabla_{(\m}\widetilde{\mc S}_{\l) abc}\e^a\e^b\e^c\right)\nn\\[4pt]
&&\qquad +\,\sfrac 12 B^{\dagger \m}B^{\dagger \l}\left((\r_a{}^\s R^\n_{\m\l\s}-\nabla_{\m}\nabla_{\l} \r_a{}^\n)({\psi}_{\n}^{\scriptscriptstyle\nabla}\e^a+\widetilde{\psi}_{\n}^{\scriptscriptstyle\nabla}A^a)-\nabla_{\m}\widetilde{\mc S}_{\l abc}A^a\e^b\e^c\right.\nn\\
& & \qquad\qquad\qquad\qquad 
 +\left.F^\s(\sfrac 23 R^\n_{\m\l\s}{\widetilde\psi}_{\n}^{\scriptscriptstyle\nabla}-\sfrac 16\nabla_{\m}(H_{\l\s a b}-R_{ab\l\s})\e^a\e^b )\right)\biggl\}\,,
 \eea 
 \bea S'_3&=&\int-\sfrac 16 B^{\dagger \m}B^{\dagger \l}B^{\dagger \s}\left(-\nabla_\s(\r_a{}^\r R_{\m\l\r}^\n-\nabla_\m\nabla_\l\r_a{}^\n)\widetilde{\psi}_\n\e^a+\sfrac 13\nabla_\s\nabla_\l \widetilde{\mc S}_{\m abc}\e^a\e^b\e^c\right. \nn\\
& & \qquad\qquad\qquad
 +\left.  \sfrac 14(\r_a{}^\r R^\n_{\m\l\r}-\nabla_{\m}\nabla_{\l} \r_a{}^\n)(H_{\n\s bc}-R_{bc \n\s})\e^a\e^b\e^c\right)\,,
\eea
where we used the following redefined antifield: 
\be\label{X dagger redef} X^{\dagger\scriptscriptstyle\nabla}_\m=X^{\dagger}_\m-\o_{\m b}^a(\e^b\e^\dagger_a+ A^bA^\dagger_a)+\mathring{\G}_{\m\s}^\n(B^{\scriptscriptstyle\nabla}_\n B^{\dagger \s}+\psi^{\scriptscriptstyle\nabla}_\n \psi^{\dagger \s}+\widetilde{\psi}^{\scriptscriptstyle\nabla}_\n \widetilde{\psi}^{\dagger \s})\,. 
\ee 
A redefinition like that is expected since this is the only antifield that has total degree 2 and participates in the same superfield as the other three fields we redefined earlier. One can also observe that the BV symplectic form after the field redefinition becomes 
\be 
\omega_{\text{\tiny{\rm{BV}}}}=\int_{\S}\d B_\m^{\scriptscriptstyle\nabla}\d B^{\dagger\,\m}+\d\psi_\m^{\scriptscriptstyle\nabla}\d \psi^{\dagger\,\m}+\d\widetilde{\psi}_\m^{\scriptscriptstyle\nabla}\d \widetilde{\psi}^{\dagger\,\m}+\d X^\m\d X^{\dagger\scriptscriptstyle\nabla}_\m+\d\e^a\d\e^\dagger_a+ \d A^a\d A^\dagger_a\,.
\ee 
A change of coordinates, as described in Section \ref{sec23}, corresponds to a canonical transformation for this symplectic form provided that the BV momentum is covariantized as in Eq. \eqref{X dagger redef}. This is true because under a change of coordinates\footnote{Here we use $\varphi '$ instead of $\widetilde{\varphi}$ to denote transformed fields in order to avoid clash of notation. } we obtain 
\be 
{X}_{\m}^{\dagger\scriptscriptstyle{\nabla}}={X'}_{\m}^{\dagger\scriptscriptstyle{\nabla}}+(\Lambda^{-1})_{\n}^\l\partial_\m \Lambda_\l^{\s}( {B'}_{\s}^{\scriptscriptstyle\nabla}{B'}^{\dagger\n}+{\psi'}_{\s}^{\scriptscriptstyle\nabla}{\psi'}^{\dagger\n}+\widetilde\psi'^{\scriptscriptstyle\nabla}_{\s}\widetilde\psi'^{\dagger\n})-(\Lambda^{-1})_{b}^c\partial_\m \Lambda_c^{a}({\e'}^b {\e'}^\dagger_a+ {A'}^b{A'}^\dagger_a)\,,
\ee 
taking into account how the connection coefficients transform. 
Notice that this becomes simple precisely because $b^{\scriptscriptstyle\nabla}$ transforms as in Eq. \eqref{b nabla trafo}. The original coordinate with transformation \eqref{b trafo} would yield a very complicated result.
Finally, we stress once more that Eqs. \eqref{S BV as sum} and \eqref{S BV as sum cov} are two ways of writing \emph{the same} master action $S_{\text{\tiny{\rm{BV}}}}$. This is possible because the solution to the master equation is up to field redefinitions and the nonmanifestly and manifestly covariant forms are related by one. 

Finally we can write the action in a completely coordinate free form term by term. The result is
\bea 
S'_0&=&\int \biggl\{-( B^{\scriptscriptstyle\nabla},F) +  \langle A,{\rm D}A\rangle-\sfrac{1}{3} {\mc T}(A,A,A)+\int_{\widehat{\S}}X^{\ast}H \biggr\} \,,\\[4pt]
S'_1&=&\int \biggl\{(X^{\dagger\scriptscriptstyle\nabla}, \r(\e)) -\langle A^\dagger, \DD\e+(\rho^\ast, \psi^{\scriptscriptstyle\nabla})-2{\mc T}^\ast(-, A,\e) \rangle \nn\\[4pt]
&+&\left(B^\dagger, \DD\psi^{\scriptscriptstyle\nabla}+(\nabla\rho(A), \psi^{\scriptscriptstyle\nabla})+(\nabla\rho(\e), B^{\scriptscriptstyle\nabla})+\widetilde{\mc S}(A,A,\e)\right.\nn\\ [4pt]
&+& \left.\sfrac{1}{2}( H(\r(\e),\r(A))-R(\e,A))(-,F)+\sfrac{1}{6} H(-,\r(\e),F,F)\right)\nn\\ [4pt]
&+& 2\langle {\e}^\dagger, (\rho, \widetilde{\psi}^{\scriptscriptstyle\nabla}) + {\mc T}(-,\e, \e) \rangle 
+ (\widetilde{\psi}^{\dagger}, (\nabla \rho(\e), \tilde{\psi}) + \sfrac{1}{3} \widetilde{\mc S}(\e, \e, \e)) \nn\\[4pt]
&+&\left( \psi^\dagger, {\rm D}\widetilde{\psi}^{\scriptscriptstyle\nabla}+(\nabla\r(A), \widetilde{\psi}^{\scriptscriptstyle\nabla})-(\nabla\r(\e), {\psi}^{\scriptscriptstyle\nabla})+\widetilde{\mc S}(A,\e,\e) \right.\nn\\[4pt]
&+& \left. \sfrac{1}{4} (H(\r(\e),\r(e))-R(\e,\e))(-,F))\right) \biggr\}\,,\\ [4pt]
S'_2&=& \int \biggl\{ -(B^\dagger, - (\nabla \rho(A^{\dagger*}), \widetilde{\psi}^{\scriptscriptstyle\nabla} )
- \widetilde{\mc S}(A^{\dagger*}, \e, \e) \nn\\[4pt]
&+& \sfrac{1}{4} (H(\r(\e),\r(\e))-R(\e,\e))(-,\r(A^{\dagger*})) )\nn\\[4pt]
&+& \sfrac{1}{2} (\psi^\dagger, (B^\dagger, (D^{LC} \rho(\e), \widetilde{\psi}^{\scriptscriptstyle\nabla}) 
+ \sfrac{1}{3} \nabla \widetilde{\mc S}(\e,\e,\e))) \nn\\[4pt]
&+& \sfrac{1}{2} (B^\dagger, (\psi^\dagger, (D^{LC} \rho(\e), \widetilde{\psi}^{\scriptscriptstyle\nabla}) 
+ \sfrac{1}{3} \nabla \widetilde{\mc S}(\e,\e,\e))) \nn\\[4pt]
&+& \sfrac{1}{2} (B^\dagger, (B^\dagger, (D^{LC} \rho(\e), \psi^{\scriptscriptstyle\nabla}) 
+ (-D^{LC} \rho(A), \widetilde{\psi}^{\scriptscriptstyle\nabla}) 
-\nabla \widetilde{\mc S}(A,\e,\e) \nn\\[4pt]
&+& \sfrac{2}{3}R^{LC}(\widetilde{\psi}^{\scriptscriptstyle\nabla}, -, -, F)
- \sfrac{1}{6}(\nabla H(\r(\e), \r(\e)) -\nabla R(\e, \e))(-,F)))) \biggr\}\,,\\[4pt]
S'_3&=& \int \biggl\{ -\sfrac{1}{6} 
(B^\dagger, (B^\dagger, (B^\dagger, \nabla D^{LC} \rho(\e), \widetilde{\psi}^{\scriptscriptstyle\nabla})) 
+ \sfrac{1}{3} \nabla \nabla \widetilde{\mc S}(\e,\e,\e) )))  \nn\\[4pt]
&-& (B^\dagger, (B^\dagger, (\sfrac{1}{4} D^{LC} \rho(\e), (H(\r(\e), \r(\e)) -R(\e, \e))(-,B^\dagger)))) \biggr\}\,, 
\eea
Here $(\cdot, \cdot)$ is the pairing between $TM$ and $T^*M$ and $\langle \cdot, \cdot \rangle$ is the pairing between $E$ and $E^*$ and contractions of two $E$'s or two $E^*$'s with the fiber metric $\eta$, Eq.\eqref{fibermetric}.
$R$ is the curvature tensor for the connection $\omega$ and $R^{LC}$ is the Riemann curvature tensor for the Levi-Civita connection $\mathring{\G}$.
$D^{LC} v := \nabla \nabla v - (R^{LC}, v)$ is the second order differential operator acting on a vector field $v \in \mathfrak{X}(M)$.
All tensors are understood in composition with the sigma model base map $X$, for example as $\rho\circ X$, $\mc T\circ X$, $\widetilde{\mc S}\circ X$ and so on, meaning that their components in a given basis are compositions with the map $X$.


\begin{thebibliography}{99}

\bibitem{Ikeda:2002wh}
N.~Ikeda,
``Chern-Simons gauge theory coupled with BF theory,''
Int. J. Mod. Phys. A \textbf{18}, 2689-2702 (2003)
doi:10.1142/S0217751X03015155
[arXiv:hep-th/0203043 [hep-th]].

\bibitem{Hofman:2002jz}
C.~Hofman and J.~S.~Park,
``BV quantization of topological open membranes,''
Commun. Math. Phys. \textbf{249} (2004), 249-271
doi:10.1007/s00220-004-1106-7
[arXiv:hep-th/0209214 [hep-th]].

\bibitem{Hofman:2002rv}
C.~Hofman and J.~S.~Park,
``Topological open membranes,''
[arXiv:hep-th/0209148 [hep-th]].

\bibitem{Roytenberg:2006qz}
D.~Roytenberg,
``AKSZ-BV Formalism and Courant Algebroid-induced Topological Field Theories,''
Lett. Math. Phys. \textbf{79} (2007), 143-159
doi:10.1007/s11005-006-0134-y
[arXiv:hep-th/0608150 [hep-th]].

\bibitem{dimaphd} 
D. Roytenberg, 
``Courant algebroids, derived brackets and even symplectic supermanifolds,'' PhD Thesis,
arXiv:math/9910078

\bibitem{Roytenberg:2002nu}
D.~Roytenberg,
``On the structure of graded symplectic supermanifolds and Courant algebroids,''
Quantization, Poisson Brackets and Beyond, Theodore Voronov (ed.), Contemp. Math., Vol. 315, Amer. Math. Soc., Providence, RI, 2002,
[arXiv:math/0203110 [math.SG]].

\bibitem{Halmagyi:2008dr}
N.~Halmagyi,
``Non-geometric String Backgrounds and Worldsheet Algebras,''
JHEP \textbf{07} (2008), 137
doi:10.1088/1126-6708/2008/07/137
[arXiv:0805.4571 [hep-th]].

\bibitem{Mylonas:2012pg}
D.~Mylonas, P.~Schupp and R.~J.~Szabo,
``Membrane Sigma-Models and Quantization of Non-Geometric Flux Backgrounds,''
JHEP \textbf{09} (2012), 012
doi:10.1007/JHEP09(2012)012
[arXiv:1207.0926 [hep-th]].

\bibitem{Chatzistavrakidis:2015vka}
A.~Chatzistavrakidis, L.~Jonke and O.~Lechtenfeld,
``Sigma models for genuinely non-geometric backgrounds,''
JHEP \textbf{11} (2015), 182
doi:10.1007/JHEP11(2015)182
[arXiv:1505.05457 [hep-th]].

\bibitem{Bessho:2015tkk}
T.~Bessho, M.~A.~Heller, N.~Ikeda and S.~Watamura,
``Topological Membranes, Current Algebras and H-flux - R-flux Duality based on Courant Algebroids,''
JHEP \textbf{04} (2016), 170
doi:10.1007/JHEP04(2016)170
[arXiv:1511.03425 [hep-th]].

\bibitem{Arvanitakis:2023dud}
A.~S.~Arvanitakis and D.~Tennyson,
``Brane wrapping, Alexandrov-Kontsevich-Schwarz-Zaboronsky sigma models, and QP manifolds,''
Phys. Rev. D \textbf{108} (2023) no.8, 086024
doi:10.1103/PhysRevD.108.086024
[arXiv:2301.02670 [hep-th]].

\bibitem{Hansen:2009zd}
M.~Hansen and T.~Strobl,
``First Class Constrained Systems and Twisting of Courant Algebroids by a Closed 4-form,''
doi:10.1142/9789814277839\_0008
[arXiv:0904.0711 [hep-th]].

\bibitem{Klimcik:2001vg}
C.~Klimcik and T.~Strobl,
``WZW - Poisson manifolds,''
J. Geom. Phys. \textbf{43} (2002), 341-344
doi:10.1016/S0393-0440(02)00027-X
[arXiv:math/0104189 [math.SG]].

\bibitem{Vaisman:2004msa}
I.~Vaisman,
``Transitive Courant algebroids,''
Int. J. Math. Math. Sci. \textbf{2005} (2005), 1737-1758
doi:10.1155/IJMMS.2005.1737
[arXiv:math/0407399 [math.DG]].

\bibitem{Batalin:1981jr}
I.~A.~Batalin and G.~A.~Vilkovisky,
``Gauge Algebra and Quantization,''
Phys. Lett. B \textbf{102} (1981), 27-31
doi:10.1016/0370-2693(81)90205-7

\bibitem{HTbook}
M.~Henneaux and C.~Teitelboim,
``Quantization of Gauge Systems,''
Princeton University Press, 1994,
ISBN 978-0-691-03769-1, 978-0-691-21386-6

\bibitem{Barnich:2000zw}
G.~Barnich, F.~Brandt and M.~Henneaux,
``Local BRST cohomology in gauge theories,''
Phys. Rept. \textbf{338} (2000), 439-569
doi:10.1016/S0370-1573(00)00049-1
[arXiv:hep-th/0002245 [hep-th]].


\bibitem{Alexandrov:1995kv}
M.~Alexandrov, A.~Schwarz, O.~Zaboronsky and M.~Kontsevich,
``The Geometry of the master equation and topological quantum field theory,''
Int. J. Mod. Phys. A \textbf{12} (1997), 1405-1429
doi:10.1142/S0217751X97001031
[arXiv:hep-th/9502010 [hep-th]].

\bibitem{Witten:1990wb}
E.~Witten,
``A Note on the Antibracket Formalism,''
Mod. Phys. Lett. A \textbf{5} (1990), 487
doi:10.1142/S0217732390000561


\bibitem{Schwarz:1992nx}
A.~S.~Schwarz,
``Geometry of Batalin-Vilkovisky quantization,''
Commun. Math. Phys. \textbf{155} (1993), 249-260
doi:10.1007/BF02097392
[arXiv:hep-th/9205088 [hep-th]].




\bibitem{Ikeda:2019czt}
N.~Ikeda and T.~Strobl,
``BV and BFV for the H-twisted Poisson sigma model,''
Annales Henri Poincare \textbf{22} (2021) no.4, 1267-1316
doi:10.1007/s00023-020-00988-0
[arXiv:1912.13511 [hep-th]].

\bibitem{Chatzistavrakidis:2022wdd}
A.~Chatzistavrakidis, L.~Jonke, T.~Strobl and G.~\v{S}imuni\'c,
``Topological Dirac sigma models and the classical master equation,''
J. Phys. A \textbf{56} (2023) no.1, 015402
doi:10.1088/1751-8121/acb09a
[arXiv:2206.14258 [hep-th]].

\bibitem{ChSI}
A.~Chatzistavrakidis, N.~Ikeda and G.~\v{S}imuni\'c,
``The BV action of 3D twisted R-Poisson sigma models,''
JHEP \textbf{10} (2022), 002
doi:10.1007/JHEP10(2022)002
[arXiv:2206.03683 [hep-th]].

\bibitem{Jotz}
M. Jotz Lean, R. A. Mehta, T. Papantonis, 
``Modules and representations up to homotopy of Lie n-algebroids,''
Journal of Homotopy and Related Structures (2023), arXiv:2001.01101

\bibitem{Gualtieri:2007bq}
M.~Gualtieri,
``Branes on Poisson varieties,''
in Oscar Garcia-Prada, Jean Pierre Bourguignon, and Simon Salamon (eds), The Many Facets of Geometry: A Tribute to Nigel Hitchin (Oxford, 2010), https://doi.org/10.1093/acprof:oso/9780199534920.003.0018
[arXiv:0710.2719 [math.DG]].

\bibitem{Chatzistavrakidis:2023otk}
A.~Chatzistavrakidis and L.~Jonke,
``Basic curvature and the Atiyah cocycle in gauge theory,''
[arXiv:2302.04956 [hep-th]].

\bibitem{Blaom}
A.~D.~Blaom, 
``Geometric structures as deformed infinitesimal symmetries,'' Trans.~Am.~Math.~Soc.~{\bf 358} (2006) 3651.

\bibitem{Crainic} 
C.~A.~Abad and M.~Crainic,
``Representations up to homotopy of Lie algebroids,'' 
J. reine angew. Math. 663 (2012), p. 91—126,  DOI:10.1515/CRELLE.2011.095

\bibitem{CF} 
M.~Crainic and R.~
L.~Fernandes,
``Secondary Characteristic Classes of Lie Algebroids,'' 
In: Carow-Watamura, U., Maeda, Y., Watamura, S. (eds) Quantum Field Theory and Noncommutative Geometry. Lecture Notes in Physics, vol 662. Springer, Berlin, Heidelberg. https://doi.org/10.1007/113427869

\bibitem{GSM} 
A.~Gracia-Saz and R.~A.~Mehta, ``Lie algebroid structures on
double vector bundles and representation theory of Lie algebroids,'' 
Adv. Math., 223 no.4 (2010),
1236–1275 
doi:10.1016/j.aim.2009.09.010 
[arXiv:0810.0066 [math.DG]]

\bibitem{Cattaneo:2010re}
A.~S.~Cattaneo and F.~Schaetz,
``Introduction to supergeometry,''
Rev. Math. Phys. \textbf{23} (2011), 669-690
doi:10.1142/S0129055X11004400
[arXiv:1011.3401 [math-ph]].

\bibitem{Qiu:2011qr}
J.~Qiu and M.~Zabzine,
``Introduction to Graded Geometry, Batalin-Vilkovisky Formalism and their Applications,''
Archivum Math. \textbf{47} (2011), 143-199
[arXiv:1105.2680 [math.QA]].

\bibitem{Ikeda:2012pv}
N.~Ikeda,
``Lectures on AKSZ Sigma Models for Physicists,''
doi:10.1142/9789813144613\_0003
[arXiv:1204.3714 [hep-th]].

\bibitem{Vaintrob}
 A. Yu. Vaintrob, 
 ``Lie algebroids and homological vector fields,''
 Russ. Math. Surv. 52 428 (1997)

\bibitem{Chatzistavrakidis:2021nom}
A.~Chatzistavrakidis,
``Topological field theories induced by twisted R-Poisson structure in any dimension,''
JHEP \textbf{09}, 045 (2021)
doi:10.1007/JHEP09(2021)045
[arXiv:2106.01067 [hep-th]].

\bibitem{Severa:2001qm}
P.~\v{S}evera and A.~Weinstein,
``Poisson geometry with a 3 form background,''
Prog. Theor. Phys. Suppl. \textbf{144} (2001), 145-154
doi:10.1143/PTPS.144.145
[arXiv:math/0107133 [math.SG]].

\bibitem{BG} 
A.~J.~Bruce, J.~Grabowski,
``Pre-Courant algebroids,''
Journal of Geometry and Physics
Volume 142 (2019) Pages 254-273


\bibitem{preCA} 
Zhangju Liu, Yunhe Sheng, Xiaomeng Xu,
``The Pontryagin class for pre-Courant algebroids,''
Journal of Geometry and Physics,
Volume 104,
2016,
Pages 148-162,
ISSN 0393-0440,
https://doi.org/10.1016/j.geomphys.2016.02.007.

\bibitem{shifted1} 
T.~Pantev, B.~To\"en, M.~Vaqui\'e, G.~Vezzosi, 
``Shifted symplectic structures,'' Publ. Math. IHES 117:1 (2013) 271–328 [arXiv:1111.3209 [math.AG]]

\bibitem{shifted2} 
B.~Pym, P.~Safronov, 
``Shifted Symplectic Lie Algebroids,'' 
International Mathematics Research Notices, Volume 2020, Issue 21, November 2020, Pages 7489–7557, https://doi.org/10.1093/imrn/rny215

\bibitem{Grutzmann:2014hkn}
M.~Gr\"utzmann and T.~Strobl,
``General Yang\textendash{}Mills type gauge theories for $p$-form gauge fields: From physics-based ideas to a mathematical framework or From Bianchi identities to twisted Courant algebroids,''
Int. J. Geom. Meth. Mod. Phys. \textbf{12} (2014), 1550009
doi:10.1142/S0219887815500097
[arXiv:1407.6759 [hep-th]].

\bibitem{Liu:1995lsa}
Z.~Liu, A.~Weinstein and P.~Xu,
``Manin Triples for Lie Bialgebroids,''
J. Diff. Geom. \textbf{45} (1997) no.3, 547-574
[arXiv:dg-ga/9508013 [math.DG]].

\bibitem{Boffo:2019zus}
E.~Boffo and P.~Schupp,
``Deformed graded Poisson structures, Generalized Geometry and Supergravity,''
JHEP \textbf{01} (2020), 007
doi:10.1007/JHEP01(2020)007
[arXiv:1903.09112 [hep-th]].

\bibitem{Kotov:2016lpx}
A.~Kotov and T.~Strobl,
``Lie algebroids, gauge theories, and compatible geometrical structures,''
Rev. Math. Phys. \textbf{31}, no.04, 1950015 (2018)
doi:10.1142/S0129055X19500156
[arXiv:1603.04490 [math.DG]].

\bibitem{GJ}
C.~J.~Grewcoe and L.~Jonke,
``Courant Sigma Model and $L_\infty$-algebras,''
Fortsch. Phys. \textbf{68} (2020) no.6, 2000021
doi:10.1002/prop.202000021
[arXiv:2001.11745 [hep-th]].

\bibitem{RW} 
D.~Roytenberg, A.~Weinstein, 
``Courant Algebroids and Strongly Homotopy Lie Algebras,'' Letters in Mathematical Physics 46, 81–93 (1998). https://doi.org/10.1023/A:1007452512084

\bibitem{Hohm:2017pnh}
O.~Hohm and B.~Zwiebach,
``$L_{\infty}$ Algebras and Field Theory,''
Fortsch. Phys. \textbf{65} (2017) no.3-4, 1700014
doi:10.1002/prop.201700014
[arXiv:1701.08824 [hep-th]].

\bibitem{Grigoriev:2023lcc}
M.~Grigoriev and D.~Rudinsky,
``Notes on the L\ensuremath{\infty}-approach to local gauge field theories,''
J. Geom. Phys. \textbf{190} (2023), 104863
doi:10.1016/j.geomphys.2023.104863
[arXiv:2303.08990 [hep-th]].

\bibitem{Chatzistavrakidis:2018ztm}
A.~Chatzistavrakidis, L.~Jonke, F.~S.~Khoo and R.~J.~Szabo,
``Double Field Theory and Membrane Sigma-Models,''
JHEP \textbf{07} (2018), 015
doi:10.1007/JHEP07(2018)015
[arXiv:1802.07003 [hep-th]].

\bibitem{Figueroa-OFarrill:2005vws}
J.~M.~Figueroa-O'Farrill and N.~Mohammedi,
``Gauging the Wess-Zumino term of a sigma model with boundary,''
JHEP \textbf{08} (2005), 086
doi:10.1088/1126-6708/2005/08/086
[arXiv:hep-th/0506049 [hep-th]].

\bibitem{Ikeda:2021rir}
N.~Ikeda,
``Higher Dimensional Lie Algebroid Sigma Model with WZ Term,''
Universe \textbf{7} (2021) no.10, 391
[arXiv:2109.02858 [hep-th]].




\bibitem{Mayer:2009wf}
C.~Mayer and T.~Strobl,
``Lie Algebroid Yang Mills with Matter Fields,''
J. Geom. Phys. \textbf{59} (2009), 1613-1623
doi:10.1016/j.geomphys.2009.07.018
[arXiv:0908.3161 [hep-th]].

\bibitem{Bojowald:2004wu}
M.~Bojowald, A.~Kotov and T.~Strobl,
``Lie algebroid morphisms, Poisson sigma models, and off-shell closed gauge symmetries,''
J. Geom. Phys. \textbf{54} (2005), 400-426
doi:10.1016/j.geomphys.2004.11.002
[arXiv:math/0406445 [math.DG]].



\bibitem{Ikeda:2020eft}
N.~Ikeda and T.~Strobl,
``From BFV to BV and spacetime covariance,''
JHEP \textbf{12} (2020), 141
doi:10.1007/JHEP12(2020)141
[arXiv:2007.15912 [hep-th]].

\bibitem{Barnich:2000me}
G.~Barnich,
``Classical and quantum aspects of the extended antifield formalism,''
Ann. U. Craiova Phys. \textbf{10} (2000) no.II, 1-93
[arXiv:hep-th/0011120 [hep-th]].

\bibitem{Ikeda:2013wh}
N.~Ikeda and X.~Xu,
``Canonical functions, differential graded symplectic pairs in supergeometry, and Alexandrov-Kontsevich-Schwartz-Zaboronsky sigma models with boundaries,''
J. Math. Phys. \textbf{55} (2014), 113505
doi:10.1063/1.4900834
[arXiv:1301.4805 [math.SG]].

\bibitem{Cattaneo:1999fm}
A.~S.~Cattaneo and G.~Felder,
``A Path integral approach to the Kontsevich quantization formula,''
Commun. Math. Phys. \textbf{212} (2000), 591-611
doi:10.1007/s002200000229
[arXiv:math/9902090 [math]].

\bibitem{Bonechi:2010tbl}
F.~Bonechi, P.~Mnev and M.~Zabzine,
``Finite dimensional AKSZ-BV theories,''
Lett. Math. Phys. \textbf{94} (2010), 197-228
doi:10.1007/s11005-010-0423-3
[arXiv:0903.0995 [hep-th]].

\bibitem{Gomis:1994he}
J.~Gomis, J.~Paris and S.~Samuel,
Phys. Rept. \textbf{259} (1995), 1-145
doi:10.1016/0370-1573(94)00112-G
[arXiv:hep-th/9412228 [hep-th]].

\bibitem{Ikeda:2001fq}
N.~Ikeda,
``Deformation of BF theories, topological open membrane and a
generalization of the star deformation,''
JHEP \textbf{07} (2001), 037
[arXiv:hep-th/0105286 [hep-th]].

\bibitem{Batalin:1983ggl}
I.~A.~Batalin and G.~A.~Vilkovisky,
``Quantization of Gauge Theories with Linearly Dependent Generators,''
Phys. Rev. D \textbf{28} (1983), 2567-2582
[erratum: Phys. Rev. D \textbf{30} (1984), 508]
doi:10.1103/PhysRevD.28.2567

\bibitem{Bonechi:2022aji}
F.~Bonechi, A.~S.~Cattaneo and M.~Zabzine,
``Towards equivariant Yang-Mills theory,''
J. Geom. Phys. \textbf{189} (2023), 104836
doi:10.1016/j.geomphys.2023.104836
[arXiv:2210.00372 [hep-th]].

\bibitem{Birmingham:1991ty}
D.~Birmingham, M.~Blau, M.~Rakowski and G.~Thompson,
``Topological field theory,''
Phys. Rept. \textbf{209} (1991), 129-340
doi:10.1016/0370-1573(91)90117-5


\bibitem{Grigoriev:2020xec}
M.~Grigoriev and A.~Kotov,
``Presymplectic AKSZ formulation of Einstein gravity,''
JHEP \textbf{09} (2021), 181
doi:10.1007/JHEP09(2021)181
[arXiv:2008.11690 [hep-th]].

\bibitem{Grigoriev:2022zlq}
M.~Grigoriev,
``Presymplectic gauge PDEs and Lagrangian BV formalism beyond jet-bundles,''
[arXiv:2212.11350 [math-ph]].

\end{thebibliography}
\end{document}